\documentclass[a4paper,fleqn,12pt,twoside]{article}
\usepackage{caption,sidecap}
\usepackage{graphicx}
\usepackage[figuresright]{rotating}
\usepackage[numbers,sort&compress]{natbib}
\usepackage{times,mathptmx}
\usepackage{boldgreek}
\usepackage[pubnote]{camera-ready}

\newcommand{\CCS}{CeCu$_2$Si$_2$}
\newcommand{\CCG}{CeCu$_2$Ge$_2$}
\newcommand{\CSG}{CeCu$_2$(Si$_{1-x}$Ge$_x$)$_2$}
\newcommand{\CNG}{CeNi$_2$Ge$_2$}
\newcommand{\CNP}{Ce(Ni$_{1-x}$Pd$_x$)$_2$Ge$_2$}

\newcommand{\CRS}{CeRh$_2$Si$_2$}
\newcommand{\CPS}{CePd$_2$Si$_2$}
\newcommand{\CRP}{Ce(Rh$_{1-x}$Pd$_{x}$)$_{2}$Si$_{2}$}

\newcommand{\CIN}{CeIn$_3$}
\newcommand{\CMI}{Ce$_{n}$M$_{m}$In$_{3n+2m}$}
\newcommand{\LRI}{LaRhIn$_{5}$}
\newcommand{\CRCI}{CeRh$_{1-x}$Co$_{x}$In$_{5}$}
\newcommand{\CRII}{CeRh$_{1-x}$Ir$_{x}$In$_{5}$}

\newcommand{\CPSi}{CePt$_3$Si}

\newcommand{\CRI}{CeRhIn$_5$}
\newcommand{\CCI}{CeCoIn$_5$}
\newcommand{\CII}{CeIrIn$_5$}

\newcommand{\CCIn}{Ce$_2$CoIn$_8$}
\newcommand{\CRIn}{Ce$_2$RhIn$_8$}
\newcommand{\CIIn}{Ce$_2$IrIn$_8$}
\newcommand{\LRIn}{La$_2$RhIn$_8$}

\newcommand{\YRS}{YbRh$_2$Si$_2$}

\newcommand{\UPD}{UPd$_2$Al$_3$}
\newcommand{\UND}{UNi$_2$Al$_3$}
\newcommand{\UBE}{UBe$_{13}$}
\newcommand{\UBT}{U$_{1-x}$Th$_{x}$Be$_{13}$}
\newcommand{\UPT}{UPt$_{3}$}
\newcommand{\URU}{URu$_{2}$Si$_{2}$}

\newcommand{\la}{\langle}
\newcommand{\ra}{\rangle}

\newcommand{\bk}{\bf k\rm}
\newcommand{\bd}{\bf d\rm}

\newcommand{\bq}{\bf q\rm}
\newcommand{\bQ}{\bf Q\rm}
\newcommand{\bv}{\bf v\rm}
\newcommand{\bu}{\bf u\rm}
\newcommand{\bH}{\bf H\rm}
\newcommand{\br}{\bf r\rm}

\renewcommand{\bm}{\bf m\rm}
\newcommand{\bg}{\bf g\rm}
\newcommand{\bx}{\bf x\rm}
\newcommand{\by}{\bf y\rm}

\newcommand{\De}{$\Delta$(\bk)}
\newcommand{\boldDelta}{{\bm \Delta}}
\newcommand{\boldchi}{{\bm \chi}}

\newcommand{\boldsigma}{{\bm \sigma}}
\newcommand{\boldepsilon}{{\bm \epsilon}}
\newcommand{\boldPhi}{{\bm \Phi}}

\author{P. Thalmeier$^1$, G. Zwicknagl$^2$, O. Stockert$^1$
G. Sparn$^1$ and F. Steglich$^1$\\[0.5cm]
$^1$Max-Planck-Institut f\"ur Chemische Physik fester Stoffe,\\
01187 Dresden, GERMANY\\[0.25cm]
$^2$Institut f\"ur Mathematische Physik,\\
Technische Universit\"at Braunschweig,\\
38106 Braunschweig, GERMANY}

\title{Superconductivity in Heavy Fermion Compounds}

\begin{document}
\maketitle


\section{Introduction}
\label{sect:introduction}

The heavy electron state in intermetallic lanthanide and actinide
compounds has its origin in the interplay of strong Coulomb repulsion
in the 4f- and  5f- shells and their hybridisation with conduction band
states. Heavy quasiparticles have mostly been observed in Ce- and
U-based intermetallics which are therefore at the focus of this
review. Ideally the ground state in these strongly correlated electron
compounds may be described as a Landau Fermi liquid (LFL) state with
large enhancement of the effective mass and associated large values of
the linear specific heat coeffient $\gamma$, Pauli susceptibility
$\chi_0$ and A-coefficient of the resistivity. It has recently become
clear that the microscopic origin of mass enhancement is quite
different in Ce- and U-compounds, as described by the Kondo lattice model and
dual model respectively.

The renormalised LFL state is however 
prone to instabilities at low temperatures due to residual heavy 
quasiparticle interactions and the sharpness of the Fermi
distribution. The most common instabilities are the spontaneous
appearance of (spin-)density wave (SDW) and superconducting (SC) order
parameters which break at least some of the underlying spatial or
internal symmetries like time reversal or gauge symmetry. Via their
associated gap functions both type of order parameters are reflected
in the modified energies and densities of low lying excitations. This
also changes low temperature properties in the ordered state and may
in turn be used as a means to obtain information on the symmetry class
of the order parameter. Both SC and SDW gap functions \De~may be
conventional or unconventional depending on whether their
\bk-dependence has the same or a lower spatial symmetry as the Fermi
surface. In the latter case particles (SC) or particles and holes
(SDW) pair preferentially at neighboring sites and avoid the strong
on-site repulsion of quasiparticles. For this reason
the SC states in Ce- and U-based HF compounds are usually of the
unconventional type. Their intriguing properties have lead to a rich
and flourishing field of reserarch, not only in the genuine HF
compounds but also in oxides, ruthenates and organic solids.

The stability of the ordered phase may be influenced
by changing microscopic control parameters via application of pressure
or by chemical substitution. In this way the SDW state may be tuned to
a magnetic quantum critical point (QCP) where the staggered moment
vanishes. On theoretical grounds it has long been suspected that
unconventional SC pairing is favored by magnetic instabilities, triplet
pairing in the ferromagnetic and singlet pairing in the
antiferromagnetic case. This correlation, however, cannot always be upheld
for real HF materials.

The accumulated evidence in the Ce122 (e.g. \CCS), Ce115 (e.g. \CCI)
and Ce218 (e.g. \CCIn) classes of superconductors seems to vindicate
this picture,
since the SC appears mostly in small 'domes' around the QCP. The
connection between critical spin fluctuation properties and the
stability of SC pair states has been analysed within strong coupling
Eliashberg type theories.
The QCP has additional implications beyond SC. In the normal state
above the SC dome thermodynamic and transport behaviour show 
distinctly anomalous non-Fermi liquid (NFL) behaviour as function of
temperature and field. This is thought to be the result of a dressing
of quasiparticles with soft spin fluctuations leading to entirely
different scaling exponents for specific heat, resistivity etc. as
compared to the LFL state. This scenario is accepted for most NFL
anomalies in Ce-HF compounds.  Alternatively they may be caused by the
existence of a 'pseudo-gap' associated with a 'hidden order'
parameter, e.g. an unconventional SDW. This may indeed play a role in
\CCI~and \CII~in a similar way as invoked for the pseudo-gap phase of
underdoped cuprates. 

In U-based HF superconductors quantum critical behaviour does not play
an important role with the possible exception of \UBE. Instead of
appearing close to the destruction of SDW order, SC in U-HF compounds
is mostly embedded within a stable AF phase of reduced (sometimes very
small) ordered moments. Like the different origin of mass enhancement
this may be connected with the multi-orbital structure and multiple
occupation of 5f shells. Instead of the Kondo lattice picture, a dual
model with partly localised and partly itinerant electrons caused by a
strongly orbital dependent effective hybridisation is more
appropriate for U-HF compounds. In turn this suggests a new SC
mechanism: Pair formation is caused by the exchange of magnetic
excitons which are CEF excitations of the localised
5f-electrons that have acquired dispersion due to
intersite-exchange. Contrary to the critical spin fluctuations in
Ce-compounds these propagating bosonic modes are not
overdamped. This mechanism has been vindicated in \UPD~by the
complementary results of quasiparticle tunneling and inelastic neutron
scattering (INS) experiments. An equivalent experimental support for
the spin fluctuation mechanism is still lacking.

This article reviews the present understanding of Ce-based and some of
the U-based HF superconductors, with emphasis on the former. We will
not discuss the physics of HF superconductors containing other 4f or
5f elements like Pr skutterudites or trans-uranium based superconductors.
For Ce-compounds we focus on the connection to magnetic quantum
critical behaviour and the implications of the dual
5f-electron nature are discussed extensively for U-compounds.  
In sect.~\ref{sect:QPmech} we give the theoretical foundation of the
different heavy quasiparticle origin in Ce- and U- compounds and
present a brief summary of NFL properties in the normal state close to
a QCP. In sect.~\ref{sect:cesc} we discuss the known classes of Ce-HF
superconductors with emphasis on the Ce122 compounds. We also discuss
the new non-centrosymmetric SC \CPSi. In sect.~\ref{sect:SCmech} we give a
description of the different microscopic pairing mechanisms present in
the Ce- and U- intermetallics. In sect.~\ref{sect:ursc} we discuss
only those two U-based HF superconductors where new results have been
obtained recently, namely \UBE~and \UPD. Finally
sect.~\ref{sect:summary} gives the summary. 

Of the many already
existing review articles and monographies on HF systems and their
superconducting state we mention here a small selection: The normal
state properties of HF metals are at the focus in
\cite{Fulde88,Grewe91,Kuramotobook,Hewsonbook}. General reviews on HF
superconductors are given in
\cite{Sigrist91,Thalmeier03b} and the monography \cite{Mineevbook}. The
exceptional case of the multicomponent SC \UPT~is discussed in detail in
\cite{Sauls94,Joynt02}.

\section{Theories for the normal HF state}
\label{sect:QPmech}

The theoretical understanding of superconductivity and magnetism in
the heavy fermion systems is still in the state of rather schematic
or illustrative models. A major difficulty is that the normal state
quasiparticles can sofar be described only within effective single-particle
renormalised band pictures with empirical input parameters. For some
compounds like UBe$_{13}$ and Ce- compounds close to the quantum
critical point the SC transition may even take place in a state where
the low energy excitations are not simple LFL quasiparticles but are
dressed by soft spin fluctuations. This is witnessed by the observation of non-Fermi
liquid (NFL) behaviour in thermodynamic and transport quantities. A
fully microscopic description in the case
where inter-site effects become important is not available. Typical
lattice effects are the formation of coherent heavy
quasiparticle bands whose Fermi surfaces were observed experimentally.
We first discuss the renormalised band theory which provides a way
to describe the coherent heavy quasiparticle bands within a Fermi
liquid approach. The latter can be calculated from single-particle
Hamiltonians where the effective potential
is constructed to account for many-body effects. The residual
interaction among the quasiparticles eventually leads to the instability
of the normal Fermi liquid phase.

\subsection{Renormalised band theory for HF C\lowercase{e}-compounds}
\label{sect:RBtheory}

The Landau theory assumes that there exists a one-to-one correspondence
between the states of the complex interacting system and those of
a gas of independent fermions,  which may move in an external potential
\cite{Landau56,Landau57,Landau58}, \cite{Abrikosov75}. 
The single-particle orbitals and energies are
determined from an effective Hamiltonian. The characteristic properties
of a system are reflected in an effective and not necessarily local
potential which describes the field of the nuclei and the
modifications arising from the presence of the other electrons. The
essential many-body aspects of the problem are then contained in the
prescription for constructing the effective potentials which have
to be determined for specific problems.

The Landau theory of Fermi liquids is a phenomenological theory. The
characteristic properties of the quasiparticles which can hardly be
calculated from microscopic theories are expressed in terms of parameters
which are determined from experiment. The quasiparticle energies in a crystal
\begin{equation}
\epsilon(\mathbf{k})=\mathbf{v}_{F}(\hat{\mathbf{k}})\cdot (\mathbf{k}-
\mathbf{k}_{F}(\hat{\mathbf{k}}))
\end{equation}
are given in terms of the (anisotropic) Fermi wave vector $\mathbf{k}_{F}$
and the Fermi velocity $\mathbf{v}_{F}$ which depend upon the direction 
${\hat{\bf k}}$. The key idea of the renormaled band method is to determine 
these quantities by computing the band structure
for a given effective potential which accounts for the many-body effects.
The periodic potential leads to multiple-scattering processes involving
scattering off the individual centers as well as the propagation between
the centers. The characteristic properties of a given material
enter through the information about single center scattering. They
can be expressed in terms of properly chosen set of phase shifts 
$\{\eta _{\nu }^{i}(\epsilon)\}$
specifying the change in phase of a wave incident on site i with energy
E and symmetry $\nu $ with respect to the scattering center. Within
the scattering formulation of the band structure problem the values
of the phase shifts at the Fermi energy $\{\eta _{\nu }^{i}(\epsilon_{F})\}$
together with their derivatives $\left\{ \left(d\eta _{\nu
}^{i}/d\epsilon\right)_{\epsilon_{F}}\right\} $
determine the Fermi wave vectors ${\textbf k}_{F}$ and the Fermi
velocity ${\textbf v}_{F}$. The scattering formulation of the effective
band structure problem provides a highly efficient parametrisation
scheme for the quasiparticles. To further reduce the number of
phenomenological parameters we refer to the microscopic model for
Ce-based HF compounds.

The similarities in the behavior of Ce-based heavy-fermion systems
to that of dilute magnetic alloys have led to the assumption that
these systems are ''Kondo lattices'' where the observed anomalous
behavior can be explained in terms of periodically repeated resonant
Kondo scattering. This ansatz provides a microscopic model for the
formation of a singlet groundstate and the existence of heavy quasiparticles.
An extensive discussion is given in \cite{Hewsonbook}. Direct evidence
for the Kondo scenario comes from photoelectron spectroscopy. The
characteristic features of a Kondo system can be summared as follows
\cite{Allen92,Malterre96}: At high temperatures, the combined PES/BIS
spectra from photoemission and inverse photoemission exhibit two distinct
peaks below and above the Fermi energy. These two features correspond
to the valence transitions f$^{n}\rightarrow $ f$^{n\mp 1}$, respectively.
The changes in the occupation of the Ce 4f- shells are associated
with energies of order eV. The high-temperature state can be modelled
by weakly correlated conduction electrons which are weakly coupled
to local f-moments. The f-derived low-energy excitations are those
of a system of local moments. The direct manifestation of the low-energy
scale is the appearance of a sharp peak near the Fermi energy at low
temperatures. In Ce systems, this many-body feature known as 
{}``Abrikosov-Suhl''
or {}``Kondo'' resonance, is centered at $\epsilon_{F}$+kT$^{*}$ slightly
above the Fermi edge $\epsilon_{F}$. The {}``Kondo temperature'' 
T$^{*}$ determines the energy scale of the dynamical screening of
the impurity spin by conduction electrons \cite{Hewsonbook}. The
evolution of the Kondo
resonance with temperature was recently observed by high-resolution
photoemission experiments \cite{Reinert01}.

The resonance is a genuine many-body feature reflecting the small
admixture of f$^{0}$ configurations to the ground state and the low-lying
excitations which are mainly built from f$^{1}$-configurations. At
sufficiently low temperatures T $\ll $ T$^{*}$, the contribution of
the narrow resonance peak to the thermodynamic and transport properties
can be described in terms of a Landau theory with heavy fermionic
quasiparticles \cite{Wilson75}. Based on the corresponding effective
Hamiltonian Nozi{\`e}res \cite{Nozieres74} introduced a resonant
phase shift to account for the impurity contribution to the low-energy
properties.

The novel feature observed in stoichiometric Ce-compounds is the formation
of narrow coherent bands of low-energy excitations.
The heavy fermions arise from a decoherence-coherence
crossover occurring at low temperatures.

The calculation of realistic quasiparticle bands in Ce-based Heavy
Fermion compounds proceeds in several steps. For a detailed description see
\cite{Zwicknagl92}. The first step is a standard LDA band structure
calculation by means of which the effective
single-particle potentials are self-consistently generated. The calculation
starts, like any other ab-initio calculation, from atomic potentials
and structure information. In this step, no adjustable parameters
are introduced. The effective potentials and hence the phase shifts
of the conduction states are determined from first principles to the
same level as in the case of ``ordinary'' metals.The f-phase shifts
at the lanthanide and actinide sites, on the other hand, are described
by an empirical resonance type expression
\begin{equation}
\eta _{\hat{f}}\simeq \tan^{-1}
\frac{\Gamma_{\hat{f}}}{\epsilon-\epsilon_{\hat{f}}}
\label{eq:efren}
\end{equation}
charactered by the position $\epsilon_{\hat{f}}$ and the
width $\Gamma_{\hat{f}}$
of the many-body resonance. One of these free parameters is eliminated
by imposing the condition that
the charge distribution is not significantly altered as compared to
the LDA calculation by introducing the renormalation. The renormalised
band method devised to calculate the quasiparticles in heavy-fermion
compounds thus is essentially a one-parameter theory. We mention that
spin-orbit and CEF splittings can be accounted for in a straight-forward
manner \cite{Zwicknagl93}.

The strong local correlations in Kondo lattices lead to
an observable many-body effect, i. e., the change with temperature
of the volume of the Fermi surface.
At high temperatures, the $f$-degrees of freedom appear as localised
magnetic moments, and the Fermi surface contains only the itinerant
conduction electrons. At low temperatures, however, the $f$ degrees
of freedom are now tied into itinerant fermionic quasiparticle excitations
and accordingly, have to be included in the Fermi volume following
Luttinger's theorem. Consequently the Fermi surface is strongly modified.
This scenario \cite{Zwicknagl93a} was confirmed experimentally by
measurements of the de Haas-van Alphen (dHvA) effect
\cite{Lonzarich88,Aoki93,Tautz95}.

\subsection{Dual model for U-based systems}
\label{sect:DUALtheory}

The Kondo picture, however, does not apply in the case of the actinide
compounds. The difficulties with this model have been discussed in
\cite{CoxBook}. The difference between the Ce-based heavy-fermion
compounds and their U-counterparts can be seen directly from the
photoemission
spectra \cite{Allen92}. In U-based heavy-fermion compounds, the fingerprint
character of the transitions f$^{n}\rightarrow $ f$^{n\mp 1}$ is
lost. Instead of the well defined f-derived peaks familiar from
the Ce systems we encounter a rather broad f-derived feature. This
fact shows that the f-valence in the actinide heavy-fermion systems
is not close to integer value as it is the case in Ce-based HF compounds.
In fact, the f-valence of the U ions has been discussed rather
controversially.

Increasing experimental and theoretical evidence points towards a
dual nature of the $5f$-electrons in actinide-based intermetallic
compounds, with some of the $5f$ electrons being localised and others
delocalised. The dual model which assumes the coexistence of both
delocalised and localised $5f$-electrons provides a scheme for microscopic
calculations of the heavy quasiparticle bands in actinide compounds. The
concept is summarised as follows: The ``delocalised'' $5f$-states
hybridise with the conduction states and form energy bands while their
``localised'' atomic-like counterparts form multiplets to reduce
the local Coulomb repulsion. The two subsystems interact which leads
to the mass enhancement of the delocalised quasiparticles. The situation
resembles that in Pr metal where a mass enhancement of the conduction
electrons by a factor of 5 results from virtual crystal field (CEF)
excitations of localised 4$f^{2}$ electrons \cite{White81}. The
underlying hypothesis is supported by a number of experiments. For
an extensive discussion we refer to \cite{Zwicknagl03a}.
Calculations based on the dual model as described above reproduce
the dHvA data in UPt$_{3}$ \cite{Zwicknagl02} and UPd$_{2}$Al$_{3}$
\cite{Zwicknagl03}. The results of the latter calculation will be
discussed in sect.~\ref{ssect:UPd2Al3}.

The dual nature of the 5$f$ states which is found in many actinide
intermetallic compounds is a consequence of the interplay between
local Coulomb correlations  and hybridisation with the conduction
electrons. The
underlying microscopic mechanism is an area of active current research (see
\cite{Efremov04} and references therein). LDA calculations show that the effective hopping
matrix elements for different $5f$ orbitals are unequal. But it is of interest
to understand why only the largest among them is important and why
the others are suppressed.

The reason lies in the competition between anisotropic hybridisation
and angular correlations in the 5f shell which can be seen by 
exact diagonalisation of small clusters modelling the U sites in heavy fermion compounds
\cite{Efremov04}. Keeping  only the degrees of freedom of the 5f shells while accounting
for the conduction states by (effective) anisotropic
5f-intersite hopping leads to the model Hamiltonian
\begin{equation}
H=-\sum _{\langle nm\rangle ,j_{z}}\, t_{j_{z}}\left(c_{j_{z}}^{\dagger }(n)\, 
c_{j_{z}}(m)+h.c.\right)+H_{\mathrm{C}} 
\label{eq:ModelHamiltonianU}
\end{equation}
The first sum is over neighboring sites $\langle nm\rangle $.
Furthermore $c_{j_{z}}^{\dagger }(n)$ ($c_{j_{z}}(n)$), creates
(annihilates) an electron at site $n$ in the $5f\, \, j=5/2$ state
with $j_{z}=-5/2,\dots ,5/2$.  The effective hopping between sites results 
from the hybridisation of the 5$f$ states with the orbitals of the ligands and 
depends generally on the crystal structure. Rather than trying to exhaust all possible
different lattice symmetries, we shall concentrate here on the special
case that hopping conserves $j_{z}$. While this is certainly an idealisation,
it allows us to concentrate on our main topic, i.~e., a study
of the influence of atomic correlations on the renormalisation of
hybridisation matrix elements. The parameters $t_{j_{z}}(=t_{-j_{z}})$
are chosen in accordance with density-functional calculations for
bulk material which use $jj_{z}$ basis states. The local Coulomb
interactions can be written in the form
\begin{equation}
H_C=\frac{1}{2}\sum_n \sum_{j_{z1},\ldots ,j_{j4}}
U_{j_{z1}j_{z2}j_{z3}j_{z4}}c_{j_{z1}}^\dagger(n) c_{j_{z2}}^\dagger(n)  
c_{j_{z3}}(n) c_{j_{z4}}(n)
\end{equation}
where the Coulomb matrix elements $U_{j_{z1}j_{z2}j_{z3}j_{z4}}$ are expressed
in terms of the expectation values $U_J$ of the repulsion between electron pairs in
states with total angular momentum $J$.
The actual calculations $U_{J}$ values used in the calculation were determined 
from LDA wavefunctions for UPt$_{3}$
\cite{Zwicknagl02}, i.~e.,~$U_{J=4}$ = 17.21 eV, $U_{J=2}$ =
18.28 eV, and $U_{J=0}$ = 21.00 eV. We expect $U_{J=4}<U_{J=2}<U_{J=0}$
always to hold for Coulomb interactions, independently of the chemical
environment. In contrast, the relative order of the hopping matrix
elements will vary strongly from one compound to the next. The average
Coulomb repulsion is irrelevant for the low-energy physics of the
model. It simply restricts the relevant configurations to states such
that each site is occupied either by 2 or 3 f electrons. The low-energy
sector is exclusively determined by the differences of the $U_{J}$
values, which are of the order of 1eV and thus slightly larger than
typical bare $f$~- band widths. The latter are obtained, e.g., from
LDA calculations for metallic uranium compounds like UPt$_{3}$.  
To mimic the situation in the U-based heavy-fermion
compounds we consider the intermediate valence regime. 
%
\begin{figure}
\includegraphics[width=6.0cm,angle=0,clip]{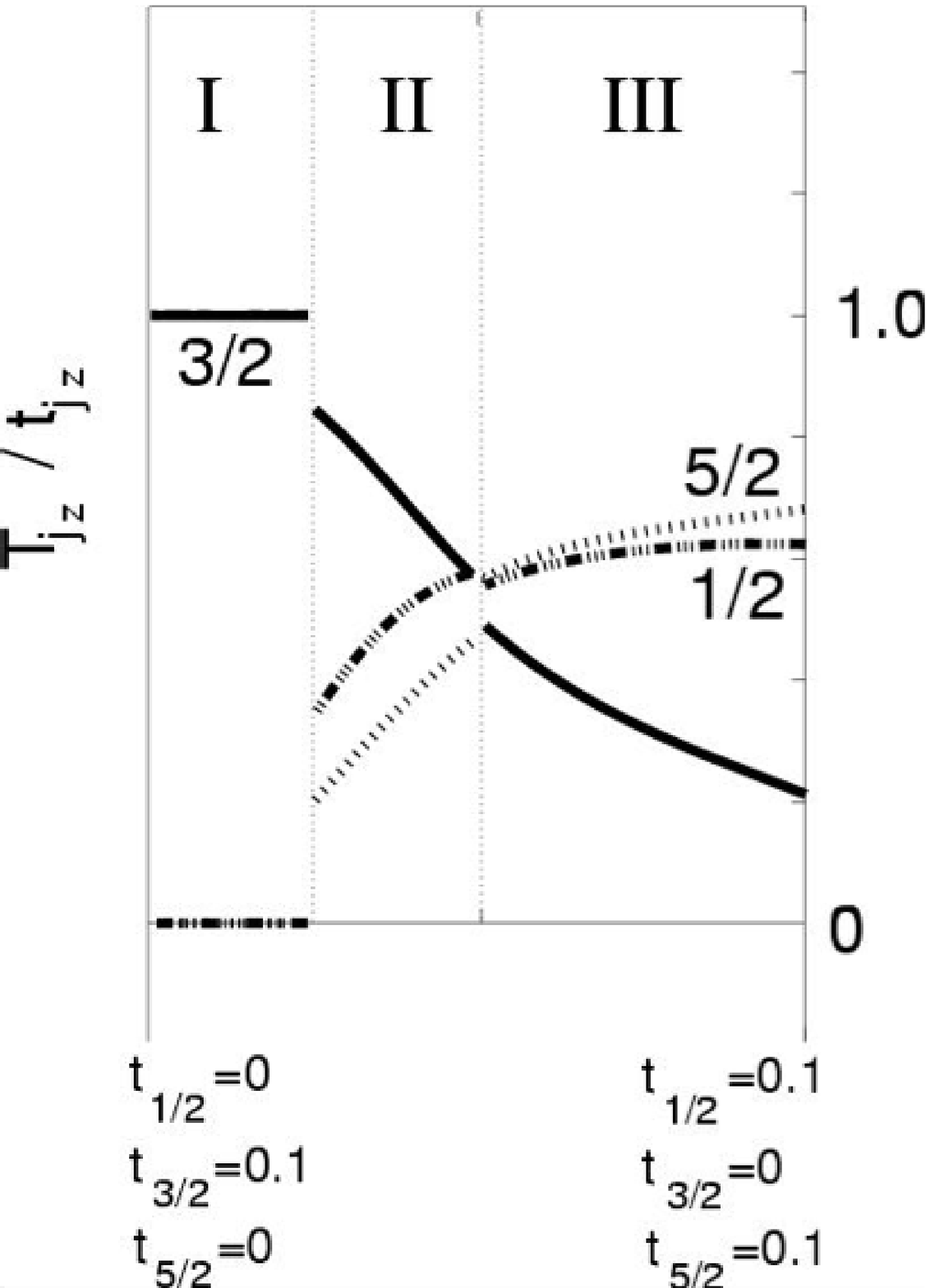}\hfill
\raisebox{1cm}
{\includegraphics[width=8cm,height=7.8cm,angle=0,clip]{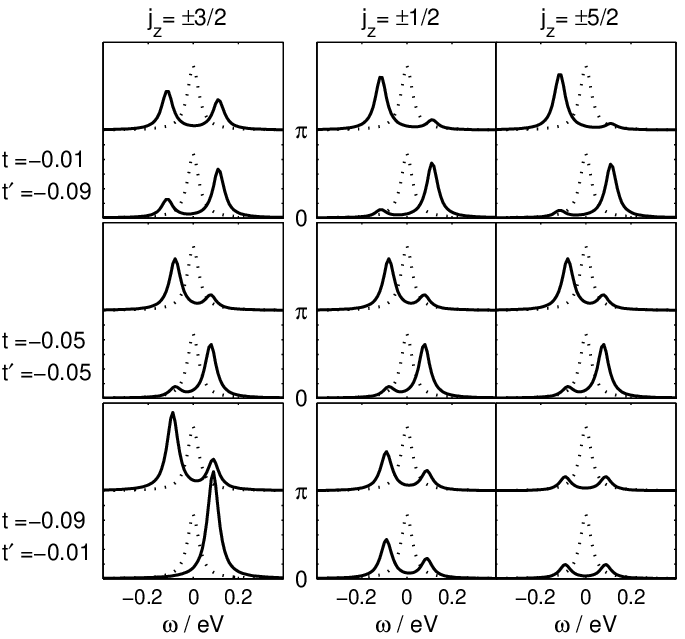}}
\caption{Left panel: Enhancement of hopping anisotropies due to
intra-atomic correlations in a two-site cluster.
The orbital-projected expectation value of the kinetic energy $T_{j_{z}}/t_{j_{z}}$ 
along the lines connecting linearly the points is written below the figure. 
Regions with $\mathcal{J}_{z}=$15/2, 5/2 and 3/2, are labeled I, II and III respectively. 
Right panel: Dispersion of the low-energy peak in the spectral functions 
($A_{j_{z}}(\mathbf{k},\omega )$ + $A_{-j_{z}}(\mathbf{k},\omega )$) for strongly anisotropic systems 
$\left| t_{3/2}\right|=\left| t\right| {\ll \atop \gg} \left|
t'\right|= \left| t_{1/2}\right|= \left| t_{5/2}\right|$)
as well as in the isotropic limit
($\left| t_{3/2}\right|=\left| t\right| = \left| t'\right|= \left| t_{1/2}\right|= \left| t_{5/2}\right|$). For 
each parameter set of hopping parameters, the spectral functions for k=0 and k=$\pi$ are compared.
The dotted curve denotes the isotropic peak in the atomic limit. }
\label{fig:SpectralFunctionTwoSiteCluster}
\end{figure}
%
The Hamiltonian eq.~(\ref{eq:ModelHamiltonianU}) conserves $\mathcal{J}_{z}=\sum _{n}J_{z}(n)$
where $\mathcal{J}_{z}$ is the z-component of the total angular momentum
of the system and the $J_{z}(n)$ refer to angular momentum projections
on individual sites. We shall therefore characterise the eigenstates
by their $\mathcal{J}_{z}$ value.  Strong on-site correlations result
in a considerable enhancement of anisotropies in the bare hopping
matrix elements. This can lead to a localisation of electrons in orbitals
with relatively weak hybridisation. The latter is effectively reduced
to zero in those cases.

The degree of localisation or, alternatively, of the reduction of hopping of a given 
$j_{z}$ orbital by local correlations, is quantified by the ratio of the $j_{z}$- projected 
kinetic energy $T_{j_{z}}$ and the bare matrix element $t_{j_{z}}$ 
\begin{equation}
\frac{T_{j_{z}}}{t_{j_{z}}}=
\sum _{\langle nm\rangle ,\pm }\langle \Psi _{\mathrm{g}s}|
(c_{\pm j_{z}}^{\dagger }(n)\, c_{\pm j_{z}}(m)+h.c.)|\Psi _{\mathrm{g}s}\rangle \, .
\end{equation}
The ground-state wavefunction $|\Psi _{\mathrm{g}s}\rangle $ contains
the strong on-site correlations. A small ratio of $T_{j_{z}}/t_{j_{z}}$
indicates partial suppression of hopping for electrons in the $\pm j_{z}$
orbitals. Two kinds of correlations may contribute to that process.
The first one is based on the reduction of charge fluctuations to
atomic $f^{2}$ and $f^{3}$ configurations. This is a result for
large values of $U_{J}$ and can be studied by setting all $U_{J}$
equal to a value much larger than the different $t_{j_{z}}$. The
second one is due to differences in the $U_{J}$ values, i.~e.~,
$U_{J=4}<U_{J=2}<U_{J=0}$. The differences in the $U_{J}$ values
are the basis of Hund's rules. Hopping counteracts Hund's rule correlations
and vice versa. What we want to stress is the fact that those correlations
can lead to a complete suppression of hopping channels except for
the dominant one which shows only little influence.

Results for the ratios $T_{j_{z}}/t_{j_{z}}$ are shown in 
fig.~\ref{fig:SpectralFunctionTwoSiteCluster}
for a two-site model \cite{Efremov04}. As the relevant correlations are local the general results
qualitatively agree with those found for a three-site cluster and
four-site clusters \cite{PollmannDiss}. We can distinguish three different
regimes with $\mathcal{J}_{z}=$ 15/2, 5/2 and 3/2, labeled I, II and
III respectively. One observes that in region I only the dominant
hybridisation of the $j_{z}=3/2$ orbital survives while that of the
$j_{z}=1/2$ and $j_{z}=5/2$ orbitals is completely suppressed. On
the other hand in regions II and III the correlation effects on different
orbitals are are not very different. These findings demonstrate that
in particular Hund's rule correlations strongly enhance anisotropies
in the hopping. For a certain range of parameters this may result
in a complete suppression of the effective hopping except for the
largest one, which remains almost unaffected. This provides a microscopic
justification of partial localisation of 5$f$ electrons which is
observed in a number of experiments on U compounds and which is the
basis for further model calculations described later on.

Many aspects of partial localisation in the ground state are described appropriately by 
both a Gutzwiller type variational wave function and by a treatment which keeps only those 
interactions which are present in LDA+U calculations \cite{Runge04}. This finding 
should encourage further applications of LDA+U and related approaches as SIC-LDA for 
ground-state properties of actinide heavy-fermion materials. The subtle angular correlations which 
determine the magnetic character are accounted for only in limiting cases. In addition, we cannot 
expect these schemes to reproduce the small energy scales responsible for the heavy fermion 
character of the low-energy excitations. 
The dual nature of the 5$f$ states is manifest in the spectral functions 
$A_{j_{z}}(\mathbf{k},\omega )$ and the density of states (DOS)
$N(\omega)$ given by
\begin{equation}
N(\omega)=\sum_{\bf k}\;\sum_{j_z}A_{j_{z}}(\mathbf{k},\omega ) \quad .
\end{equation}
The spectral functions which are defined by
\begin{eqnarray}
A_{j_{z}}(\mathbf{k},\omega ) & = & 
\sum _{n}\left|\left\langle \psi _{n}^{N+1}\left|c_{j_{z}}^{\dagger }(\mathbf{k})\right|\psi _{0}^{N}
\right\rangle \right|^{2}\delta \left(\omega
-\left(E_{n}^{N+1}-E_{0}^{N}\right)\right)+ \nonumber\\
 &  & +\sum _{n}\left|\left\langle \psi _{n}^{N-1}\left|c_{j_{z}}(\mathbf{k})\right|\psi _{0}^{N}\right
\rangle \right|^{2}\delta \left(\omega +\left(E_{n}^{N-1}-E_{0}^{N}\right)\right)
\end{eqnarray}
measure the weights for removing an electrons in state $({\bf k}j_z)$ from the ground state 
(negative energies) or adding on to it (positive energies). Here $\left|\psi _{0}^{N}\right.\rangle$
denotes the exact ground state of an $N$-particle system with ground state energy $E_0^{(N)}$ 
while $\left|\psi _{n}^{N+1}\right.\rangle$ and
$\left|\psi _{n}^{N-1}\right.\rangle$ are complete sets of eigenstates of the $N+1$- and
$N-1$-particle systems with energies $E_n^{(N+1)}$ and  $E_n^{(N-1)}$, respectively. 
Technically, the spectral functions are obtained from the one-particle
Green's functions according to
\begin{equation}
A_{j_{z}}(\mathbf{k},\omega )=-\frac{1}{\pi }Im G_{j_{z}}(\mathbf{k},\omega )
\quad 
\end{equation}
which are approximately calculated for a cluster following \cite{Dagotto94}. As a first step towards a 
microscopic treatment of partial localisation in extended systems the one-particle Green's functions
are calculated within cluster perturbation theory. The method combines exact diagonalisation of 
small clusters with strong-coupling perturbation theory \cite{Senechal02}. The many-particle states of 
the clusters are determined by means of the Jacobi-Davidson method
\cite{Sleijpen99}. 
%
\begin{SCfigure}
\raisebox{-0.9cm}
{\includegraphics[width=7cm,angle=0,clip]{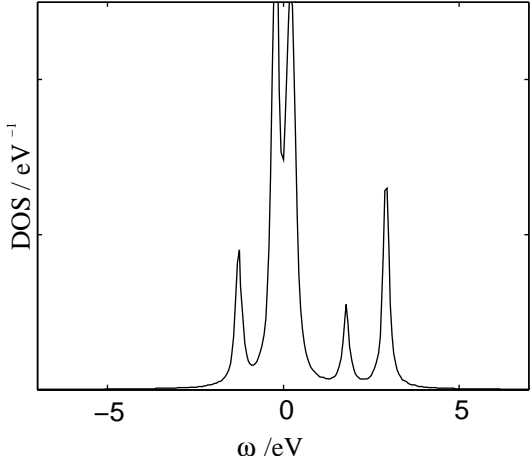}}
\caption{Density of states for a linear chain calculated within CPT. The short-range correlations are 
treated within a two-site cluster. The hopping parameters are $\left|
t_{3/2}\right|$ = 0.2 eV and 
$\left| t_{1/2}\right|=\left| t_{5/2}\right|$ = 0). The high-energy side-bands have predominantly 
j$_z$ = $\pm$5/2 character.}
\label{fig:DOSCPTTwoSite}
\end{SCfigure}
%
The formation of the heavy quasiparticles, i. e., the enhancement of the effective mass over the bare 
band mass is related to the transfer of spectral weight from low energies to high-energy satellites. 
Different kinds of processes may contribute to this phenomenon. First, there is the suppression of 
charge fluctuations which effectively restricts the ground state in the mixed-valent regime to 
$f^2$ and $f^3$ configurations. Neglecting angular correlations, the spectral function will have a 
low-energy peak from the transitions $f^2 \to f^3$ and $f^3 \to f^2$ whose weight will be reduced 
due to the atomic valence transitions $f^{2}\rightarrow f^{1}$ and $f^{3}\rightarrow f^{4}$. These 
transition appear in the spectrum at energies well separated from the Fermi level. Second, there are 
the angular correlations as described by Hund's rules. They transfer spectral weight to atomic side 
bands which appear at energies $\sim 1-2$ eV below (or above) the Fermi edge. These structures 
can be seen from the DOS displayed in fig.~\ref{fig:DOSCPTTwoSite}. 
Experimentally, the ``Hund's rule structures'' are observed as ``humped features'' in photoemission
experiments \cite{Fujimori99}.  Due to the atomic transitions the overall width is much 
larger than the value predicted by standard band structure calculations.

Suppression of charge fluctuations and angular correlations reduce the weight of the low-energy
peak to 0.41. This
reduction corresponds to an isotropic enhancement of the quasiparticle
mass by $\sim 2.5$. The quasiparticle weight is further reduced by
hopping between the sites.  The latter introduces dispersion effects and
orbital-dependence as can be seen from
fig.~\ref{fig:SpectralFunctionTwoSiteCluster} (right panel).
The spectral weight of the isotropic ``atomic'' peak is distributed
over a rather broad energy range. The strong local correlations result
in a considerable enhancement of anisotropies in the bare hopping
matrix elements. The $j_{z}$ channels with the dominant hopping exhibit
well-defined dispersive quasiparticle peaks in the low-energy regime
whereas the remaining channels contribute a non-dispersive incoherent
background. The results are consistent with previous results the
orbital-projected expectation values of the kinetic energy.

\subsection{The NFL state and its scaling laws}
\label{sect:NFLtheory}

In the vicinity of a second order phase transition the behavior of
various physical properties exhibit singularities which can be characterised
by critical exponents \cite{MaBook}. The latter are universal in the
sense that they do not depend upon the detailed microscopic nature
of the system. They are determined by the dimensionality of the system
and the degrees of freedom associated with the long-range correlations
in the ordered phase. As a consequence, the critical exponents have
been successfully used to classify the critical behavior in classical
systems. The universality reflects the fact that the characteristic
length and time scales, i.e., the correlation length $\xi $ and the
relaxation time $\tau $ diverge like
\begin{equation}
\xi \sim \left|g\right|^{-\nu }\textrm{ and }\tau \sim
\left|g\right|^{-\nu z}\textrm{ }
\label{NFLcorr}
\end{equation}
for $\left|g\right|\to 0$ (see e. g. \cite{SachdevBook, ContinentinoBook}) where
$g$ = t, r measures the distance to the critical point according to
\begin{equation}
t=\frac{T-T_{c}}{T_{c}},\qquad  r=\frac{p-p_{c}}{p_{c}}\qquad \textrm{ or } 
\qquad  r=\frac{H-H_{c}}{H_{c}}
\end{equation}
Here $T$, $p$ and $H$ are temperature, pressure and field while
$T_{c}$, $p_{c}$ and $H_c$ are their critical values. It is important
to note that the critical exponents are not independent. They are
related by scaling relations and, in addition, they are connected
with the dimension $d$ of the system by hyperscaling relations. 
On approaching T$_c$ from above (t$\rightarrow$ 0$^+$) the growing 
amplitude of thermally excited collective modes drives the system to 
a new state with spontaneously broken symmetry characterised by an order
parameter.

On the other hand quantum phase transitions QPT) which take place
at zero temperature (T$_c$=0) are driven by the contribution of
quantum fluctuations to the ground state energy which depends on
a control parameter r $\rightarrow$ 0$^+$ associated with pressure or field.
In this case the effective dimensionality $d_{eff}$ is given by 
\begin{equation}
d_{eff}=d+z
\end{equation}
where $z$ is the dynamical exponent introduced above. Due to the
effective increase in dimensionality the latter may reach the upper
critical dimension. As a consequence, the critical exponents associated
with the QPT may assume mean-field values \cite{Hertz76}.
In general, the characterisation of a continuous quantum phase transition
is more subtle than that of a classical one. 

Besides specific heat, susceptibility and resistivity the Gr\"uneisen
ratio $\Gamma $ has turned out to be a sensitive quantity to
charaterise the vicinity of the QCP. Zhu et al. \cite{Zhu03} have shown that the
$\Gamma $ ratio defined in terms of the molar specific
heat $c_{p}=\frac{T}{N}\left(\partial S/\partial T\right)_{p}$
and the thermal expansion $\alpha =\frac{1}{V}\left(\partial V/\partial T\right)_{p,N}=
\frac{1}{V}\left(\partial S/\partial p\right)_{T,N}$, namely
\begin{equation}
\Gamma =\frac{\alpha }{c_{p}}=-\frac{1}{V_{m}T}
\frac{\left(\partial S/\partial p\right)_T}{\left(\partial S/\partial T\right)_p}
\label{eq:GrueneisenRatioDef}
\end{equation}
diverges at any QCP. Here $S$ is the entropy and $V_{m}=\frac{V}{N}$
is the molar volume. Away from the QCP when the low-temperature
behavior is characterised by a single energy scale $E^{*}$ 
the Gr\"uneisen ratio in eq.~(\ref{eq:GrueneisenRatioDef})
reduces to 
\begin{equation}
\Gamma\simeq\frac{1}{V_{m}E^*}\frac{\partial E^{*}}{\partial p}
\end{equation}
This form was already introduced and discussed in 
\cite{Takke81,Thalmeier86,Fulde88a,Thalmeier91a}.
Instead of pressure, an external magnetic field $H$ may be used as
control parameter. In this case, the ratio
\begin{equation}
\Gamma _{H}=-\frac{1}{c_{H}}\left(\frac{\partial M}{\partial\mathrm{T}}\right)_{H}=
-\frac{1}{T}\frac{\left(\partial S/\partial H\right)_{T}}
{\left(\partial S/\partial\mathrm{T}\right)_{H}}
=-\frac{1}{T}\left(\frac{\partial T}{\partial H}\right)_{S}
\end{equation}
can be determined directly from the magneto-caloric effect.
The Gr\"uneisen ratio $\Gamma$ and its magnetic counterpart
$\Gamma _{H}$ owe their importance to the fact
that at the critical value $r=0$ the temperature dependence is given by
\begin{equation}
\Gamma(T,r=0)\sim T^{-\frac{1}{\nu z}}
\label{eq:GrueneisenCriticalScaling}
\end{equation}
and consequently permits to measure $\nu z$.
Comparison with experiment usually requires to account for corrections
to the simple scaling ansatz from which eq.~(\ref{eq:GrueneisenCriticalScaling})
is derived. The corrections, however, have to be evaluated from a
microscopic model. For a discussion of the validity of
eq.(\ref{eq:GrueneisenCriticalScaling}) we refer to \cite{Continentino04,Zhu04}.

Neutron diffraction results for the A-phase in CeCu$_{2}$Si$_{2}$
suggest that the magnetic quantum phase transition in stoichiometric
heavy fermion compounds arises from the instability of the strongly
renormalised Fermi liquid with respect to the formation of a SDW.
Model studies for the SDW transition therefore deserve special attention.
Using the Ginzburg-Landau-Wilson functional for the effective action \cite{Hertz76}
\begin{eqnarray}
S\left[\phi\right]& =& \sum_{{\bf
q},i\omega_n}\left(r+q^2+\frac{|\omega_n|}{\Gamma_q}\right)
\left|\Phi_{{\bf q},i\omega_n}\right|^2+S^{(4)}\qquad \mbox{with}\qquad
\Gamma_{\bf q}=\Gamma_0q^{z-2} \nonumber\\
S^{(4)} & = & u \int_0^\beta\int d^d{\bf r} \left|\Phi({\bf r},\tau)\right|^4
\end{eqnarray}
\begin{SCfigure}
\raisebox{-1.1cm}
{\includegraphics[width=7cm,angle=0,clip]{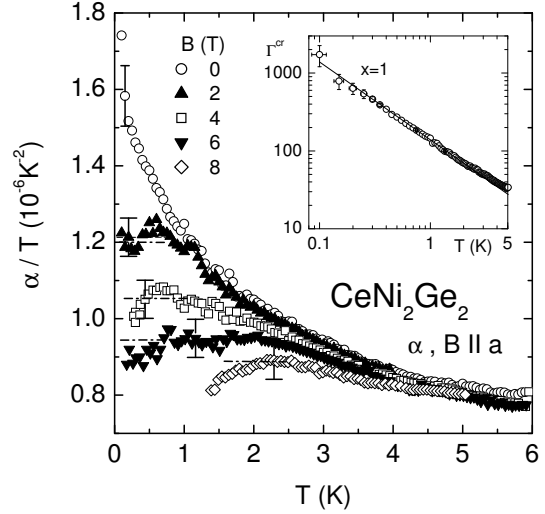}}
\caption{Thermal expansion showing the supression of NFL behaviour as
function of field. The inset shows that for B = 0 the critical
contribution to the Gr\"uneisen ratio of
\CNG~(sect.~\ref{ssect:CeNi2Ge2}) scales like $\Gamma\sim$ 1/T$^x$
with x = 1. After eq.(\ref{eq:GrueneisenCriticalScaling}) this means
(assuming z = 2 for AF SDW) a mean field correlation length exponent
$\nu$ = 1/2 which is in agreement with d$_{eff}$ = d+z = 5 for the
effective dimension \protect\cite{Kuechler03}.}
\label{fig:CeNi2Ge2Gruen}
\end{SCfigure}
the thermal expansion and Gr\"uneisen ratio were calculated
on the nonmagnetic side of the phase diagram \cite{Zhu03}. Here $\Phi$ is the order parameter that 
fluctuates in space and imaginary time, $\beta=1/kT$ is the inverse temperature and 
$\omega_n=2n\pi T$ are bosonic Matsubara frequencies. The parameter $r$ controls the 
distance from the critical point. The Landau damping which is linear
in $\left|\omega\right|$ is due to the 
scattering of quasiparticles by spin fluctuations. It is
characterised by the dynamical exponent
$z$. The theory starts on the Fermi liquid side ($r>r_c$)
which is characterised by a large Fermi surface including the f-degrees of freedom. The 
low-energy excitations are fermionic quasiparticles and their collective excitations. Close to the QCP
the static susceptibility is assumed to diverge at a specific wave vector ${\bf Q}$ which 
signifies the transition into the SDW state. The quasiparticles are strongly scattered along 
``hot lines'' which are connected by ${\bf Q}$ \cite{Rosch99}. This strong scattering modifies the 
low-temperature thermodynamic and transport properties which exhibit anomalous 
scaling relations close to QCP. They differ from those familiar from Fermi liquids. The 
scaling relations are derived adopting the renormalisation scheme of Millis 
\cite{Millis93,Zuelicke95} which proceeds in close analogy to the spin-fluctuation theory of
ferromagnetism \cite{MoriyaBook}. The scaling behavior obtained for various models and many
non-Fermi liquid compounds are summarised in \cite{Stewart01}. 

The model calculations in \cite{Zhu03} classify transitions according
to various types of  magnetic order. Two-dimensional (d=2) as well as three-dimensional (d=3) systems are considered
assuming the values $z=2,3$ for the dynamical exponents. Comparison of
calculated Gr\"uneisen ratios with experiments in CeNi$_2$Ge$_2$ \cite{Kuechler03} shows that the
QCP in this stoichiometric HF compound are consistent with the SDW scenario
for three-dimensional critical AF spin fluctuations (fig.~\ref{fig:CeNi2Ge2Gruen}).


\section{Cerium-based HF Superconductors}
\label{sect:cesc}

\begin{SCfigure}
\raisebox{-0.3cm}
{\includegraphics[angle=0,width=4.5cm]
{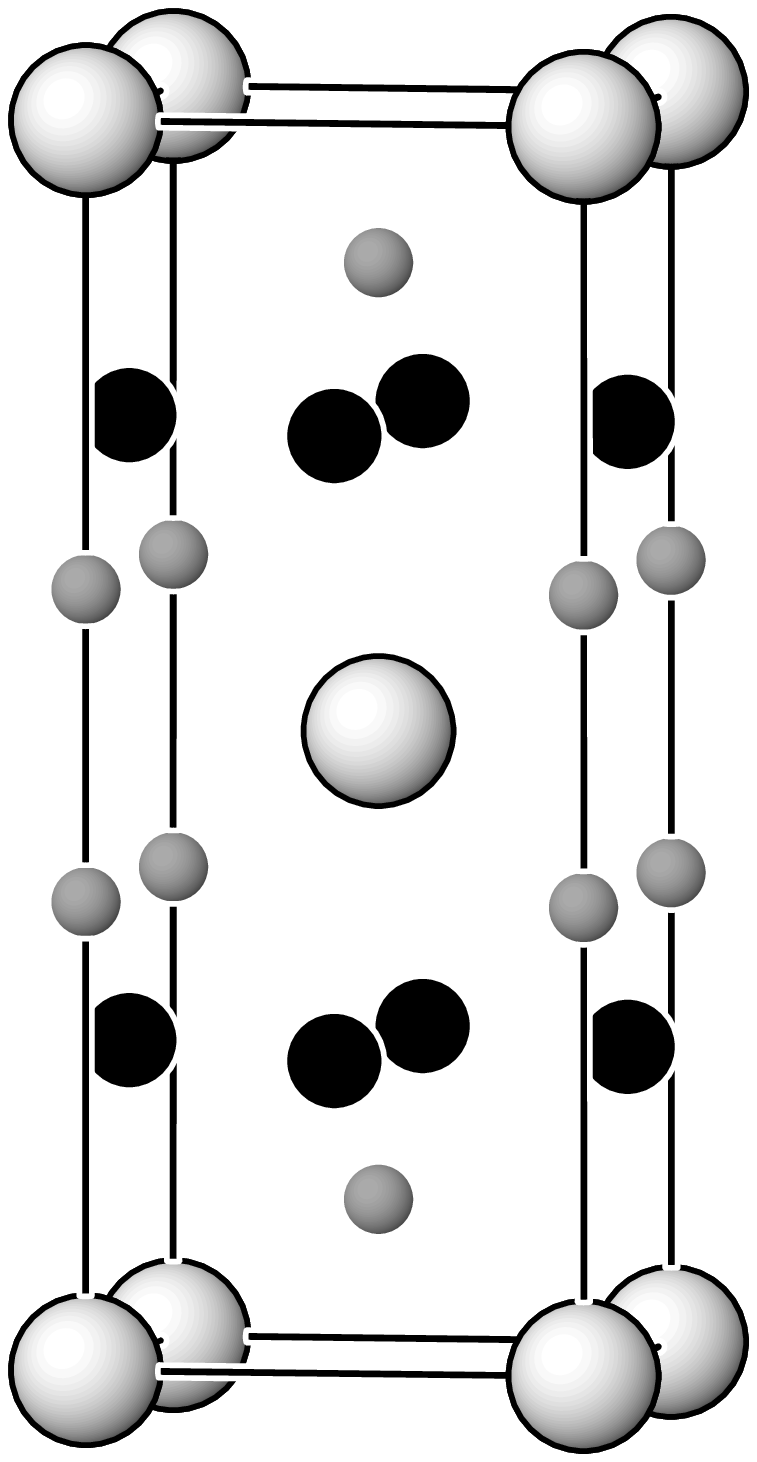}}\hfill
\raisebox{-0.3cm}
{\includegraphics[angle=0,width=4.5cm]{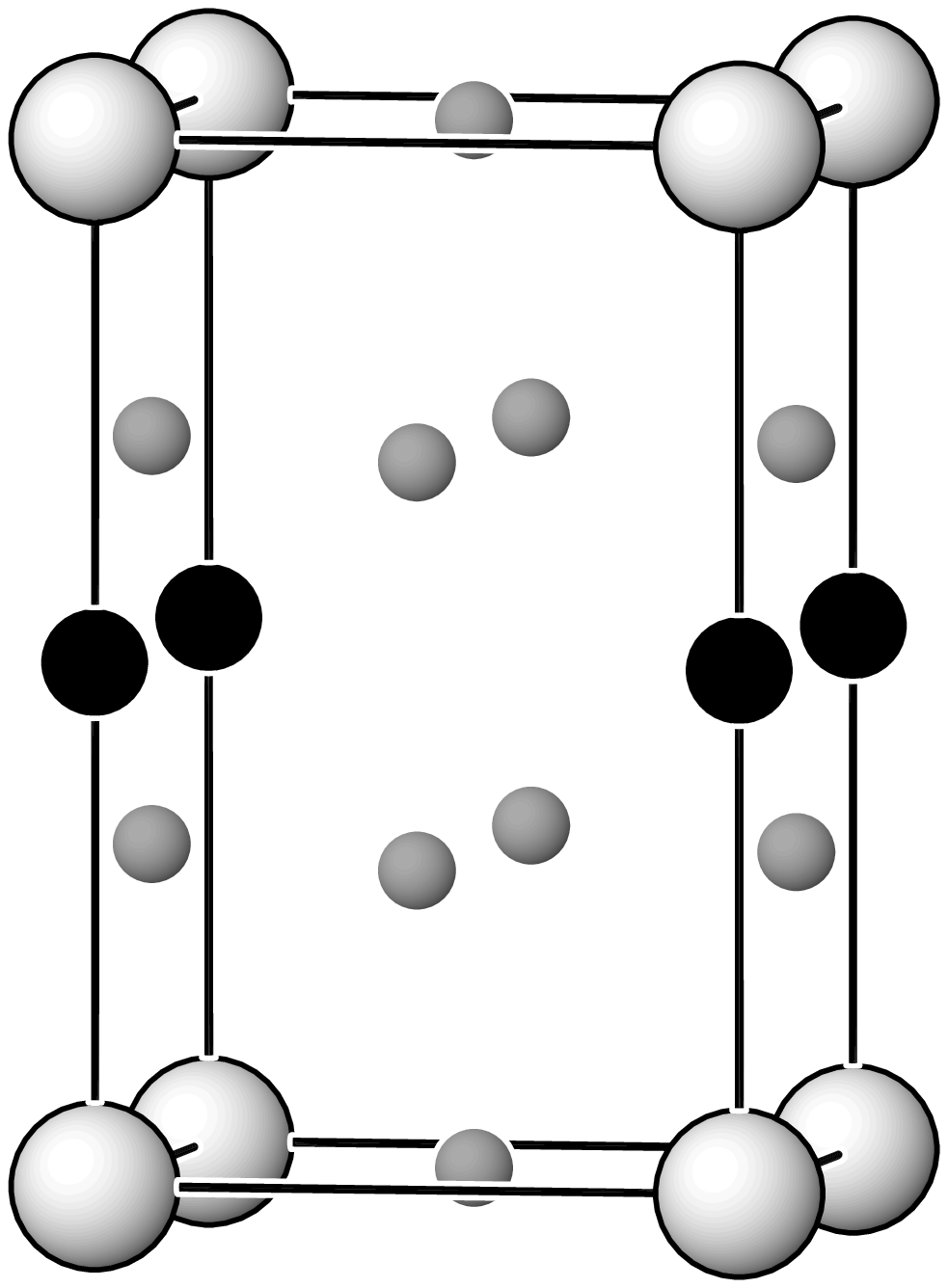}}
\caption{Left: Tetragonal unit cells of the ThCr$_{2}$Si$_{2}$ and
CeM$_2$X$_2$ (Ce122) structure where M = Cu, Ni, Ru, Rh, Pd, Au, .. (black
circles); X = Si, Ge (small grey circles) and Ce (large grey circles). 
Right: Conventional unit cell of the CeMIn$_5$ (Ce115) structure whith
M = Co, Ir, Rh (black circles); In (small grey circles) and Ce (large
grey circles).}
\label{fig:ThCr2Si2Structure}
\end{SCfigure}

Since the discovery of heavy electrons and their superconducting state
\cite{Steglich79} in \CCS~ the Ce-based tetragonal 112-compounds
(space group I4/mmm, see fig.~\ref{fig:ThCr2Si2Structure}) have been
at the focus of investigation. This class of
materials offers a unique stage to watch the interplay of 
quantum-criticality, heavy Landau Fermi liquid (LFL) and non- Fermi liquid
(NFL) behaviour as well as unconventional 
superconductivity. Although it has a long history, major aspects of
the parent compound \CCS~ of this class where only understood very
recently: Firstly the long mysterious 'A-phase' associated with the
low pressure quantum critical point, has finally been identified as an
incommensurate (IC) SDW phase  caused by the nesting of heavy
quasiparticle FS sheets \cite{Stockert03}. In fact it is now possible to
follow the continuous evolution of the SDW wave vector and moment size of
magnetic phases in \CSG~ from the small moment IC-SDW in \CCS~ to the
atomic moment commensurate antiferromagnetism in \CCG~\cite{Trovarelli97}.

Secondly the anomalous stability of SC in \CCS~ for pressures far above
the A-phase QCP \cite{Jaccard99} which did not exhibit the typical
dome-shape around the QCP pressure has now been understood. Likewise
\CCG~ does not have a dome-shaped SC T$_c$(p) curve once AF order
is suppressed. Employing a negative pressure shift caused by the
larger Ge- radius it looks rather similar to that of \CCS. Artificial
reduction of the SC  condensation energy in \CCS~ by Ge- alloying and applying
positive shifts with hydrostatic pressure leads to a breakup of the SC
region into two isolated domes which are associated with the previous
A-phase magnetic QCP and a new high pressure valence transition QCP
respectively \cite{Yuan03b}. This also suggests that SC phases and pairing
mechanisms in the isolated SC
regimes are different. Quantum criticality in the \CSG~ system is
connected with distinct NFL anomalies that can only partly be
explained within the standard nearly AF Fermi liquid picture, notably
the discrepancy of the temperature scaling of resistivity and $\gamma$-
coefficients at higher fields close to the A- phase QCP may require
different concepts. 

The HF compounds \CRS~and \CPS~
also become superconducting only under application of
pressure like \CCG. As in \CCG~AF order in \CRS~vanishes in a first order
phase transition at the critical pressure p$_c$. SC in the latter appears
only in an extremely narrow region around p$_c$, unlike in the former
compound. Due to the first order transition both \CCG~and \CRS~
are in a LFL state close to p$_c$. 
For \CPS~ on the other hand a QCP where AF order vanishes continuously
leads to a SC p-T phase diagram which is of the canonical single dome
shape centered at p$_c$ where clear cut NFL behaviour in the resistivity is
observed.  Finally \CNG~is quite unique because it is not magnetic but
rather displays almost ideal NFL behaviour at ambient pressure
although incipient superconductivity is found at very low temperatures
in highest-purity samples.

A comparison of the various Ce122 compounds leads one to conclude 
that the existence of an AF QCP with associated NFL behaviour is
sufficient to induce superconductivity around p$_c$. Indeed the AF spin
fluctuation theory (sect.~\ref{sect:SCmech})
predicts a stabilisation of the SC state when the QCP is
approached. However, as the examples of \CCG~and \CRS~show, SC in
Ce122 compounds may just as well appear in the normal LFL state.

It took more than 20 years to discover an additional class of Ce-based HF
superconductors with the general formula Ce$_n$M$_m$In$_{3n+2m}$ (M =
Co, Ir and Rh). This belongs to the tetragonal layer type
materials derived from the 'infinite layer' parent compound \CIN~by
stacking n `\CIN' and m `MIn$_2$' subunits along the tetragonal axis. While \CCI~and
\CII~are single layer (n=1) HF compounds which are superconducting at
ambient pressure \cite{Petrovic01}, \CRI~is a local moment AF which
becomes SC only under pressure. Likewise in the bilayer (n=2)
compounds SC appears both at ambient pressure for \CCIn~and \CIIn~and
at finite pressue for \CRIn. 
Especially \CCI~has turned out to be of great interest. Its
comparatively large T$_c$ has allowed to determine the d-wave symmetry
of the SC gap function by field-angle resolved
magnetothermal conductivity \cite{Izawa01} and specific
heat \cite{Aoki04} measurements. High-field specific heat
\cite{Bianchi03} and ultrasonic attenuation measurements
\cite{Watanabe03} have indicated that firstly the SC transition evolves to
first order at larger fields, and secondly evidence for the long-sought FFLO
phase with SC pairing at finite momentum exists.
As in the Ce122 class SC appears preferably in the vicinity of magnetic QCPs and
the corresponding AF/SC phase diagrams have been determined in
pressure experiments. Likewise NFL phenomena above the AF QCP have been observed.

Another exotic HF Ce- superconductor, \CPSi~has now been discovered
which is the first with a non-centrosymmetric structure
\cite{Bauer04}. Superconductivity in crystals without inversion center was discussed
theoretically first in \cite{Bulaevskii76}.
The lack of inversion symmetry raises fundamental
questions of classification of the SC states according to odd and even
parity \cite{Frigeri03}. It remains to be seen whether this compound
is the first member of a new class of HF SC.
%
%

\subsection{Quantum critical behaviour and superconductivity in \CCS
and the alloy series \CSG}

\label{ssect:CeCu2Si2}

The first Ce- based HF metal and superconductor \cite{Steglich79} has
for a considerable time served as a most fruitful model for
investigating strongly correlated electron systems. Until rather
recently it was also the only HF Ce compound exhibiting
superconductivity. Most importantly in \CCS~ Si can be continuously subsituted
by Ge while the compound changes from a HF superconductor to a normal
antiferromagnet \cite{Trovarelli97}. The Ge substitution acts like a
negative chemical pressure, likewise fully substituted \CCG~ under
positive hydrostatic pressure behaves similar as \CCS~ under ambient pressure.
Ge- substitution and application of  hydrostatic pressure therefore
allows to vary the coexistence and competition behaviour of
superconducting and other order parameters in the alloy series \CSG~.  

\begin{figure}[t b h]
\raisebox{0.2cm}{\includegraphics[width=7cm]{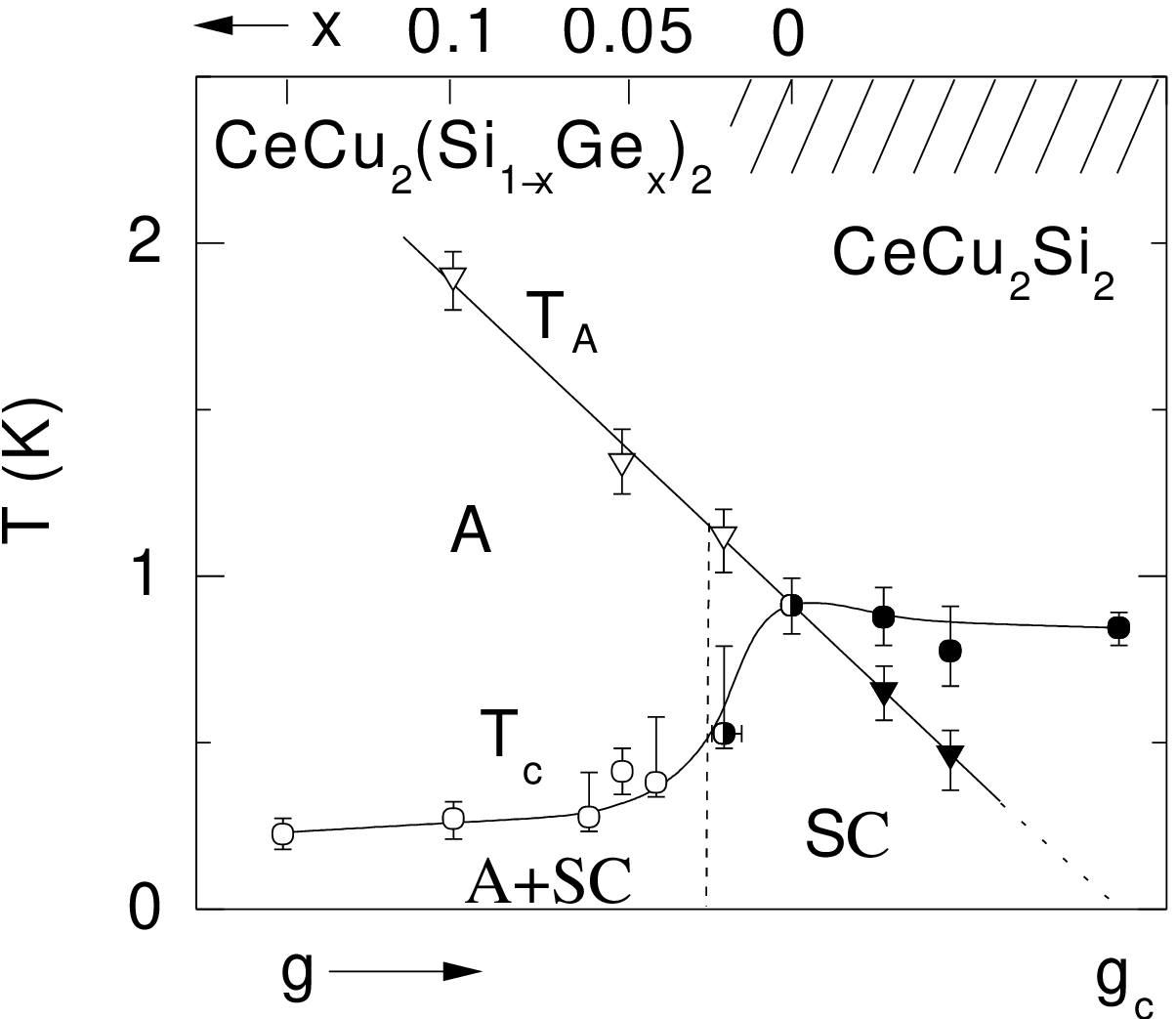}}
\hfill
\includegraphics[width=7cm,height=5.92cm]{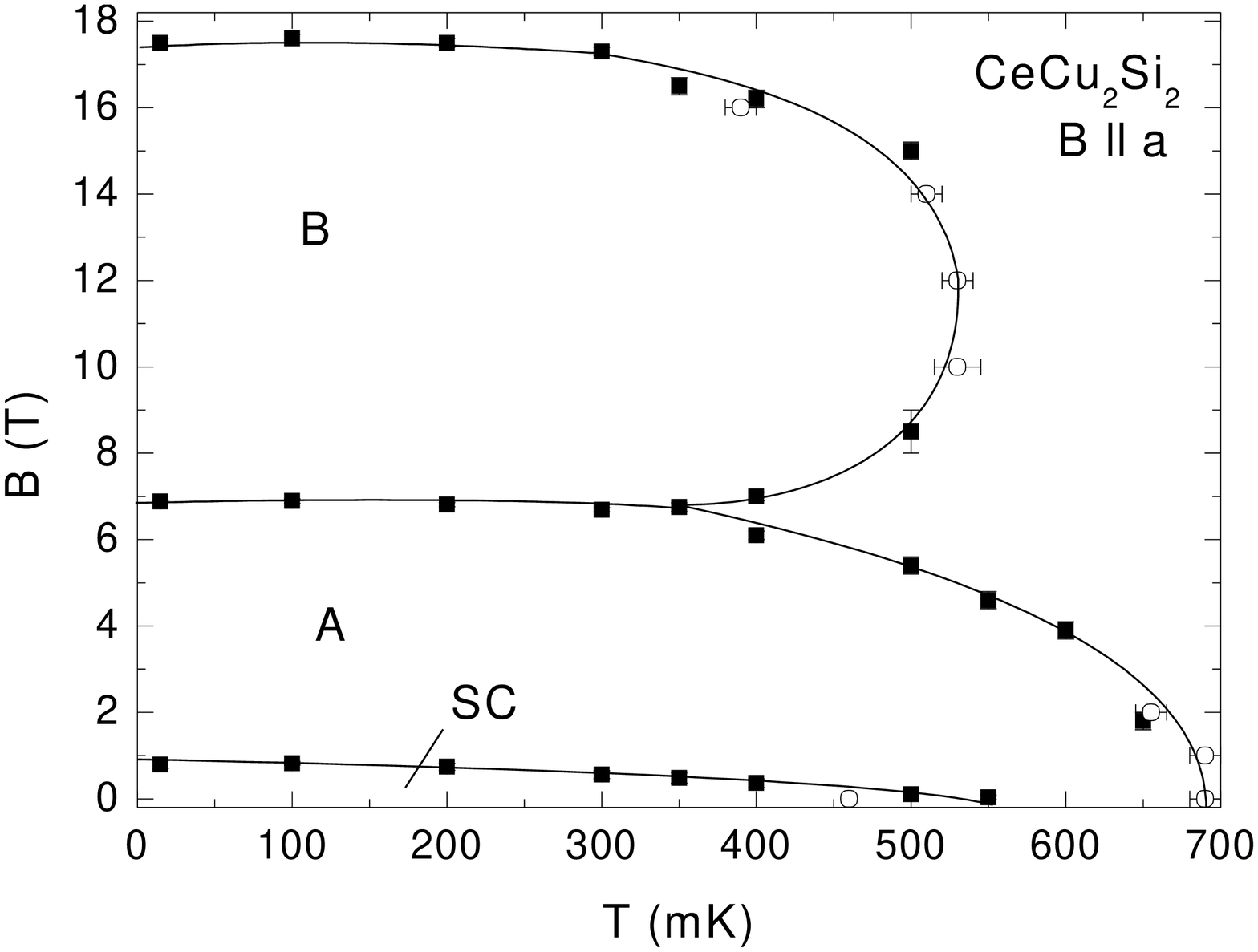}
\caption{Left panel: Coexistence/competition phase diagram of SC and
magnetic A phases close to the magnetic QCP1. 
Phase boundaries are obtainded from susceptibility and resistivity
measurements and composed of alloying (negative pressure) and
hydrostatic pressure results, separated by a dashed line. Here g
$\sim$ 1-x on the left and g $\sim$ p on the right \protect\cite{Steglich01b}
(c.f. theoretical results in fig.~\ref{fig:CeCu2Si2DahmM}).
Right panel: B-T phase diagram of CeCu$_2$Si$_2$ for B $\parallel$
a. Original version in \protect\cite{Bruls94}, completed version from
\protect\cite{Weickert03}. For this sample the A-phase is expelled from
the SC region (no coexistence).}
\label{fig:CeCu2Si2BTPhaseDiag}
\end{figure}

\begin{SCfigure}
\includegraphics[clip,angle=0,width=7cm]{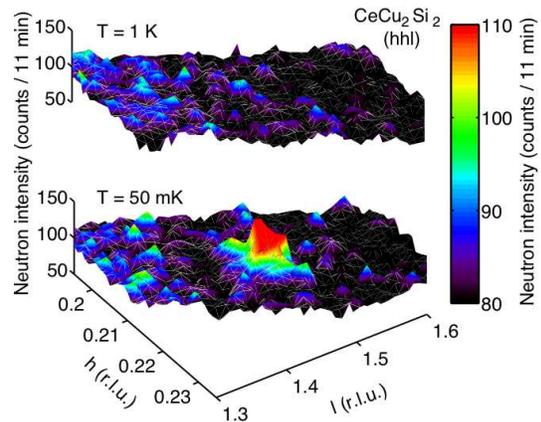}
\caption{Neutron diffraction intensity in reciprocal (hhl) plane in
CeCu$_2$Si$_2$ at temperatures above and below the A-phase transition
temperature T$_A$. The magnetic peak appears at (0.215,0.215,1.470)
corresponding to an incommensurate
SDW modulation vector \bQ~= (0.215,0.215,0.530) 
\protect\cite{Stockert03}.} 
\label{fig:CeCu2Si2StockertND}
\end{SCfigure}

Indeed already stoichiometric \CCS~ at ambient pressure
shows an additonal 'A-phase' which envelops SC in the B-T phase diagram
\cite{Bruls90} (fig.~\ref{fig:CeCu2Si2BTPhaseDiag}). When T$_c$ and
T$_A$ are close
they do not coexist and the A-phase is expelled from the SC
region of the B-T plane. Ge-substition or suitable deficit
of Cu however stabilises the A-phase (T$_A>$ T$_c$) and coexistence of
both phases results. Hydrostatic pressure and Cu excess on the other
hand reduces T$_A$ to zero at the quantum critical point (QCP). Even
after single crystals
became available \cite{Sun90} the nature of the A-phase has remained
mysterious. Later, neutron diffraction on weakly Ge- substituted \CSG~
samples have shown that the magnetic order observed in the stoichiometric \CCG~
compound is still present, although with continuously decreasing
moment \cite{Stockert04b}. The
slightly incommensurate (IC) modulation vector did not change very
much as function of x (Ge concentration). Finally IC magnetic order
with \bQ~ = (0.215, 0.215, 0.530) (r.l.u) has been found in the
stoichiometric compound \CCS~ by neutron diffraction
\cite{Stockert03}, which identified the A-phase as a conventional IC-SDW
phase with rather low moment $\sim$ 0.1 $\mu_B$ per Ce site
(fig.~\ref{fig:CeCu2Si2StockertND}). The ordering wave vector nicely
agrees with the nesting wave vector of heavy FS sheets in \CCS~ found
by Zwicknagl et al. \cite{Zwicknagl93,Stockert03}. It is clear now
that the A-phase of
\CCS~ evolves continuously from the magnetic phase of \CCG~, when x is
reduced from x = 1 progressively .
Taken together the pressure and Ge- substitution dependences 
at first sight do not support the idea that SC appears around the QCP
of the A-phase because the T$_c$(x,p) curve (full dots in
fig.~\ref{fig:CeCu2Si2QCP2}) lacks the typical 'dome
shape'. Specifically T$_c$(p) strongly increases for increasing
hydrostatic pressure p \cite{Jaccard99} and reaches its maximum far away
from the QCP (fig.~\ref{fig:CeCu2Si2QCP2}). This feature is very
different from the conventional SC/AF QCP phase diagram which applies for
other Ce- based HF superconductors like \CPS~and
\CRS~(fig.~\ref{fig:CePdRh2Si2p}). However it is now clear
\cite{Yuan03b} that this anomalous high pressure maximum in T$_c$ is
due to the presence of an additional high pressure QCP
associated with an only weakly first order valence phase transition of
Ce which does not break any symmetry \cite{Holmes04}. The charge
fluctuations associated with the valence transition have been proposed
to mediate the SC pairing around the high pressure maximum
\cite{Onishi00} which is distinct from both spin fluctuation and
magnetic exciton mechanism invoked for other Ce- and U-HF compounds
respectively. 
The appropriate dome shape of two separated QCPs has been revealed by
using simultaneous tuning with Ge-substitution and hydrostatic
pressure. The latter compensates the effect of the former, but the
mean free path is decreased which destabilises the unconventional SC
order parameter and reduces it's T$_c$. In this way the large
stability region of stoichiometric \CCS~may be broken into the two
domes associated with the magnetic and valence transition QCPs
(fig.\ref{fig:CeCu2Si2QCP2}). It seems now that \CCS~and substitutes
also fit into the generic QCP scenario of Ce-HF superconductors albeit
with the surprise of an additional qualitatively new QCP.
\begin{table}
\begin{center}
\begin{tabular}{l|c|c|c|c|l}
\hline
~ & $\gamma$ [mJ/molK$^2$] & T$_N$[K] & $\mu$[$\mu_B$] 
~ & T$_c$(p=0)[K] & T$_c^m$(p$_m$)[K] \\
\hline
\CCS  & 700  & 0.8  &0.10  & 0.7 & 2.3 (3 GPa) \\
\CCG  &1200$^{(1)}$  & 4.3  &1.05  & -   & 1.7 (16 GPa) \\
\CNG  & 350   & -    &  -  &-  &  0.4 (2.2 GPa)\\
\hline
\end{tabular}
\end{center}
\caption{Material parameters of Ce- based HF superconductors. $^{(1)}$
Value extrapolated to T = 0 by fitting the S=1/2 Bethe ansatz result
to data for T $>$ T$_N$ using T$^*$ = 6K.}
\label{tab:CECUSI}
\end{table}
%
%
\subsubsection{Quantum critical points and the B-T phase diagram}

\begin{SCfigure}
\raisebox{-1.2cm}
{\includegraphics[width=7cm,height=7cm]{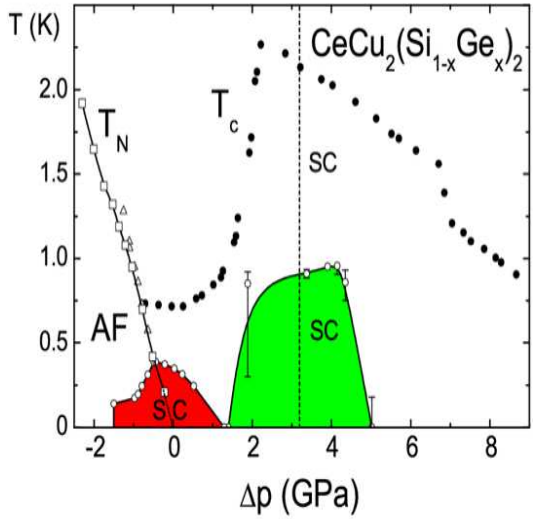}}
\caption{p-T phase diagram of \CSG~composed from pressure (p)
dependence and Ge-doping (x) dependence of $\rho$(T), C(T) and
$\chi$(T). For different substitution level x, the pressure is shifted
by the value of its corresponding critical pressure p$_{c1}$(x) for
the magnetic QCP1 ($\Delta$p = p - p$_{c1}$(x)).
Two SC domes are associated with QCP1(p$_{c1}$) (left dome) of the A-
phase and QCP2(p$_{c2}$) of the integer valence (Kondo) phase (right dome)
\protect\cite{Yuan03b}. The dashed line indicates the valence
transition in \CCG~and T$_c$ (dots \protect\cite{Thomas93})
corresponds to stoichiometric \CCS.}  
\label{fig:CeCu2Si2QCP2}
\end{SCfigure}

Like the other Ce122 compounds the \CSG~ series crystallises in the 
tetragonal body centered ThCr$_2$Si$_2$ structure
(fig.~\ref{fig:ThCr2Si2Structure}). The 
stability of A and SC phases is extremely sensitive to the 
stoichiometry of the constituent elements. The homogeneity range of
\CCS~in the ternary chemical Ce-Cu-Si phase diagram is quite limited and 
comprises three regions where either A or SC alone are present or 
A/SC coexist \cite{Steglich01b}. This agrees with the observation that 
in the combined Ge- doping and pressure  phase diagram shown in
(fig.~\ref{fig:CeCu2Si2BTPhaseDiag}) 
the crossing of T$_{c}$ and T$_{A}$ curves is located around the 
stoichiometric case at ambient pressure. Around the first low 
pressure magnetic QCP where T$_{A}\rightarrow$ 0 NFL behaviour with 
$\Delta\rho(T)\sim T^{3/2}$ and $\gamma(T)=\gamma_{0}-\alpha T^{1/2}$ was observed 
for moderately low temperatures and moderately high fields 
\cite{Gegenwart98}. This low temperature phase diagram which has been 
composed of different transport and susceptibility measurements 
already shows that T$_{c}$ saturates in a plateau after the T$_{A}$- 
crossing instead of forming the canonical QCP dome shape. Indeed on 
increasing pressure even further T$_{c}$ increases steeply reaching a 
maximum around 3 GPa \cite{Thomas93} to 4 GPa \cite{Jaccard99}
(fig.~\ref{fig:CeCu2Si2QCP2}).
The delicate competition and coexistence behaviour for stoichiometric 
single crystals of \CCS~ has already been obvious from earlier 
investigations on the B-T phase diagram obtained with ultrasonic
methods and dilatometry \cite{Bruls94,Weickert03} which is shown in 
fig.~\ref{fig:CeCu2Si2BTPhaseDiag} (right panel). For this sample T$_{c}\leq$
T$_{A}$ and A/SC phases do not 
coexist. The expulsion of the A-phase from the SC region in 
fig.~\ref{fig:CeCu2Si2BTPhaseDiag} can be clearly seen from sound
velocity anomalies across 
the A and SC phase boundaries. The still poorly understood B phase 
might be another SDW phase with different propagation vector \bQ~ 
and/or SDW moment polarisation.  

To unravel the puzzling anomalous stability of T$_{c}$ at high 
pressure systematic studies with slightly Ge doped single crystal 
\CCS~ were performed \cite{Yuan03b}. It has been long suspected that 
the anomalous T$_c$(p) behaviour is associated with the presence of a
second QCP connected with a 
Ce-valence transition \cite{Holmes04}. To support this scenario one 
has to prove that the region around T$_{c}\simeq$ T$_{A}$ 
and around the maximum T$_{c}$ correspond  to distinct SC phases. 
This is achieved by the following strategy \cite{Yuan03b}: i) Due to
larger Ge ionic radius slight Ge doping 
corresponds to application of internal negative pressure which may be 
compensated again by application of external positive pressure, 
returning to zero effective pressure. ii) Because the electronic mean 
free path has decreased due to scattering from Ge 
impurities, the condensation energy and hence T$_{c}$ of the 
anisotropic SC phase will be decreased, destabilising SC. 
In this way the continuous SC region of stoichiometric \CCS~ indeed 
breaks up into two separate SC domes as shown in
fig.~\ref{fig:CeCu2Si2QCP2}. The first 
one centered at p$_{c1}$(x) is the magnetic QCP associated with 
the A-phase and its critical line T$_{N}$($\Delta p$), the second QCP 
at p$_{c2}$(x) is supposed to
describe the Ce$^{3+}$/Ce$^{4+}$ valence transition where the 4f electron 
of Ce$^{3+}$ becomes delocalised. This transition does not break any 
symmetry and therefore its critical line T$_{V}$($\Delta p$) (not
shown in fig.~\ref{fig:CeCu2Si2QCP2}) is assumed to end in a critical
endpoint. The critical line and its endpoint, however, have sofar not
been confirmed for these doped  \CSG~ systems by independent methods. 
However for pure \CCG~a weak, symmetry-conserving valence transition
was observed via x-ray diffractometry \cite{Onodera02} near p $\simeq$
15 GPa, where
also T$_c$(p) assumes it's maximum value \cite{Jaccard99}. Certainly a large 
volume collapse signifying the 4f delocalisation as in the 
$\alpha-\gamma$ Ce transition would not be compatible with the 
presence of the SC phase around p$_{c2}$(x).
%
%
\subsubsection{The A-phase and electronic structure of \CCS}

The incommensurate magnetic order of the A-phase suggests a SDW origin
associated with the itinerant heavy electron FS
in \CCS. To support this conjecture the static magnetic susceptibility
$\chi$(\bq) has to be calculated within the context of renormalised
band theory \cite{Zwicknagl93}. In this approach
(sect.~\ref{sect:QPmech}) resonant phase
shifts $\eta_{\hat{f}m}$ for the 4f-states are introduced empirically
into the LDA scheme to generate the heavy bands at the Fermi
level while charge neutrality and proper f-count are preserved. For
the three CEF split Kramers doublets of Ce$^{3+}$
(J=$\frac{5}{2}$) the ansatz for the phase shift is

\begin{equation}
\eta_{\hat{f}m}(\epsilon)=\tan^{-1}\frac{\Gamma_{\hat{f}}}
{\epsilon-\epsilon_{\hat{f}_m}}\qquad
\epsilon_{\hat{f}_m}= \epsilon_{\hat{f}}$+$\Delta_m
\end{equation} 

\begin{figure}[t b h]
\raisebox{0.4cm}{\includegraphics[width=7.5cm]
{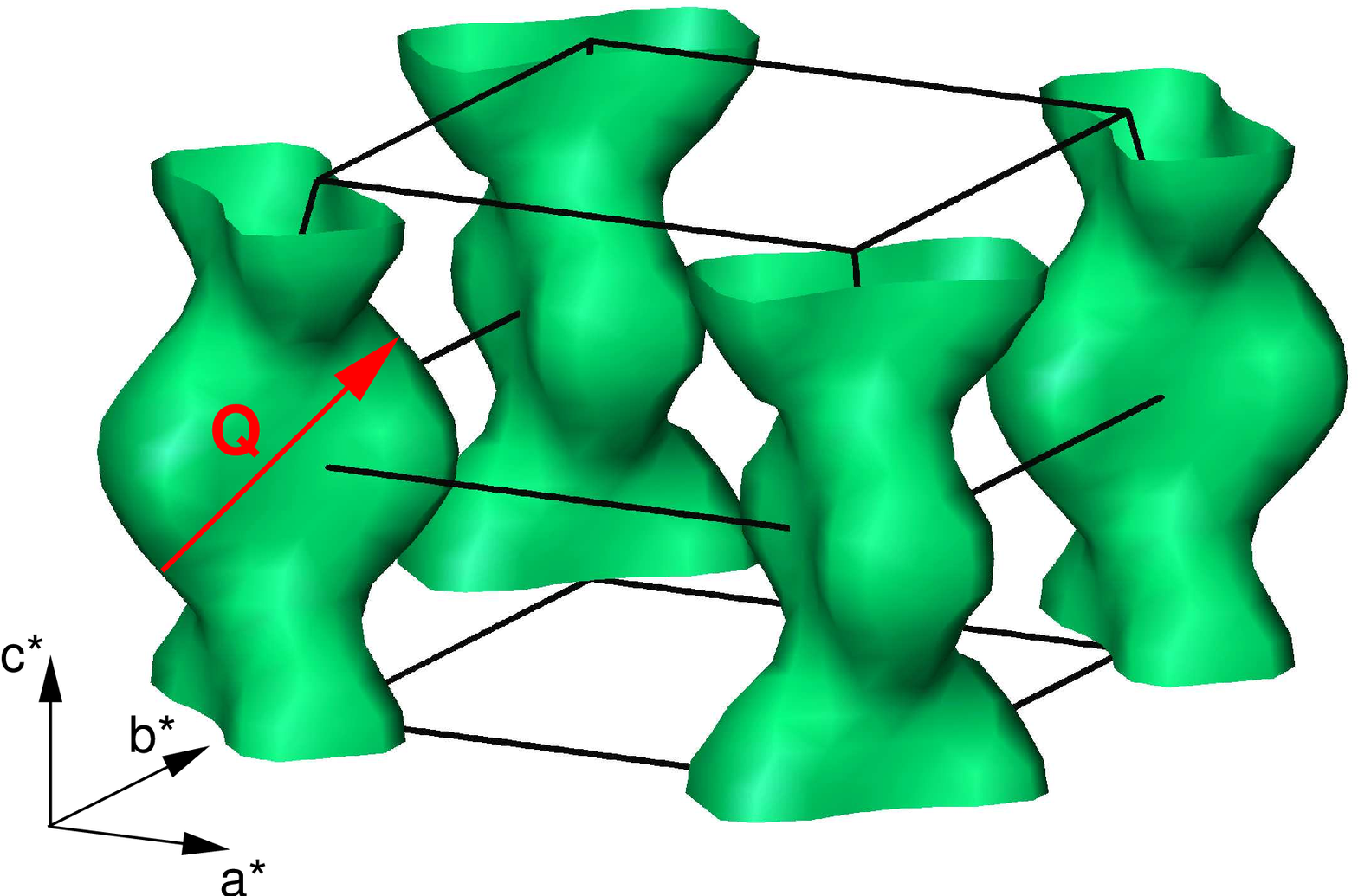}}\hfill
\includegraphics[width=7cm]{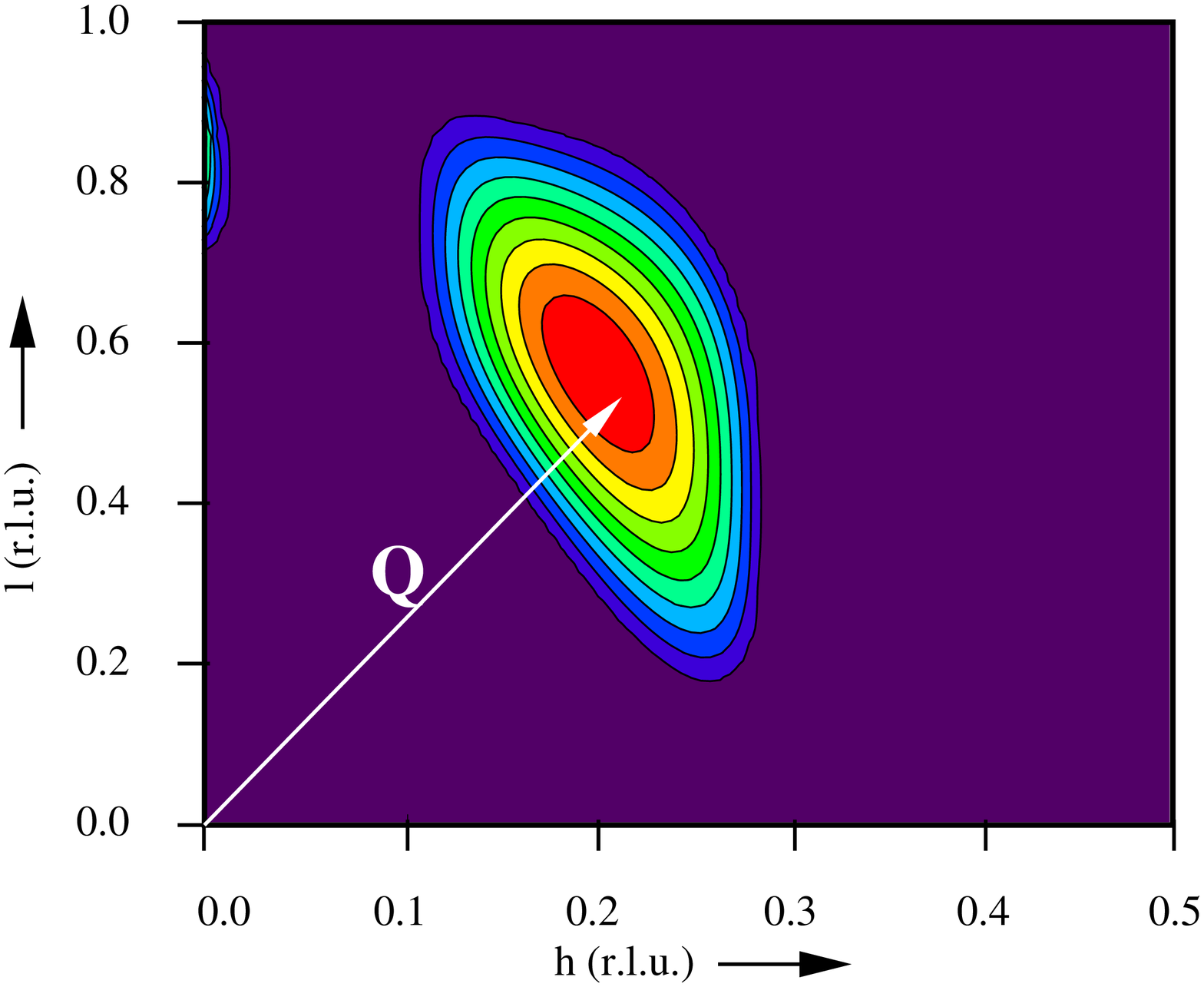}
\caption{Left panel: 
Main heavy FS sheet in \CCS~ where columnar nesting with wave vector
\bQ~ is indicated. Parameters for renormalised band calculations are:
T$^*\sim$ 10 K and $\Delta_{\mbox{CEF}}$ = 330 K overall CEF splitting.
Right panel: Comparison of experimental propagation vector
(fig.~\ref{fig:CeCu2Si2StockertND}) and contour map of theoretical
$\chi$(\bq) in (hhl) plane \protect\cite{Stockert03}.} 
\label{fig:CeCu2Si2FScolumns}
\end{figure}

Here $\epsilon_{\hat{f}_m}$ defines
the centers of heavy bands with  $\epsilon_{\hat{f}}$ denoting the
renormalised 4f- level and $\Delta_m$ the energies of the three
Kramers doublets taken from neutron scattering. The resonance width
$\hat{\Gamma}_f$ is an empirical parameter which determines the width
kT$^*$ or mass m$^*$ of the heavy electron band. It is adjusted to reproduce
the proper experimental value of $\gamma$ = C/T. The main heavy electron
sheet obtained in \cite{Zwicknagl92} is shown in
fig.~\ref{fig:CeCu2Si2FScolumns}. It consists of stacked
columns along c. Obviously there are large flat parts on the columns
with a nesting vector \bQ~ as indicated. Indeed this leads to a pronounced
maximum of $\chi$(\bq) at \bq~= \bQ~as shown in the contour plot of
fig.~\ref{fig:CeCu2Si2FScolumns}. The experimental IC ordering vector of the
A-phase (fig.~\ref{fig:CeCu2Si2StockertND}) agrees well with the
maximum position of $\chi(\bq)$ giving credence to the SDW picture for
the A phase.
%
%
\subsubsection{The high pressure mixed valent phase transition}

\begin{SCfigure}
\raisebox{-0.8cm}
{\includegraphics[width=7.5cm]{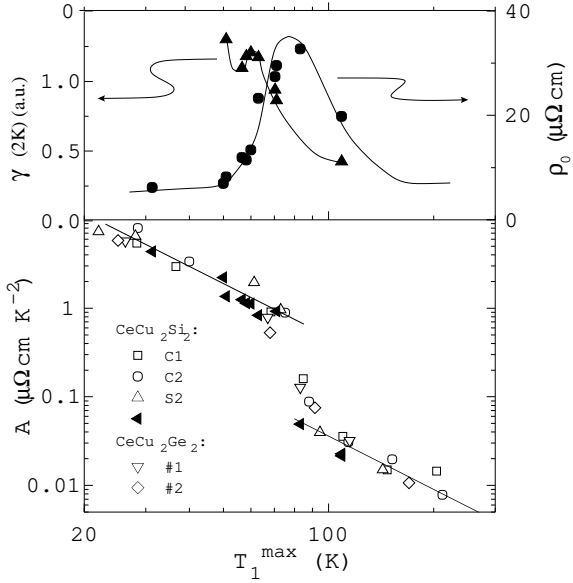}}
\caption{Behaviour of $\gamma$-coefficient, residual
resistivity $\rho_0$ and A-coefficient in $\rho$(T) = $\rho_0$+AT$^2$ against
T$_1^{max}\sim$ T$^*$ which scales monotonously with pressure
\protect\cite{Holmes04}.} 
\label{fig:CeCu2Si2QCP2Val}
\end{SCfigure}

The present experimental evidence for the second QCP at high pressure
connected with Ce-valence change was described in
\cite{Holmes04}. A valence transition does not break any spatial
symmetry but charge fluctuations and volume strain are strongly coupled
which commonly leads to a first order valence transition as in the 
famed $\gamma$-$\alpha$ transition of Ce. No such volume change
has been found at p$_{c2}$ of \CCS~making it a rather exceptional
case. On the other hand this feature may provide a link to
superconductivity as discussed in the next section. The only direct
evidence for a small valence change was obtained earlier in L$_{\mbox{III}}$-
x-ray absorption experiments \cite{Roehler88}. There is however
considerable indirect support for the valence change. The gradual
delocalisation of the 4f$^1$- electron under pressure leads to a less
strongly correlated electron state, i.e. the effective mass will be
reduced according to
\begin{eqnarray}
\frac{m^*}{m}=\frac{1-n_f/2}{1-n_f}
\end{eqnarray}
when the f-level occupation n$_f$ is reduced significantly below its
value in the Kondo limit n$_f\leq$1. The linear specific heat
coefficient $\gamma$ = C(T)/T should reflect this decrease. Likewise the
Kadowaki-Woods ratio A/$\gamma^2$ should decrease accross the valence
transition \cite{Holmes04}. If it is not of first
order, critical valence fluctuations should also lead to a strong
increase in the residual resistivity around the transition. These expected
fundamental features of a valence transition are indeed present in
\CCS~ as shown in fig.~\ref{fig:CeCu2Si2QCP2Val}.
%
%
Theoretical models for the present case of a rapid but continuous
valence transition under pressure are based on the extended periodic
Anderson model (PAM) including a Coulomb repulsion U$_{fc}$ between
conduction and f- electrons, in addition to the on-site U$_{ff}$
Coulomb term in the common PAM. Within the more simple impurity model
this term has the tendency to decrease the f-occupation and when the
energy of the 4f$^1$ state, $\epsilon_f$+U$_{fc}$ approaches $\epsilon_F$,
i.e. the energy of the  4f$^0$ hole state, n$_f$ is rapidly reduced to
values much less than one (fig.~\ref{fig:CeCu2Si2Miyake}). External
pressure is thought to tune
primarily U$_{fc}$ which is an inter-site Coulomb integral and thus
cause the valence transitions. For the lattice model the valence
change as function of U$_{fc}$ has been calculated within 
slave boson fluctuation approximation \cite{Onishi00}. 

\subsubsection{Unconventional superconductivity and the two QCP's}

The breakup of the SC region into two independent domes in
fig.~\ref{fig:CeCu2Si2QCP2} allows one to speculate that the SC
pairing mechanism at
each QCP involves exchange of fluctuations of the associated order
parameters, magnetic (A-phase transition) or charge fluctuations
(valence transition) respectively. This may possibly imply a different
type of superconducting gap symmetry, although not much is known about
the SC state in both regimes, especially for the high pressure SC
region. In the magnetic QCP regime Cu NQR experiments under pressure
for polycrystalline 
\CSG~ \cite{Kawasaki04} and under ambient pressure for Cu and Ce off-
stoichiometric polycrystalline  samples \cite{Ishida99} show
i) 1/T$_1\sim$ T$^3$ behaviour indicating a SC state with gap nodes
and ii) the presence of soft magnetic
fluctuations close to a pressure or doping regime where the SC is
stabilised. This suggests that the magnetic QCP scenario for
unconventional HF-SC in \CCS~ may be correct. Further knowledge about
the nature of the SC gap function can only come from field-angle resolved
pure single crystal investigations, e.g. C(\bH,T) measurements at very
low temperatures.\\
%
\begin{SCfigure}
\raisebox{-0.9cm}
{\includegraphics[width=7cm]{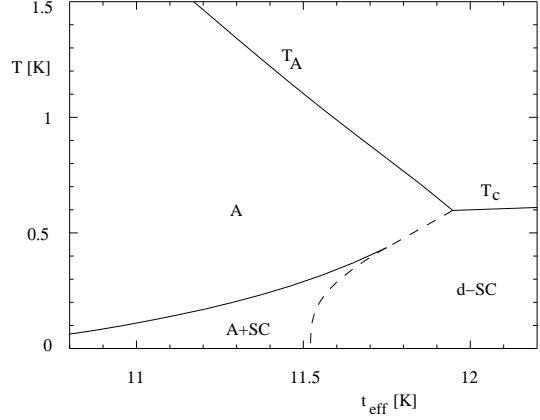}}
\caption{Theoretical phase diagram for A and d$_{xy}$- SC order from
a two band model. Full and dashed line correspond to second and first
order transitions. A+SC is the coexistence region. The effective
tight binding hopping integral t$_{eff}$ scales with p or 1-x
(c.f. experimental results in fig.~\ref{fig:CeCu2Si2BTPhaseDiag})
\protect\cite{Steglich01b}} 
\label{fig:CeCu2Si2DahmM}
\end{SCfigure}
%

\noindent
{\em Model for SC at the magnetic A- phase QCP:}\\
A 2D two-band model for SDW-SC states in \CSG~ has been studied
in some detail \cite{Steglich01b}. The microscopic origin of the
pair potential is not specified in the BCS type model.
A tight binding (TB) band centered at the $\Gamma$(0,0) point and
with \bQ~= ($\frac{1}{2}$,$\frac{1}{2}$) nesting feature represents a
simplified version of the FS
columns in fig.~\ref{fig:CeCu2Si2FScolumns} whose nesting properties
are known to be responsible
for the SDW A-phase. In addition spherical electron pockets which
represent the remaining FS sheets of \CCS~ are included. The former
carries both SDW and SC order parameters, the latter only SC. Therefore a
d$_{xy}$ SC state is favored. The A/SC competition-coexistence phase
diagram resulting from the coupled gap equations is shown in
fig.~\ref{fig:CeCu2Si2DahmM}. There t$_{eff}$ is the hopping element
of the TB band which is thought to scale linearly with the control
parameter g which is 1-x (x = Ge doping) or the pressure p. The
comparison of the theoretical (fig.~\ref{fig:CeCu2Si2DahmM}) and
experimental phase diagram (fig.~\ref{fig:CeCu2Si2BTPhaseDiag})
shows a striking resemblance.
In this model the tail of the coexistence A+SC regime is stabilised by
the SC pairing on the spherical FS sheets. A more elaborate discussion
of this issue using the real FS of \CCS~is given in sect.~\ref{ssect:FSCOEX}.\\

\begin{SCfigure}
\raisebox{-0.7cm}
{\includegraphics[width=7cm]{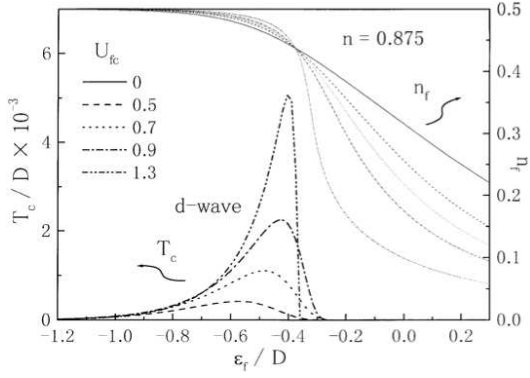}}
\caption{Valence (n$_f$) transition in the extended (impurity) Anderson model
as function of the f-level position $\epsilon_f$ (D = conduction band
width, $\epsilon_F\equiv$ 0). Associated T$_c$ dependence of d-wave SC is also
shown. \protect\cite{Onishi00}} 
\label{fig:CeCu2Si2Miyake}
\end{SCfigure}

\noindent
{\em Model for SC at the valence transition QCP:}\\
A more microscopic model for the SC mechanism around the high pressure
valence transition has been proposed in \cite{Onishi00}. The extended
PAM mentioned above leads to an electron pairing mainly via the
exchange of c$\leftrightarrow$f charge fluctuations with small
wavelength. In weak
coupling approximation a d-wave state is found to be stabilised in the
regime where appreciable charge fluctuations, controled by the pressure
dependent U$_{fc}$- parameter, set in but the quasiparticle DOS (the
mass enhancement) is still sizable. The maximum in T$_c$ scales
with the steepness of the continuous valence transition as seen in
fig.~\ref{fig:CeCu2Si2Miyake}. This does not
incorporate realistic FS features but assumes a single spherical
conduction band. Also the real change in n$_f$ across the valence
transition will certainly be much less than in the theoretical
model (fig.~\ref{fig:CeCu2Si2Miyake}). Therefore it should be taken as
an illustration of the mechanism and not literally as explanation for
high pressure SC in \CSG.
%
%
\subsubsection{non-Fermi liquid state and quantum critical behaviour}

\begin{figure}[tbh]
\includegraphics[width=8cm]{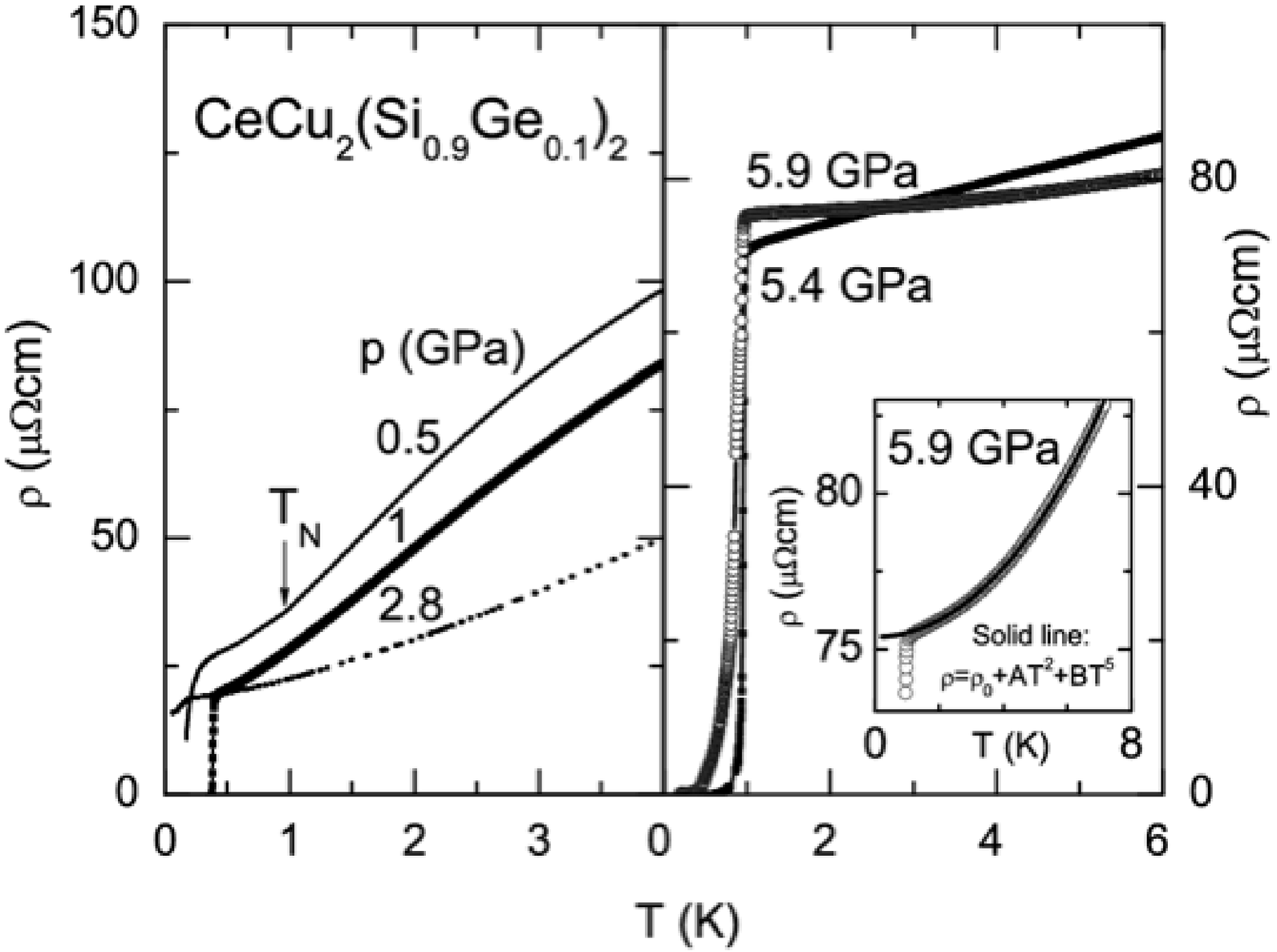}\hfill
\raisebox{0.8cm}
{\includegraphics[width=7cm,angle=0,clip]{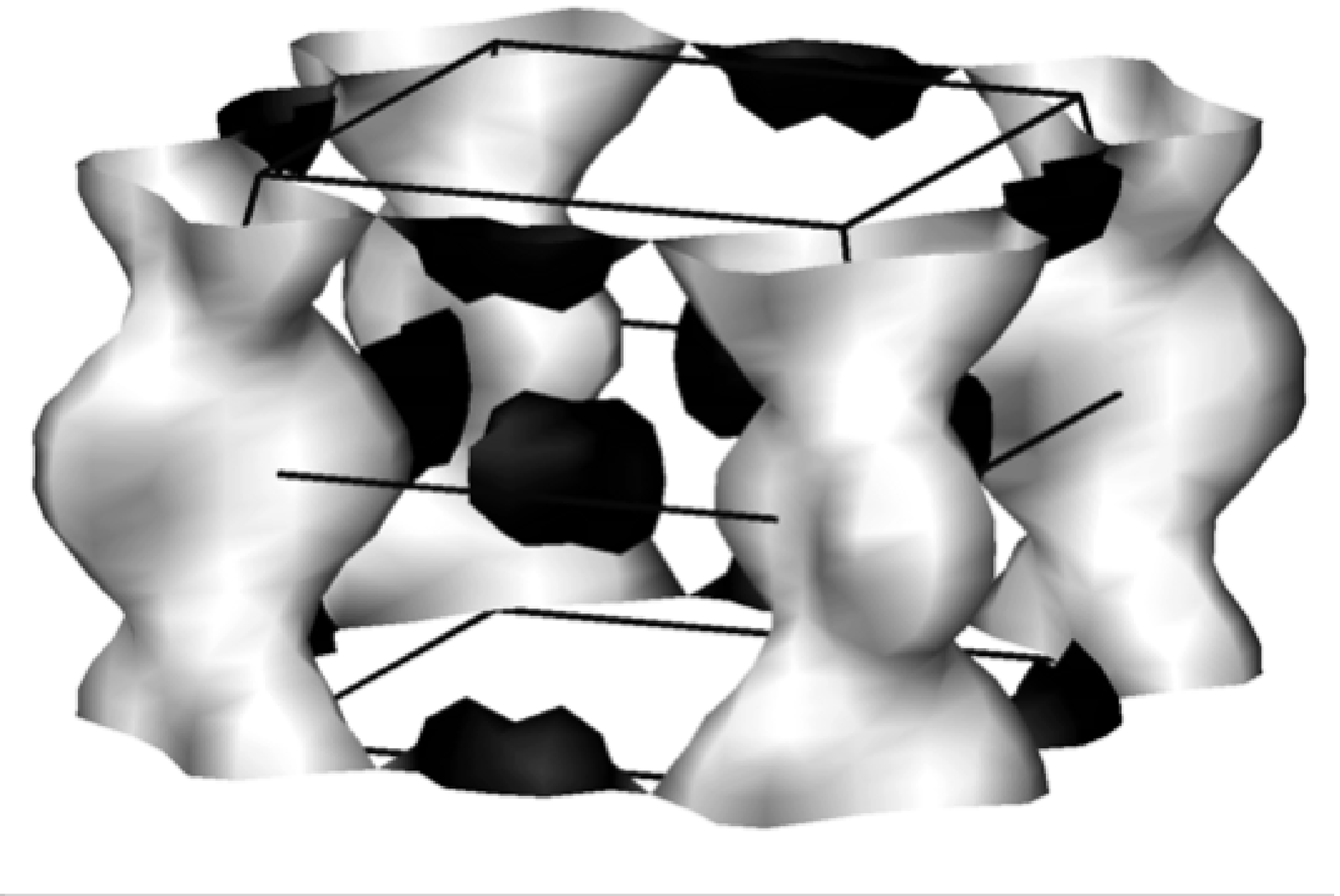}}
\caption{Left panel: NFL behaviour of $\Delta\rho$(T) in \CSG~ close to QCP1
(left) and QCP2 (right) \protect\cite{Yuan03b}.
Right panel: 
Variation of the SC gap function amplitude $\left|\Delta
(\mathbf{k})\right|/\left|\Delta _{0}(T)\right|=\left|
\phi _{\Gamma_{3}}(\mathbf{k})\right|$
for the (pseudo-) singlet wave function with $\Gamma _{3}$-symmetry
$\phi _{\Gamma _{3}}(\mathbf{k})\sim \cos k_{x}a-\cos k_{y}a$ on
the main sheet of the paramagnetic FS \cite{Neef04,Neef04a}. The
amplitude of this SC order parameter is maximal on the kidney-shaped
surfaces centered along the $\Sigma$-direction which are (almost)
unaffected by the formation of the A-phase (dark grey).
The dominant contributions to the latter come from the nesting parts
on the heavy columns where the superconducting amplitude is small
(light gray).}
\label{fig:CeCu2Si2NFL}
\end{figure}

Above the SC domes the two QCP's in \CSG~ lead to pronounced NFL
anomalies in thermodynamic and transport properties. Naturally one
would invoke here the quasiparticle scattering from low energy
fluctuations of the associated order parameters \cite{Rosch99} as one possible
mechanism behind NFL behaviour. For the A-phase QCP a
detailed comparison of experimentally observed T,B-scaling laws for
$\rho$(T) = $\rho_0$ + $\Delta\rho$(T), $\chi$(T) and $\gamma$(T) for
single crystal \CCS~ and polycrystalline samples of various Ge-dopings
has been performed
\cite{Gegenwart98,Steglich01}. It was found that for samples slightly
above T$_A$(x) = 0 the expected NFL scaling laws at a 3D AF QCP
\begin{equation}
\Delta\rho(T)=\beta T^\frac{3}{2}\qquad \gamma(T)=
\gamma_0-\alpha T^\frac{1}{2}
\end{equation}
are indeed well fulfilled. However there are also distinct deviations,
notably the high-field behaviour of $\Delta\rho$(T) where a crossover
of the resistivity exponent from 3/2 to the FL exponent 2 is
observed. In contrast $\gamma$(T) remains qualitatively unchanged
\cite{Gegenwart98} at moderately low T and exhibits an upturn at the
lowest temperatures which is not yet understood.
Such disparities in the scaling of $\Delta\rho$(T)
and $\gamma$(T) are now also known from other NFL systems and
proposals how to account for them have been
made \cite{Gegenwart98}. The study of NFL behaviour at the high
pressure valence transition QCP on the other hand is still in its
beginning. The NFL behaviour in  $\Delta\rho$(T) above the two distinct
SC domes of fig.~\ref{fig:CeCu2Si2QCP2} is illustrated in
fig.~\ref{fig:CeCu2Si2NFL}.  
%
%
\subsubsection{Coexistence of superconductivity and A-phase in \CCS}
\label{ssect:FSCOEX}

The central goal in this section is to study the
competition-coexistence phase diagram within a more realistic model
than above. We start from the real FS sheets of \CCS~as obtained by
renormalised band structure calculations. The FS nesting 
properties determine the observed magnetic structure. Of particular
interest are the superconducting states which can either coexist with
the A-phase or expel it. The model Hamiltonian is given by
\begin{equation}
H=\sum_{\mathbf{k}s}\epsilon_{\bk}
c_{\mathbf{k}s}^{\dagger}c_{\mathbf{k}s}+H_{int}
\label{eq:HamDef}
\end{equation}
where the first term describes heavy quasiparticles and H$_{int}$
their residual interactions. The creation (annihilation) operators are
for quasiparticles with wavevector $\mathbf{k}$, (pseudo) spins
$s=\uparrow ,\downarrow $ and energy $\epsilon_{\bk}$ are denoted by
$c_{\mathbf{k}s}^{\dagger }$($c_{\mathbf{k}s}$).
The energies which are measured relative to the Fermi level are calculated
within the renormalised band scheme. The residual interactions in
the strongly renormalises Fermi liquid is assumed to be repulsive
for short separations while being attractive for two quasiparticles
of opposite momenta on neighboring sites. The former favors the formation
of a SDW while the latter gives rise to the superconducting instability.
Within mean-field approximation one has H$_{int}$ = H$_{SDW}$+H$_{SC}$ where
\begin{eqnarray}
H_{SDW}&=&-\sum _{\mathbf{k}s}s\frac{1}{2}\sum _{\mathbf{Q}_{j}}
[h\left(\mathbf{Q}_{j}\right)c_{\mathbf{k}s}^{\dagger }
c_{\mathbf{k}+\mathbf{Q}_{j}s}+h.c.]\\
H_{SC}&=&\frac{1}{2}\sum _{\mathbf{k}ss'}[\Delta
_{ss'}(\mathbf{k})c_{\mathbf{k}s}^{\dagger
}c_{-\mathbf{k}s'}^{\dagger }+h.c.]
\label{eq:HSC}
\end{eqnarray}
The periodically modulated magnetisation $h\left({\textbf Q}_{j}\right)$
associated with the eight
equivalent SDW propagation vectors $\mathbf{Q}_{j}$ defined below as
well as the superconducting pair potential $\Delta $ have to be
determined selfconsistently
\begin{eqnarray}
h\left({\textbf Q}_{j}\right)&=&\frac{U}{L}\sum _{\textbf
{k}s}\frac{s}{2}\left\langle c_{{\textbf k}+
{\textbf Q}_{j}s}^{\dagger}c_{{\textbf k}s}\right\rangle\\
\Delta _{ss'}\left(\textbf k\right)&=&
 \frac{1}{L}\sum g_{ss';s''s'''}\left(\mathbf{k},\mathbf{k}'\right)
\left\langle c_{-{\textbf k}'s''}c_{{\textbf k}'s'''}\right\rangle 
\label{eq:SelfconSC}
\end{eqnarray}
where the strength $U$ of the local Hubbard-type repulsion is of
the order of the quasiparticle band width $k_{B}T^{*}$ and
$g_{ss';s''s'''}\left(\mathbf{k},\mathbf{k}'\right)$
is the effective pair attraction. The $\textbf k$-summation runs
over the entire paramagnetic Brillouin zone and $L$ denotes the number
of lattice sites. The expectation values denoted by $\left\langle
\ldots \right\rangle $
have to be evaluated with the eigenstates of the total mean-field Hamiltonian
$H_{MF}=H_{0}+H_{SDW}+H_{SC}$, and, consequently, depend upon the
order parameters. The selfconsistency equations are therefore coupled.
The mean-field Hamiltonian implicitly assumes that the amplitudes
of both order parameters are small. In particular, we neglect here
the pairing amplitudes of the form $\left\langle c_{-{\textbf
k}s''}c_{{\textbf k}+{\textbf Q}_{j}s'''}\right\rangle $.
The latter are important when the gaps introduced by the antiferromagnetic
order into the quasiparticle spectrum are large on the scale set by
superconductivity. For a discussion of this point we refer to
\cite{Zwicknagl81,Fulde82}.
The periodically modulated magnetisation associated with the SDW acts
on the conduction electrons like a periodic spin-dependent potential
which we approximate by
\begin{equation}
h\left(\textbf x\right)=\sum _{{\textbf Q}_{j}}h\left({\textbf Q}_{j}\right)
e^{i{\textbf Q}_{j}\cdot {\textbf x}}
\label{eq:AFMolecularField}
\end{equation}
with identical amplitudes $h\left({\textbf Q}_{j}\right)=h_{0}$ for
the eight commensurate wave vectors ${\textbf Q}_{j}$ $\in$ $\{($
$\pm\frac{\pi}{2a}$, $\pm\frac{\pi}{2a}$, $\pm\frac{\pi}{c}$ $)\}$
which are used for a commensurate approximation to the experimentally
found IC SDW.
The magnetic superstructure breaks the translational invariance of
the underlying lattice but it conserves the point group symmetry.
However the mean-field Hamiltonian is invariant under translations
with the lattice vectors 
\begin{equation}
{\textbf a}'_{1}=\left(2a,2a,0\right);\;\;\;
{\textbf a}'_{2}=\left(2a,-2a,0\right);\;\;\;
{\textbf a}'_{3}=\left(2a,0,c\right)
\label{eq:AFLattice}
\end{equation}
The volume of the magnetic supercell is 32 times the volume of the
paramagnetic unit cell. As a result, the Brillouin zone is reduced
and the quasiparticle states are modified by extra Bragg planes.
The opening of new gaps is important at sufficiently low temperatures
$T\ll T_{A}$ where $T_{A}$ is the SDW transition temperature.

Before proceeding with the SC/SDW coexistence analysis of \CCS~we
have to give a brief description of the symmetry classification of
unconventional SC gap functions.
The fundamental property of $\Delta_{ss'}$(\bk) is its behaviour as a
two-fermion wave function in many respects. This expresses the fact
that an ODLRO order parameter is not the thermal expectation value of
a physical observable but rather a complex pseudo- wave function describing
quantum phase correlations on the macroscopic scale of the SC
coherence length. Its phase is a direct signature of the broken gauge
invariance in the SC condensate. The Pauli principle requires
$\Delta_{ss'}$(\bk) to be antisymmetric under the interchange of particles  
\begin{equation}
{\Delta_{ss'}}( {\bf k})=-{\Delta_{s's}}{(- {\bf k})} 
\label{eq:antisymm}
\end{equation}
In addition, it transforms like a two-fermion 
wave function under rotations in position and spin space and under gauge
transformations. The transformation properties yield a general
classification scheme for the superconducting order parameter which is
represented by a $2 \times
2$-matrix in (pseudo-) spin space. It can be decomposed into an 
antisymmetric (s) and a symmetric (t) contribution according to
${{\boldDelta}}( {\bf k})={{\boldDelta}}_s
( {\bf k})+{{\boldDelta}}_t( {\bf k})$ with 
\begin{equation} 
{\boldDelta}_s({\bf k})  = 
\phi( {\bf k}) i\boldsigma_{2} \qquad 
{\boldDelta}_t( {\bf k})  = 
\sum_{{\mu}=1}^3 d_\mu( {\bf k})\boldsigma_{\mu} i\boldsigma_{2}
\label{eq:Psi_decomp}
\end{equation}
where $\boldsigma_{\mu}$  denote the Pauli matrices. Antisymmetry 
${\boldDelta}( {\bf k})\;=\;-\;{\boldDelta}^T(- {\bf k})$ 
requires
\begin{equation}
\phi( {\bf k})\;=\;\phi(- {\bf k})  
\qquad \textrm{and} \qquad
d_\mu( {\bf k})\;=\;-d_\mu(- {\bf k})
\end{equation} 
for the complex orbital functions $\phi( {\bf k})$ and 
$ d_\mu( {\bf k})$ ($\mu$ = 1-3). For brevity we will frequently
write $\Delta({\bf k})$ for $\phi( {\bf k})$ or $|{\bf d}({\bf k})|$.

The order parameter can be chosen 
either as purely antisymmetric (${\boldDelta}_s$) or purely 
symmetric (${\boldDelta}_t$) when spin-orbit interaction can be neglected.
In the 4f- and 5f- based heavy fermion superconductors spin-orbit
interaction is strong. As a consequence classification according to
physical pair spins cannot be used. If their high-temperature crystal
structures, however, have an inversion center classification
according to parity is still possible as is the case in \CCS.

The general classification scheme for superconducting order parameters
proceeds from the behavior under the transformations of the symmetry
group ${\cal G}$ of the Hamiltonian. It consists of the crystal point
group G, the spin rotation group SU(2), the
time-reversal symmetry group ${\cal K}$, and the gauge group U(1). 
The appropriate choice of rotations corresponding to weak or strong
spin- orbit coupling case is determined by microscopic considerations.
Using the above transformation properties the singlet and triplet gap
functions $\phi({\bf k})$ and ${\bf d}({\bf k})$ respectively may
further be decomposed into basis functions 
$\phi^n_\Gamma({\bf k})$ or ${\bf d} ^n_\Gamma({\bf k})$ of the irreducible
representations $\Gamma$ (degeneracy index n) of G$\times$SU(2) (weak
s.o. coupling) or G (strong s.o. coupling). In the latter case the
pseudo spin associated with Kramers degeneracy replaces the physical
conduction electron spin.

The occurrence of long-range order at a phase transition          
described by an order parameter is most frequently associated with
spontaneous symmetry breaking. The simplest superconductors where only gauge 
symmetry is broken are called {\em conventional}. In this case the SC
order parameter has the same spatial symmetry as the underlying   
crystal, i.e. it transforms as a fully symmetric even parity singlet
$\Gamma_1$ representation of G. It should be noted, however, that
conventional is not a synonym for isotropic, for any G one can form
$\Gamma_1$ representations from angular momentum orbitals of higher
order \it l\rm, for example \it l\rm~$\geq$ 2 for tetragonal and
hexagonal symmetry and \it l\rm~$\geq$ 4 for cubic symmetry. On the
other hand, a superconductor
with additional broken symmetries besides gauge symmetry is called
{\em unconventional}. It can have either parity. 
A recent summary is found in the monography \cite{Mineevbook}. 

NMR experiments \cite{Ishida99,Kawasaki04} in slightly Ge-doped or
Cu-off-stoichiometric samples suggest that
the SC gap is unconventional and has line nodes. In our discussion we
restrict ourselves to a singlet pair state as revealed by strong Pauli
limiting seen in early H$_{c2}$ studies \cite{Rauchschwalbe82}.
The singlet state is characterised by a scalar order parameter
\begin{equation}
\Delta \left(\textbf k\right)_{ss'}=
\phi (\mathbf{k})\left(i\sigma _{2}\right)_{ss'}
\label{eq:DeltaEvenParity}
\end{equation}
 The anisotropic effective pair interaction in the (pseudo-) spin
singlet channel can be expanded according to
\begin{eqnarray}
g\left({\textbf k}s,-{\textbf k}s';-{\textbf
k}'s'',{\textbf k}'s'''\right)\rightarrow 
\left(i\sigma _{2}\right)_{ss'}\left(i\sigma _{2}\right)_{s''s'''} 
\times \sum _{j}\frac{1}{2}g\left(\Gamma ^{(j)}\right)
\sum _{\kappa =1}^{d^{(j)}}
\phi _{\kappa }^{(j)}\left(\textbf k\right)\phi _{\kappa }^{(j)*}
\left({\textbf k}'\right)
\label{eq:SymmetrizedCoupling}
\end{eqnarray}
where $\phi _{\kappa }^{(j)}\left(\textbf k\right)$ is a basis
function which belongs to the $\kappa $th row, $\kappa =1,\ldots ,d^{(j)}$,
of the $d^{(j)}$-dimensional representation $\Gamma ^{(j)}$ of the
symmetry group. In principle we have to classify the order parameters
with respect to the antiferromagnetic lattice. Since the two ordering
temperatures are so close we focus on pair states which are compatible
with the translational symmetry of the paramagnetic lattice. The corresponding
basis functions are listed in \cite{Sigrist91,Konno89,Ozaki89,Zwicknagl92}.
In the explicit calculations, we restrict ourselves to one-dimensional
representations for simplicity. The generalisation to multi-dimensional
representations is rather straightforward \cite{Neef04}.

\begin{figure}[h t b]
\includegraphics[width=7cm,angle=0,clip]
{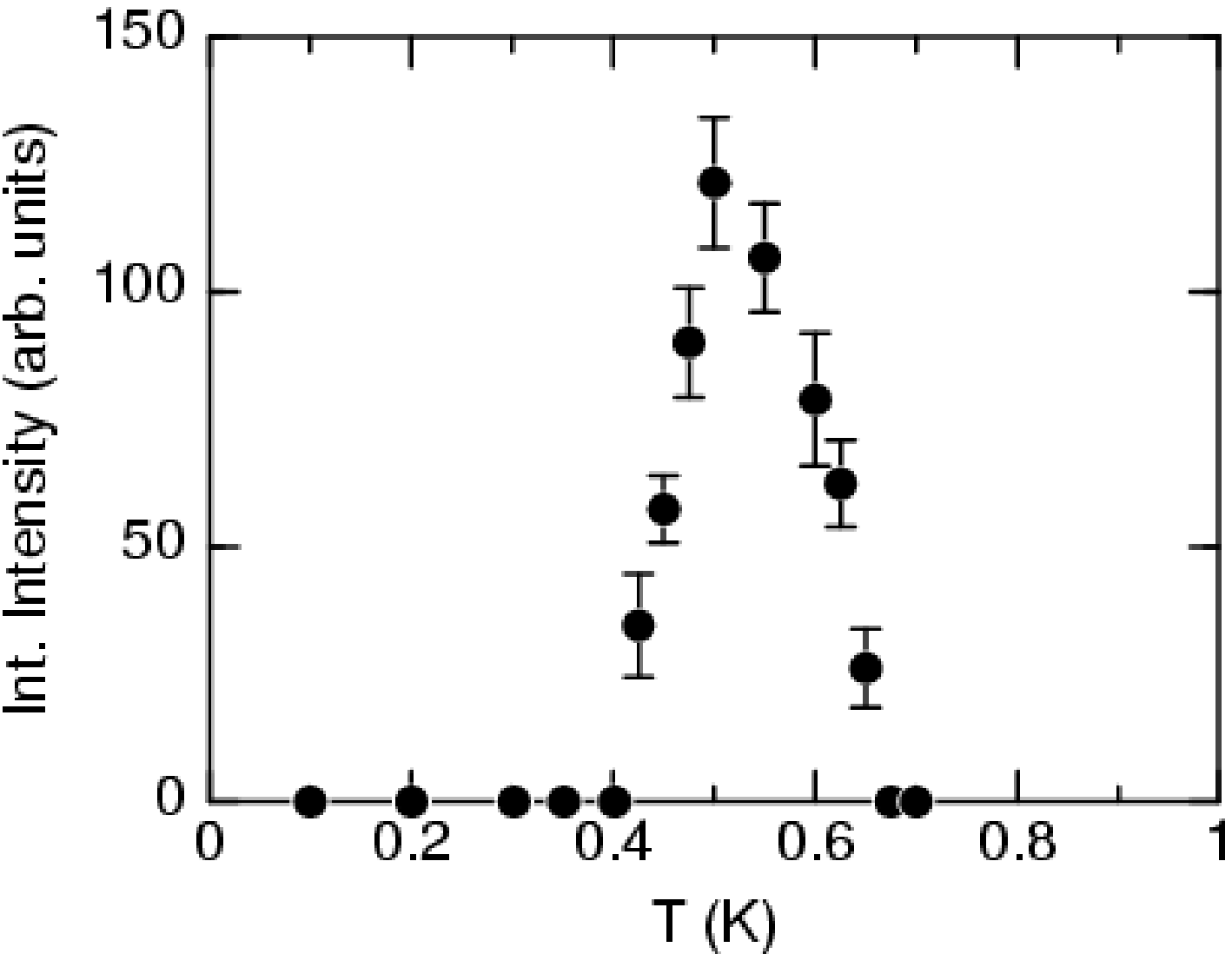} \hfill
\raisebox{0.0cm}
{\includegraphics[width=7.5cm,angle=0,clip]
{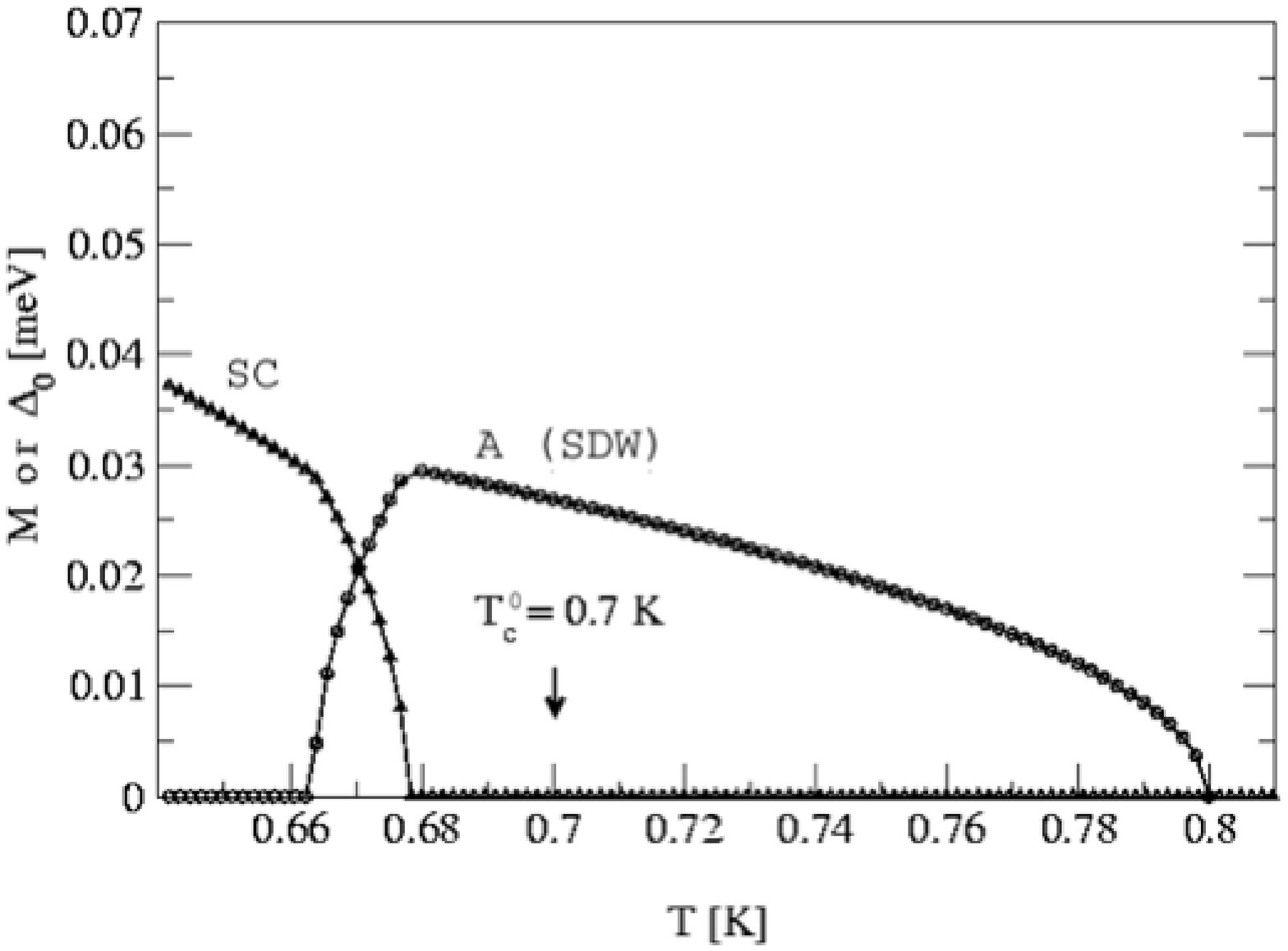}}
\caption{Left panel: Integrated neutron diffraction intensity of the
A-phase SDW satellite corresponding to \bQ~$\simeq$
(0.215,0.215,0.530) as a function of temperature. It is proportional to the
square of the A-phase order parameter in \CCS~\protect\cite{Stockert04}. 
Right panel: Variation with temperature of the A-phase (SDW) and
superconducting order parameters with $\Gamma _{3}$-symmetry, $\phi
_{\Gamma _{3}}(\mathbf{k})\sim \cos k_{x}a-\cos k_{y}a$
calculated for $T_{A}^{(0)}=0.8K>T_{c}^{(0)}=0.7K$ 
\protect\cite{Neef04,Neef04a}.
For this choice of parameters, the magnetic A-phase initially suppresses
the superconducting state which forms at $T_{c}<T_{c}^{(0)}$. The
two ordering phenomena coexist in a narrow temperature range below
$T_{c}$. The superconducting correlations, however, suppress the
long-range magnetic order which disappears continuously at
$T_{A}'<T_{c}<T_{A}^{(0)}$.} 
\label{fig:CeCu2Si2ASCoexistence}
\end{figure}

We further simplify the problem by focussing on the symmetry $\Gamma $
which yields the strongest quasiparticle attraction $g_{\Gamma }$.
Assuming that this most stable order parameter is non-degenerate yields
the separable interaction kernel
\begin{equation}
g\left({\textbf k}s,-{\textbf k}s';-{\textbf {k'}}s'',{\textbf
k}'s'''\right)
\rightarrow \left(i\sigma _{2}\right)_{ss'}\left(i\sigma
_{2}\right)_{s''s'''}\frac{g_{\Gamma }}{2}\, 
\phi _{\Gamma }\left(\textbf k\right)\phi _{\Gamma }^{*}\left(\textbf
{k'}\right)
\label{eq:FinalPairingInteraction}\end{equation}
 and the selfconsistency condition
\begin{equation}
\Delta ({\textbf k})=-g_{\Gamma }\phi _{\Gamma }({\textbf
k})\frac{1}{L}\sum _{{\textbf k}'}\phi _{\Gamma }^{*}({\textbf k}')
\left\langle c_{-{\textbf k}'\downarrow }c_{{\textbf k}'\uparrow }\right\rangle 
\end{equation}
We adopt the Nambu formalism to diagonalise the mean-field Hamiltonian in
eq. (\ref{eq:HSC}) which allows us to reduce it to single-particle
form:
\begin{equation}
H=\sum _{\bk}^{AFBZ}\Psi _{\textbf k}^{\dagger }\left\{
\boldepsilon(\bk)\hat{\tau }_{3}+\mathbf{h}\hat{1}
+\boldDelta\left(\textbf k\right)\hat{\tau }_{1}\right\} \Psi _{\textbf k}
\label{eq:HamiltonNambu}
\end{equation}
where we explicitly exploited the fact that the (pseudo) spin is
conserved. The Nambu spinors $\Psi _{\textbf k}$ have 32 components
and are defined as
\begin{equation}
\Psi _{\textbf k}=\left(c_{{\textbf k}\uparrow
},c_{{\textbf k}+{\textbf Q}_{1}\uparrow },
\ldots ,c_{-{\textbf k}\downarrow }^{\dagger },c_{-{\textbf k}-
{\textbf Q}_{1}\downarrow }^{\dagger },\ldots \right)
\label{eq:NambuSpinor}
\end{equation}
They account for the coherent superposition of particles and holes
which is the characteristic feature of the superconducting state.
Here $\hat{1}$, $\hat{\tau }_{1}$ and $\hat{\tau }_{3}$ denote
the unit matrix and the Pauli matrices in particle-hole space. The
16 wave vectors $\mathbf{Q}_{0}=0,\mathbf{Q}_{1},\ldots ,\mathbf{Q}_{15}$
are the reciprocal lattice vectors appearing in the antiferromagnetic
phase. The set includes the eight propagation vectors of the SDW and
their harmonics. The $\textbf k$-summation is restricted to the reduced
Brillouin zone (AFBZ) of the antiferromagnetic state defined by the
SDW. The structure of the Hamiltonian matrix in particle-hole space is
\begin{equation}
\hat{H}=
\left(\begin{array}{cc}
\boldepsilon\left(\textbf k\right)+\mathbf{h} & \boldDelta \\
\boldDelta & -\boldepsilon\left(\textbf k\right)+\mathbf{h}\end{array}
\right)
\label{eq:HamiltonParticleHole}
\end{equation}
where the $16\times 16$-diagonal matrix contains the quasiparticle
energies of the paramagnetic normal phase
\begin{equation}
\boldepsilon\left(\mathbf{k}\right)_{\mathbf{Q}_{i}\mathbf{Q}_{j}}=
\delta_{\mathbf{Q}_{i}\mathbf{Q}_{j}}\epsilon\left(\mathbf{k}+
\mathbf{Q}_{i}\right)
\end{equation}
The magnetic Umklapp scattering associated with the modulated spin
density is accounted for by the matrix
\begin{equation}
\mathbf{h}=-h_{0}\left(T\right)\mathbf{m}
\label{eq:MagFieldMatrix}
\end{equation}
 whose temperature-dependent amplitude $h_{0}\left(T\right)$ has
to be determined selfconsistently while $\mathbf{m}$ is a given matrix.
The diagonal matrix
\begin{equation}
\boldDelta\left(\textbf k\right)_{\mathbf{Q}_{i}\mathbf{Q}_{j}}=
\Delta_0(T){\boldPhi }\left(\textbf k\right)_{\mathbf{Q}_{i}\mathbf{Q}_{j}}=
\delta _{\mathbf{Q}_{i}\mathbf{Q}_{j}}\Delta _{0}(T)
\phi _{\Gamma }\left(\mathbf{k}+\mathbf{Q}_{i}\right)
\end{equation}
contains the superconducting order parameters with the given $\textbf
k$-dependent
function $\phi _{\Gamma }\left(\textbf k\right)$ and a temperature-dependent
amplitude $\Delta _{0}\left(T\right)$. The selfconsistency equations
in eq.~(\ref{eq:SelfconSC})
can be formulated in terms of the off-diagonal elements of the
$32\times 32$-matrix Green's function
\begin{equation}
\hat{G}\left(i\epsilon _{n},{\textbf k}\right)=\left(i\epsilon _{n}\hat{1}-\hat{H}\left(\textbf k\right)\right)^{-1}
\label{eq:FullNambuGreenFunction}
\end{equation}
 according to
\begin{eqnarray}
h_{0}(T) & = & \frac{U}{L}T\sum _{i\epsilon _{n}}^{\epsilon _{c}}\sum
_{\textbf k}^{AFBZ}\frac{1}{16}\mathrm{Tr}
\left[\mathbf{m}\hat{1}{\bf G}\left(i\epsilon _{n},{\textbf k}\right)\right]\nonumber \\
\Delta _{0}(T) & = & -\frac{g_{\Gamma }}{L}T\sum _{i\epsilon
_{n}}^{\epsilon _{c}}\sum _{{\textbf k}'}^{AFBZ}\frac{1}{2}
\mathrm{Tr}\left[\boldPhi \left({\textbf
k}'\right)\hat{\tau}_{1}\hat{\tau}_{3}{\bf G}\left(i\epsilon _{n},{\textbf
k}'\right)\hat{\tau}_{3}\right]
\label{eq:SelfenergySelfcon}
\end{eqnarray}
Here $i\epsilon_{n}=\pi T\left(2n+1\right)$ denote the T-dependent
Matsubara frequencies and $\epsilon _{c}$ is the energy cut-off required
in weak-coupling theory. The coupling constants $U$ and $g_{\Gamma }$
as well as the cut-off $\epsilon_{c}$ are eliminated in the usual
way in favor of the decoupled transition temperatures $T_{A}^{(0)}$ and
$T_{c}^{(0)}$. The temperature dependence of SDW (A-phase) and SC
order parameters is shown in the right panel of
fig.~\ref{fig:CeCu2Si2ASCoexistence}. On entering the SC regime the A-phase
is expelled in a finite temperature interval. This is nicely confirmed
by recent neutron diffraction results \cite{Stockert04} shown in the
left panel of the figure.
%

\subsection{The (almost) ideal NFL compound \CNG}
\label{ssect:CeNi2Ge2}

\begin{figure}
\raisebox{-5.1cm}
{\includegraphics[width=7cm]{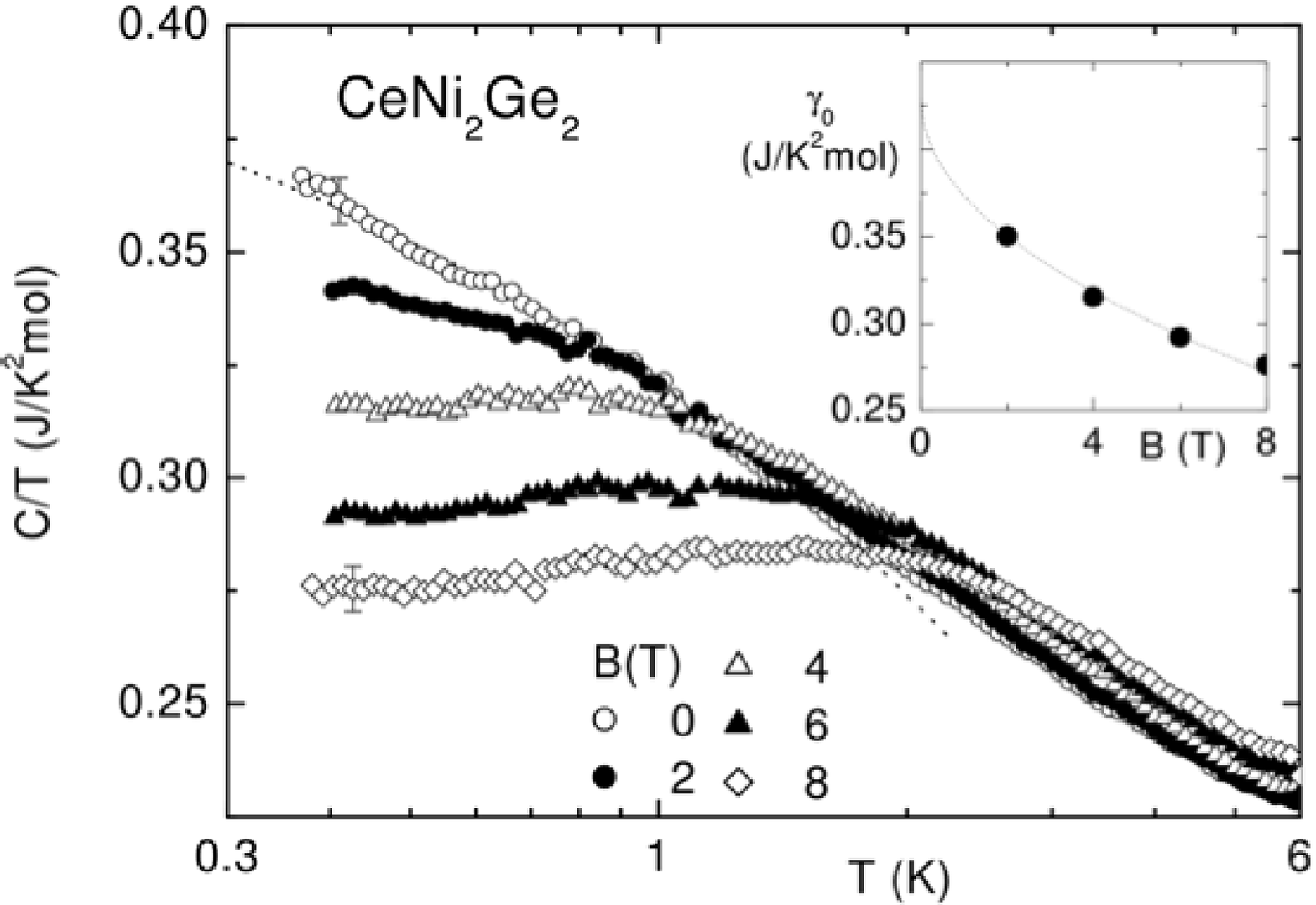}}\hfill
\includegraphics[width=5.5cm,height=7cm,angle=-90]{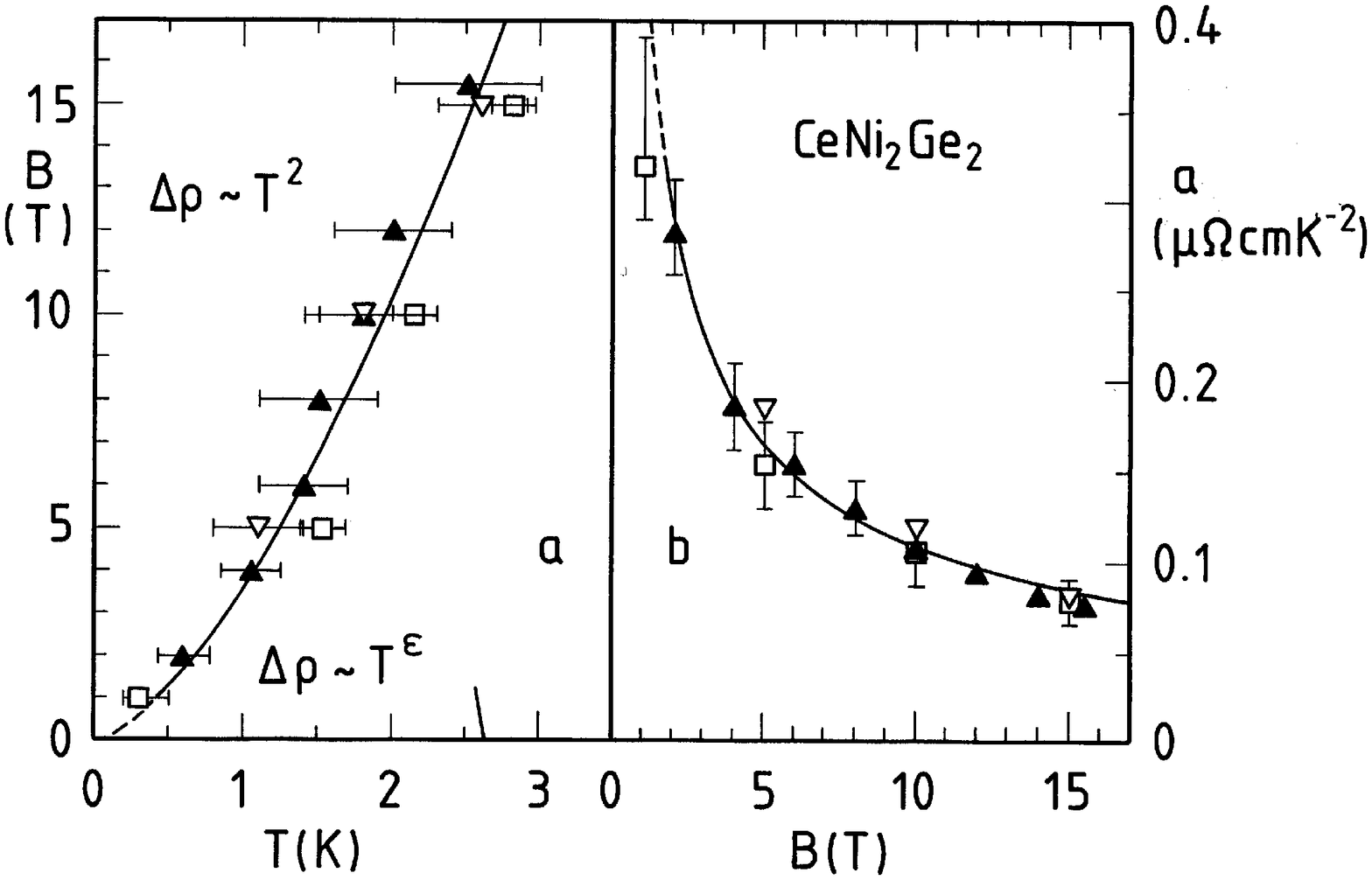}
\caption{Left panel:  Field dependence of $\gamma$ = C/T. Dotted line
(B = 0) indicates $\gamma$(T) = $\gamma_0$-$\beta\sqrt{T}$. Inset
shows $\gamma$(B) in FL (plateau) regime \protect\cite{Gegenwart03}.
Right panel: (a) B-T phase boundary separating the FL/NFL regimes; (b)
Field dependence of the A-coefficient, solid line indicates A $\sim$
B$^{-0.6}$ \protect\cite{Gegenwart03}.} 
\label{fig:CeNi2Ge2NFL}
\end{figure}

This is one of the few Ce-HF compounds which neither order
magnetically nor become superconducting, except for incipient SC
at very low temperatures in very clean samples. Instead it exhibits clear cut
NFL anomalies over more than two temperature decades. Together with the
recently discovered \YRS, \CNG~ is one of the rare cases where one is 
accidentally very close to a QCP at ambient pressure. This is
supported further by the x (Pd -concentration) dependence of the N\'eel
temperature in the alloy series \CNP~\cite{Knebel99}. It is found that T$_N$(x)
extrapolates to zero at at a very small positive concentration
x$_c$. Thus \CNG~ is (in contrast to \YRS) slightly on the nonmagnetic
side of the QCP. After some initial doubts \cite{Gegenwart98} it is
now accepted \cite{Gegenwart03} that the 3D-SDW QCP scenario analysed
in detail in
\cite{Rosch99} is realised in \CNG. This predicts a scaling
$\gamma$(T) = $\gamma_0$ - cT$^\frac{1}{2}$
which is well fulfilled in \CNG~ (fig.~\ref{fig:CeNi2Ge2NFL}) down to
temperatures around 0.3 K. For even lower temperatures sample
dependent upturns in $\gamma$(T) of unknown origin are
present. Likewise the expected NFL resistivity behaviour
$\rho$(T) = $\rho_0$ + $\beta$T$^\epsilon$ with $\epsilon\geq$ 1.5
depending on the sample quality ($\rho_0$) is found.
The very closeness of the AF QCP implies an extreme
sensitivity of the scaling to an external field. With increasing field
strength one crosses rapidly into a Fermi liquid regime as witnessed
by the appearance of a widening plateau in $\gamma$(T)
(fig.~\ref{fig:CeNi2Ge2NFL})
and a LFL type resistivity $\rho$(T) = $\rho_0$ + A(B)T$^2$. The boundary
between the T$^\epsilon$ and T$^2$ scaling marks the transition from
the NFL to the LFL regime (fig.~\ref{fig:CeNi2Ge2NFL}). As this
boundary is approached from B $>$ B$_c$ the A(B)-coefficient diverges
indicating a singular energy dependence of the scattering rate.
Thus as the QCP is approached by varying B and T the quasiparticle
mass and scattering rates in \CNG~ grow in agreement with the predictions of
the 3D SDW scenario for quantum criticality.
%
%
More recently the NFL behaviour in \CNG~has has also been found in a low
temperature divergence of the Gr\"uneisen ratio obtained from thermal expansion
as discussed in sect.~\ref{sect:NFLtheory}.
\subsection{Superconductivity and quantum criticality under
pressure in \CPS~ and \CRS~}
\label{ssect:CePd2Si2}

At ambient pressure \CCS~and \CSG~together with \CCI~and \CII~are still the
only Ce-based HF superconductors, aside from incipient
superconductivity around 100 mK in \CNG~ in exceptionally clean samples
and the recently discovered
non - centrosymmetric SC in \CPSi~\cite{Bauer04}. In the former case this may be
due to the accidental closeness of the A-SDW phase QCP at ambient pressure,
witnessed by the phase diagram in
fig.~\ref{fig:CeCu2Si2BTPhaseDiag} and the associated NFL
anomalies. To find more examples of Ce-based HF superconductors it is
therefore an obvious strategy to look for magnetically ordered Ce-HF
compounds and drive them to a QCP with applied pressure, hoping that a
SC 'dome', however small, might appear. This has been succesful for
\CCG~\cite{Jaccard92} and more recently for \CRS~\cite{Movshovich96}
and also \CIN~and \CPS~\cite{Mathur98}. Note however that strictly
speaking \CRS~is not a HF system due to its small mass enhancement
visible from the low $\gamma$-value at ambient pressure.

The magnetism of these compounds (table~\ref{tab:CePd2Si2}) and of the
Pd-alloy series \CRP~at ambient
pressure has been subject to many investigations
\cite{Mathur98,Grosche01,Demuer02}. While there
is unanimous agreement on the localised moment nature of \CPS~, the
interpretation of magnetism in \CRS~is controversial, partly favoring
the itinerant picture from resistivity and specific heat measurements
\cite{GomezBerisso02} and partly a localised picture from dHvA experiments
and their interpretation and comparison with band structure
calculations for the reference compound LaRh$_2$Si$_2$ \cite{Araki01}. 

\begin{SCfigure}
\raisebox{-0.7cm}
{\includegraphics[width=7.5cm]{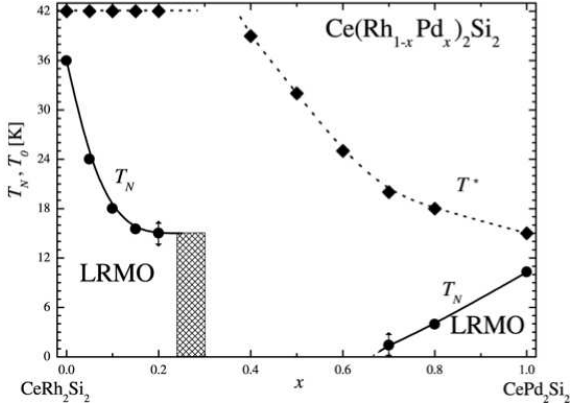}}
\caption{Magnetic x-T phase diagram of \CRP~with N\'eel temperature
T$_N$(x) (circles) and Kondo temperature T$^*$(x) (diamonds). Rh-rich
side: Itinerant SDW regime. Pd-rich side: Local moment regime with
Doniach type phase diagram \protect\cite{GomezBerisso02}.} 
\label{fig:CeRhPd2Si2Phase}
\end{SCfigure}

\begin{figure}
\raisebox{0.1cm}
{\includegraphics[width=7cm]{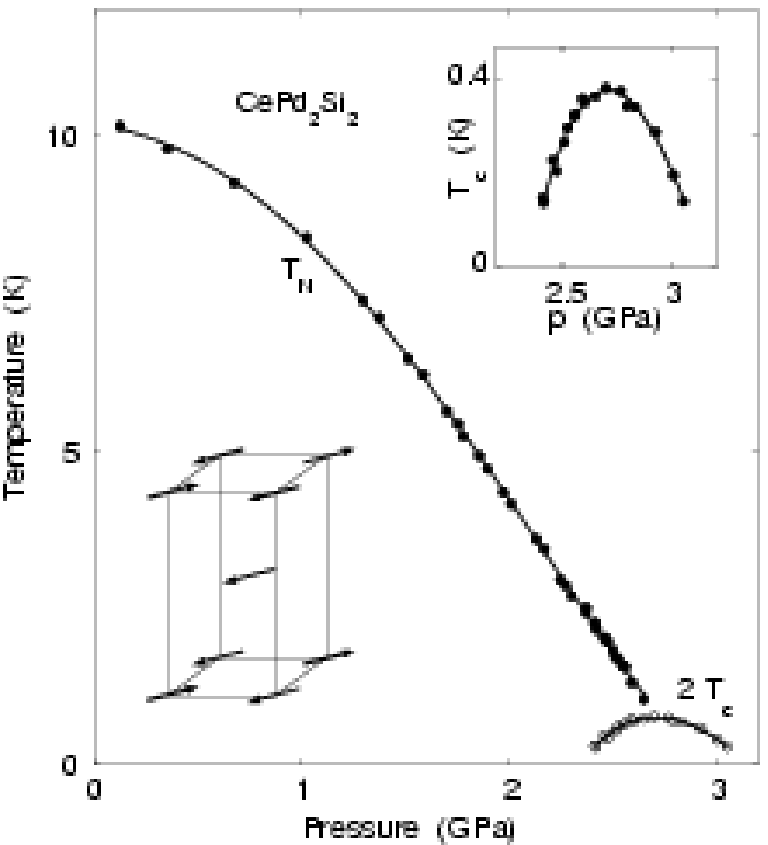}}\hfill
\includegraphics[width=7cm,height=7.7cm]{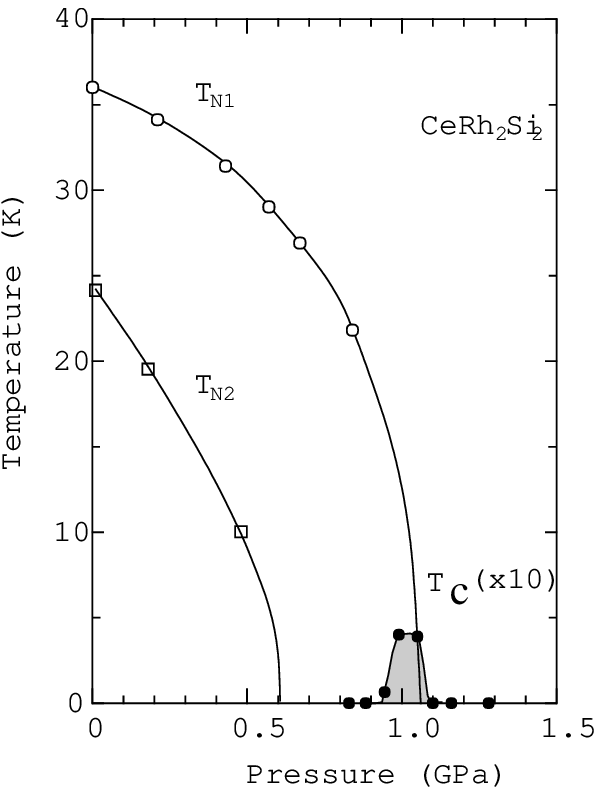}
\caption{Magnetic and SC phase boundaries under hydrostatic pressure in \CPS~
(left panel) \protect\cite{Grosche01} and \CRS~(right panel)
\protect\cite{Onuki04}; note the enlarged scales of T$_c$. In \CRS~the
AF structure changes at T$_{N2}$.} 
\label{fig:CePdRh2Si2p}
\end{figure}

The commensurate AF structure of \CPS~consists of FM (110)
sheets stacked alternatingly with a wave vector \bQ~=
($\frac{1}{2}$,$\frac{1}{2}$,0) and moments oriented along [110].
The local moment nature of \CPS~can directly be infered from
susceptibility measurements and INS experiments \cite{vanDijk00}. The
former give a tetragonal CEF level scheme of the localised Ce$^{3+}$ 4f$^1$
electrons as a sequence of three Kramers doublets $\Gamma_7^{(1)}$(0),
$\Gamma_6$(19 meV) and $\Gamma_7^{(2)}$(24 meV). Well developed AF
spin waves with an uniaxial anisotropy gap $\Delta$ = 0.83 meV and a
dispersion of 2 meV were observed which can be explained within the AF
Heisenberg model for local moments. This is further supported by the magnetic
phase diagram of the \CRP~alloy series shown in
fig.~\ref{fig:CeRhPd2Si2Phase} on the Pd-rich side: The monotonous decrease
of T$_N$ and simultaneous strong increase of the characteristic HF
temperature T$^*$ with Rh-doping (1-x) corresponds to the QCP of a
Doniach phase diagram for a Kondo-lattice type local moment
system. There the destruction of AF order is due to the compensation
of magnetic moments in the Fermi sea of conduction electrons.
\begin{table}
\begin{center}
\begin{tabular}{c|c|c|c|c|c}
\hline
~&$\gamma$(p=0) [mJ/molK$^2$] & T$_N$ [K] & $\mu$ [$\mu_B$] 
 & T$^m_c$(p$_m$) [K] & p$_m$ [GPa] \\
\hline
\CPS     & 250 & 10    & 0.62 & 0.4  & 2.71  \\
\CRS     &  23 & 36    & 1.42 & 0.4  & 1.05 \\
\hline
\end{tabular}
\end{center}
\caption{Material parameters and magnetic properties of the sister
compounds \CRS~and \CPS.}
\label{tab:CePd2Si2}
\end{table}
%
%
The magnetism in \CRP~is less well understood. The origin of the
exceptionally high
T$_{N1}$ = 36 K is a longstanding mystery which defies understanding
in the local moment system if deGennes scaling of T$_{N1}$ with respect to
GdRh$_2$Si$_2$ is applied. Like \CPS~ it orders with
\bQ~= ($\frac{1}{2}$,$\frac{1}{2}$,0), first with a single
\bq-structure at T$_{N1}$ which changes to a 4\bq-structure below
T$_{N2}$ = 24 K (fig.~\ref{fig:CePdRh2Si2p}). Also the (partly)
plateau-like behaviour of T$_N$(x) and T$^*$(x) on the Rh-rich side of
fig.~\ref{fig:CeRhPd2Si2Phase} speaks against the local moment Kondo lattice
picture and therefore an itinerant SDW origin of AF in \CRP~ has been
proposed in \cite{GomezBerisso02}. On the other hand the moment 
size (1.5$\mu_B$, \bm~$\parallel$ c) can be explained within a
CEF-split localised 4f$^1$ model for Ce which is supported by the
entropy gain
$\Delta$S(T$_N$) $\simeq$ R$\ln$2 indicating a CEF Kramers
doublet. From dHvA experiments \cite{Araki01} it was found that some Fermi
surface branches are well explained by LDA results for LaRh$_2$Si$_2$
which advocates that 4f- electrons in \CRS~ do not contribute to the
Fermi surface.
%
%
Alloying leads to an increase of the mean free path and may easily
destroy unconventional superconducting states. Therefore, unlike in
the \CSG~system, no superconductivity has been found around the
magnetic QCPs of the above T-x magnetic phase diagrams. However the
magnetic QCP may also be approached by applying hydrostatic pressure
to the pure compounds. Detailed studies of \CPS~\cite{Grosche01} have
revealed the QCP
to lie at p$_c$ = 2.86 GPa. The almost linear scaling
T$_N$(p) $\sim$ (p$_c$-p) observed above 1.5 GPa is not in agreement
with 3D spin fluctuation theory which predicts a pressure exponent
2/3. The resistivity at p$_c$ exhibits a NFL-type dependence of 
$\rho$(T) = $\rho_0$+BT$^\alpha$ with an exponent
$\alpha\simeq$ 1.2. For samples with sufficiently small $\rho_0$ an
approximately symmetric superconducting dome was indeed identified
around p$_c$ as shown in fig.~\ref{fig:CePdRh2Si2p}. From upper
critical field studies the coherence length was estimated as
$\xi_0\simeq$ 150\AA.

Likewise hydrostatic pressure application destroys AF
in \CRS~\cite{Onuki04} at p$_c$ = 1.05 GPa. The transition seems to
be of first order. Nevertheless in good polycrystalline
samples it was found that SC appears around p$_c$, albeit in a very
small pressure interval from 0.97 to 1.20 GPa, much smaller than
previously thought (fig.~\ref{fig:CePdRh2Si2p}). Again the optimal
T$_c\simeq$ 0.4 K. From the H$_{c2}$ curves \cite{Araki01}
at p$_c$ the coherence lengths $\xi_c\simeq$ 240\AA~ and $\xi_a\simeq$
310\AA~ are obtained. The anisotropy of H$^{a,c}_{c2}$(0) may be
caused by the anisotropic paramagnetic limiting
effect ($\chi_c$/$\chi_a$ = 4), its observation also suggests a SC
singlet pairing state. The dHvA results \cite{Onuki04} show a
discontinuous change of FS topology at 
p$_c$, also the effective masses increase by almost an
order of magnitude to m$^*$/m $\simeq$ 30 above p$_c$. Therefore the
transition across the QCP was interpreted as change from the low
pressure AF local moment phase with light electrons to the
paramagnetic heavy electron state above p$_c$ \cite{Onuki04}.

Although the superficial appearance of quantum critical phase
diagrams in fig.~\ref{fig:CePdRh2Si2p} is quite similar for both
compounds, there is an important difference: quite opposite to \CPS~,
\CRS~ does not show any signature of NFL behaviour around p$_c$,
$\gamma(T)$ has no significant T-depencence below 10 K and its value
even increases up to 80 mJ/molK$^2$ at p$_c$ due the suppression of AF
order \cite{Graf97}. This means there is no direct connection between
NFL behaviour in the QCP regime and the appearance of a SC dome. This
conclusion can also be drawn from the investigation of Ce155 compounds
(sect.~\ref{ssect:CeCoIn5}).

Indeed from the theoretical point of view there is no reason to expect
such a connection as can be concluded from the discussion in
sect.~\ref{sect:SCmech}. At the QCP the magnetic correlation length $\xi_m$
diverges as described by eq.~(\ref{NFLcorr}). Accordingly the spectral
maximum at an energy $\sim\xi_m^{-2}$ of critical spin fluctuations responsible
for NFL behaviour shifts to zero on approaching the QCP
(eq.(\ref{SFLkernel})). As explained in sect.~\ref{sect:SCmech}, this increases
the pair breaking effect due to low lying spin excitations and hence should lead
even to a suppression of T$_c$ close to the QCP. The soft spin
fluctuations responsible for NFL behaviour in the Ce122 compounds
close to the QCP are therefore not essential for the SC pair
formation. Therefore compounds like \CCG~ and \CRS~ which are in the
LFL state around p$_c$
may form the same kind of SC state as NFL compounds like \CPS.

\subsection{The new C\lowercase{e}115 and C\lowercase{e}218 class of
HF superconductors}
\label{ssect:CeCoIn5}

Only recently a new class of promising Ce-based HF superconductors
with the general formula \CMI~(M = Co,Ir,Rh)
has been discovered. Again the guiding principle of looking for SC
close to magnetic QCP's has been succesful, which naturally means that
for the members of the family the SC transition takes place out of a
pronounced NFL-like normal state. These tetragonal compounds (space
group P4/mmm) are composed of alternating n-fold \CIN~layers and
m-fold MIn$_2$ layers derived from the parent compound
\CIN~(n=$\infty$). The n=1, m=1 compounds
(fig.~\ref{fig:ThCr2Si2Structure})
\CCI~and \CII~are ambient pressure
superconductors and the AF \CRI~becomes superconducting under
pressure similar to the AF n=2, m=1 family members \CCIn~ and
\CRIn. Although they partly have FS sheets with 2D character like
slightly warped cylinders oriented along c, the physical
properties, e.g. resistivity and upper critical field are not
excessively anisotropic. In fact, an analysis of uniaxial pressure
effects of both normal-state and SC properties based on
high-resolution thermal expansion measurements reveals that the SC
T$_c$ is strongly affected by at least two factors: The lattice
anisotropy and the 4f-conduction electron hybridisation which is most
sensitive to c-axis lattice distortions \cite{Oeschler03b}.

Their electronic structure is well understood
and the ambient pressure superconductors \CCI~and \CII~have conduction
band states with strong 4f admixture, contrary to the ambient
pressure AF compounds which have well localised 4f states witnessed by
the similarity of their FS to the La parent compound. Large mass enhancements
and specific heat $\gamma$ coefficients are found but achieving the FL
state may require application
of a sufficiently large field. Maximum T$_c$s $\simeq$ 2.5 K are achievable
under pressure and already realised at ambient pressure for \CCI. This
is higher than in other families of Ce- or U-HF superconductors,
except for \CCS~ under pressure, and a
factor of ten larger than in the parent compound \CIN. A
compilation of physical data is given in table~\ref{tab:Ce155C}, see also
\cite{Thompson02}. Certainly \CCI~has turned out to be the most interesting
system and is also most thoroughly investigated, partly due to its
large ambient pressure T$_c$ and large H$_{c2}$ as well as
availability of excellent single crystals. Its unconventional
superconducting state with d-wave symmetry is possibly the first which
exhibits a change from a second to first order SC transition at
H$_{c2}$(T) below a temperature T$_0<$ T$_c$. Furthermore specific
heat, NMR 
and ultrasonic attenuation suggest that \CCI~is the first SC which
exhibits the elusive Fulde-Ferrell-Larkin-Ovchinnikov (FFLO) phase at
high field and low temperature.

\subsubsection{Basics of the electronic structure}

There are extensive LDA calculations and dHvA investigations of the
CeMIn$_5$ and Ce$_2$MIn$_8$ electronic structure and FS topology
\cite{Shishido02,Maehira03,Onuki04,Ueda04}. For the reference compound
LaRhIn$_5$ excellent agreement is found and the dHvA frequencies of a
few sheets exhibit clear 2D cylindrical structure. The AF compound
\CRI~has localised 4f electrons which are CEF split into three Kramers
doublets $\Gamma_7^{2}$(0), $\Gamma_7^{1}$(6.9 meV) and
$\Gamma_6$(23.6 meV) \cite{Christianson02}. Therefore \CRI~should have
a similar FS as the La parent. Its AF order introduces many
new sheets by folding into the AF BZ but the main FS sheets are indeed
in good agreement. On the other hand the almost identical FS of \CCI~and \CII~
are well explained by including the 4f states as itinerant electrons
in the FS volume which makes them highly different from the above \LRI~ and
\CRI~FS. The selfconsistently calculated f-level occupations for \CCI~and \CII~
are indeed close to one \cite{Maehira03}. The enhanced $\gamma$ values are
given in table~\ref{tab:Ce155C} and are generally larger for the itinerant
4f-electron compounds. The different size of mass enhancement for \CCI~
and \CII~was attributed to the influence of CEF effects \cite{Maehira03}.
The mass anisotropies as obtained from upper critical field measurements
\cite{Shishido02} are considerable, m$^*_c$/m$^*_a$ = 4.8 in \CII~and 
m$^*_c$/m$^*_a$ = 5.6 in \CCI, reflecting the partly 2D FS topology.
As in \CRI~the AF bilayer compound \CRIn~has largely localised 4f
states and therefore again the FS sheets are very similar to the \LRIn~
parent compounds \cite{Ueda04}, three similar sheets with 2D cylindrical
appearance have been found in dHvA experiments and LDA calculations.
\begin{table}
\begin{center}
\begin{tabular}{l|c|c|c|c|c}
\hline
~& $\gamma$ [mJ/molK$^2$] & T$_N$[K] & $\mu$[$\mu_B$] 
& T$_c$(p=0)[K] & T$_c^m$(p$_m$)[K] \\
\hline
\CIN     & 130      &10.2  & 0.65  &  -     & 0.25 (2.6 GPa)  \\
\hline
\CCI     & 350     & -    & -     &  2.3    & 2.5 (1.56 GPa)  \\
\CII     & 750     & -    & -     &  0.4    & -  \\
\CRI     & 400     & 3.8  & 0.37  &  -      & 2.2 (2.5 GPa)  \\
\hline
\CCIn    & 500     & -    & -     &  0.4    & -  \\
\CIIn    & 700     & -    & -     &  -      & -  \\
\CRIn    & 400     & 2.8  & 0.55  &  -      & 2.0 (2.3 GPa) \\
\hline   
\end{tabular}
\end{center}
\caption{Material parameters of Ce115 and Ce218 HF superconductors.}
\label{tab:Ce155C}
\end{table}
\subsubsection{AF quantum critical points and superconductivity }

When magnetic order changes to a different symmetry or vanishes as a
function of an external or internal parameter this may be interpreted as
a quantum phase transition provided the respective ground state
energies are largely determined by the contribution from quantum
fluctuations. If the transition is of second order the critical
parameter value then defines a proper magnetic quantum critical point (QCP).
In its vicinity NFL behaviour and a superconducting phase
transition may be induced by spin fluctuations of the type discussed
in sect.~\ref{sect:SCmech}. The Ce115 and Ce218 compounds give
considerable support for this scenario.

\begin{figure}[tbh]
\raisebox{-0.2cm}
{\includegraphics[clip,width=7cm]{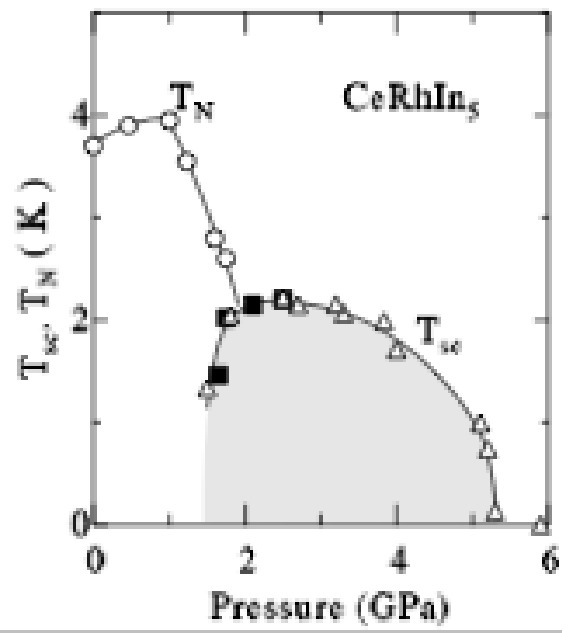}}\hfill
\includegraphics[clip,width=7cm,height=7.82cm]{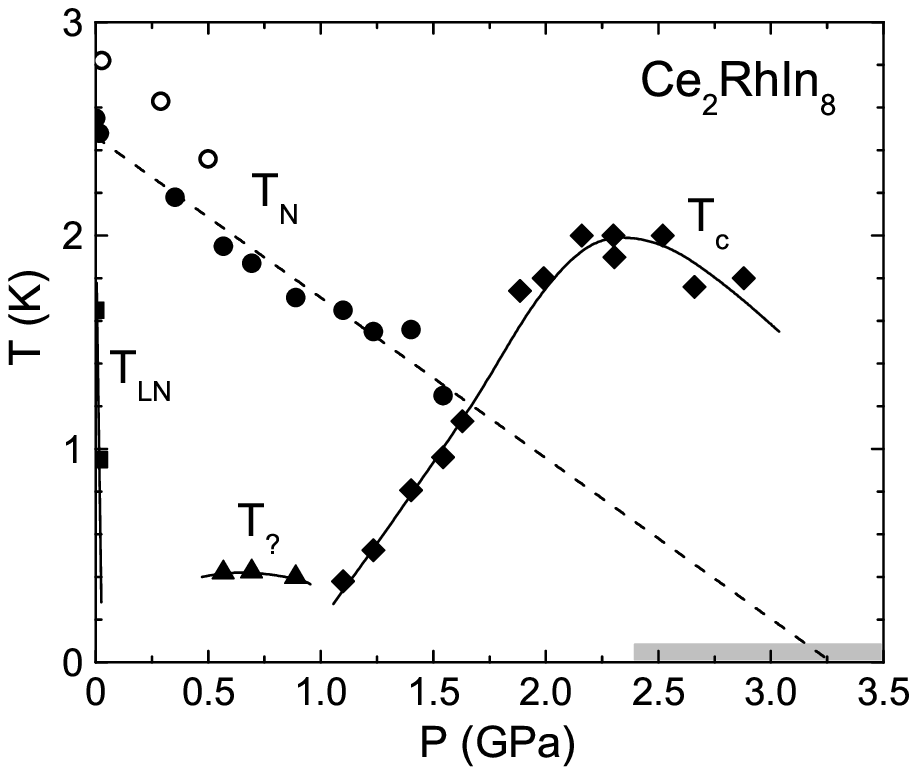}
\caption{Quantum critical p-T phase diagrams for
\CRI~\protect\cite{Onuki04} and \CRIn~\protect\cite{Nicklas03}. When
T$_N\sim$T$^m_c$ as in \CRIn, AF and SC coexist over considerable
pressure range, contrary to the situation in \CRI~where T$_N\gg$ T$^m_c$.} 
\label{fig:CeRhIn5phase}
\end{figure}

In the T-p phase diagrams of Ce115 and Ce218 and also Ce122 compounds
which are AF at ambient pressure and SC under pressure two typical
idealised situations may occur \cite{Kitaoka04}: (1) When T$_N$(p=0) $\gg$
T$_c$(p$_c$) a symmetric SC
dome around p$_c$ appears and FL behaviour is observed above T$_c$ and
for p $>$ p$_c$, AF order is suppressed inside the SC
region. This situation is realised for \CRS~and \CPS~ as seen in
fig.~\ref{fig:CePdRh2Si2p}. (2) When
T$_N$(p=0) $\geq$ T$_c$(p$_c$) the SC region is commonly asymmetric around the
QCP and AF order partly coexists with SC. Around p$_c$ and above T$_c$ an
extended region of NFL behaviour is observed. This holds for \CSG,
\CRI~and \CRCI~ (fig.\ref{fig:CeRhIn5phase}). This picture is,
however, oversimplified as is obvious from the case of \CIN~where the
conditions (1) are fulfilled but nevertheless coexistence SC/SDW has
been identified in NMR experiments\cite{Kawasaki04}. In fact these
results suggest that instead of having a QCP within the SC dome a
first order transition with AF/PM phase separation occurs. It remains
to be seen whether this also holds true for \CPS.
%
\begin{figure}[tbh]
\raisebox{0.7cm}
{\includegraphics[clip,width=7cm]{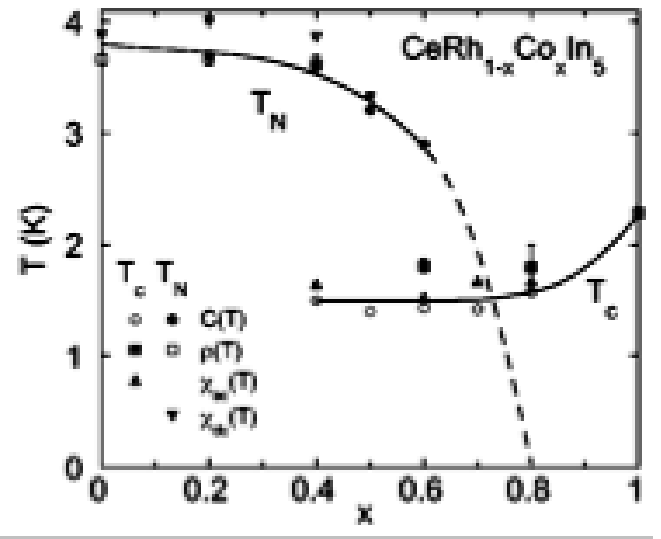}}\hfill
\includegraphics[clip,width=7cm,height=6.65cm]{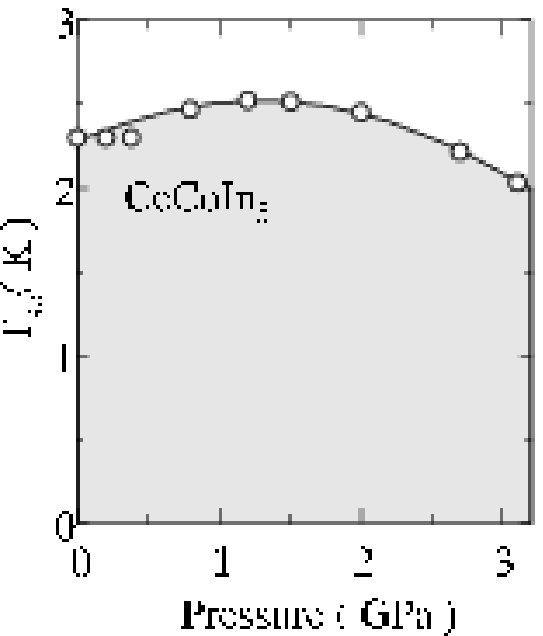}
\caption{Left panel: Coexistence of AF order and SC in \CRCI~
('negative pressure' side) \protect\cite{Zapf01}. Right panel:
Hydrostatic pressure
dependence of T$_c$ in \CCI. The critical pressure where T$_c$ = 0 is
not known \protect\cite{Onuki04}.} 
\label{fig:CeRhCoIn5phase}
\end{figure}
%
Especially the quantum criticality of \CCI~and its connection to
superconductivity has been
clarified by external field and pressure application and by internal
pressure through Rh substitution of Co. At ambient pressure \CCI~ is
nonmagnetic and the 'large' T$_c$ is almost independent in hydrostatic
pressure up to 3 GPa and no magnetic order appears. Apparently doping
with Rh, which has a larger ionic radius compared to Co, exerts a
'negative pressure' which for 1-x $>$ 0.2 Rh doping achieves AF order 
(fig.~\ref{fig:CeRhCoIn5phase}) while SC persists, albeit with
slightly reduced
T$_c$. Thus a considerable regime of AF/SC coexistence appears. Taken
together Rh-doping and hydrostatic pressure do
not present an ideal dome-shaped SC region, it is rather reminiscent
of slightly doped \CSG~under pressure. The combined
x,p-T phase diagram of the Rh-doped compound \CRII~ \cite{Pagliuso01} and
\CII~under hydrostatic pressure \cite{Muramatsu03} is quite similar
to fig.~\ref{fig:CeRhCoIn5phase}. \CRII~also develops AF order that
coexists with SC between 0.35$<$ x $<$0.5 \cite{Zheng04}. The T$_c$(p) curve of
\CII~starts at T$_c$(p=0) = 0.4 K and reaches T$_c$(p$_m$) = 1.05 K at
p$_m$ = 2.2 GPa and finally drops to zero at p$_c$ = 6 GPa. This
critical pressure for the \CCI~sister compound is not yet known. 

Other family members which are AF at ambient pressure but not SC
naturally exhibit better realisations of the canonical AF/SC quantum
critical phase diagrams under hydrostatic pressure. This is already
true for the AF parent compound \CIN~(T$_N$ = 10K) of the Ce115 family
\cite{Onuki04} with an optimum T$_c$ = 0.2 K at 2.5 GPa and especially for
the localised 4f compound \CRI~(x = 0 in
fig.~\ref{fig:CeRhCoIn5phase}) as is shown in
fig.~\ref{fig:CeRhIn5phase}. The related AF bilayer compound \CRIn~ is
another exciting example of AF/SC quantum critical phase diagram
(fig.~\ref{fig:CeRhIn5phase}). The remarkably high optimum T$_c$
rivals that of ambient pressure \CCI~ and is comparable to its own
T$_N$(p=0) = 2.8K.
%
%
\subsubsection{NFL anomalies in field and pressure induced QCP's}

The common signature of a NFL state are anomalous temperature
and field scaling exponents for thermodynamic quantities like specific
heat, susceptibility, Gr\"uneisen parameters (thermal expansion) and
transport quantities like resistivity or Hall coefficient. Their origin
and appearance has been briefly discussed in sect.~\ref{sect:NFLtheory}. 

\begin{figure}[tbh]
\raisebox{0.8cm}
{\includegraphics[clip,width=7.5cm]{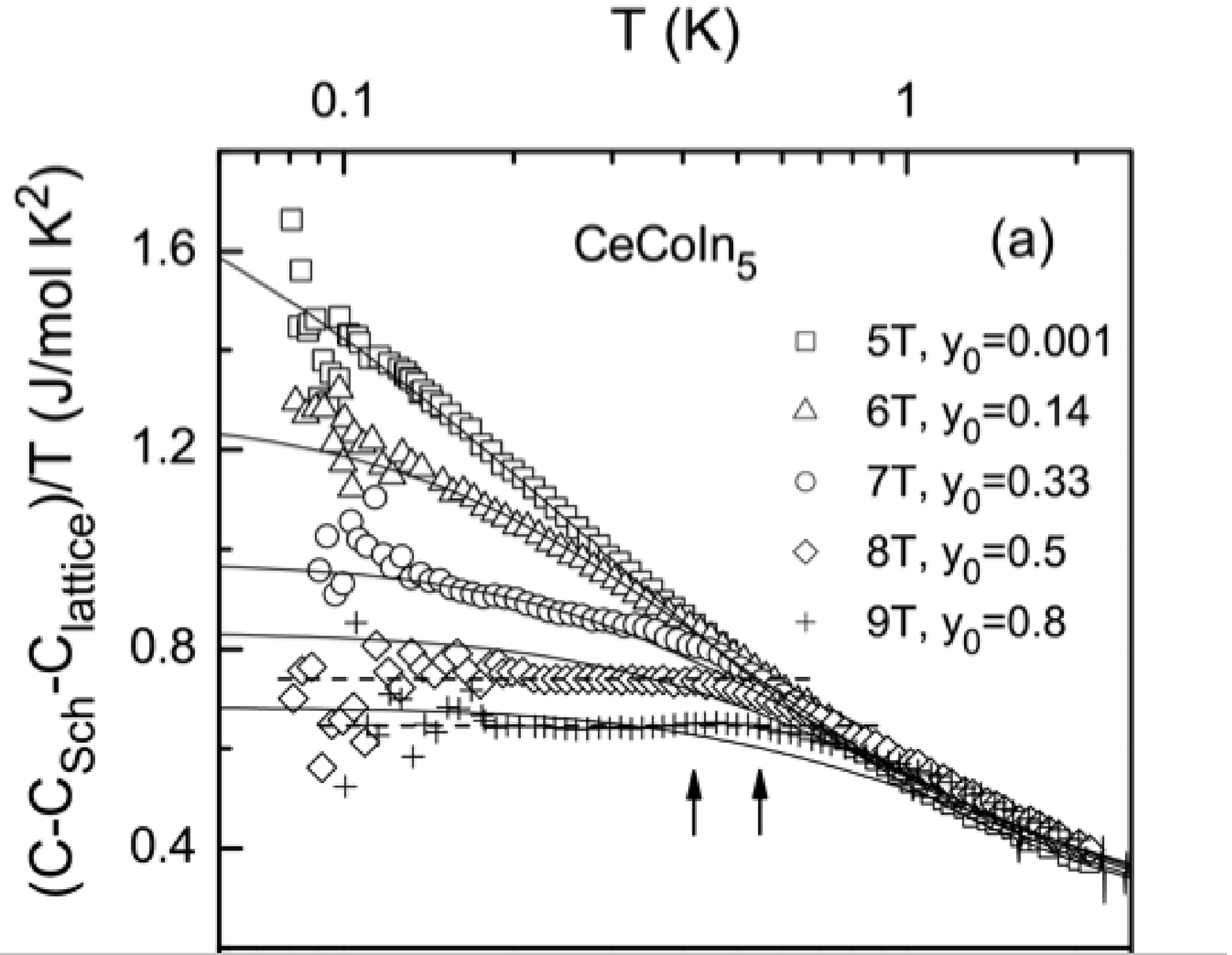}}\hfill
\includegraphics[clip,width=7.5cm,height=5.75cm]{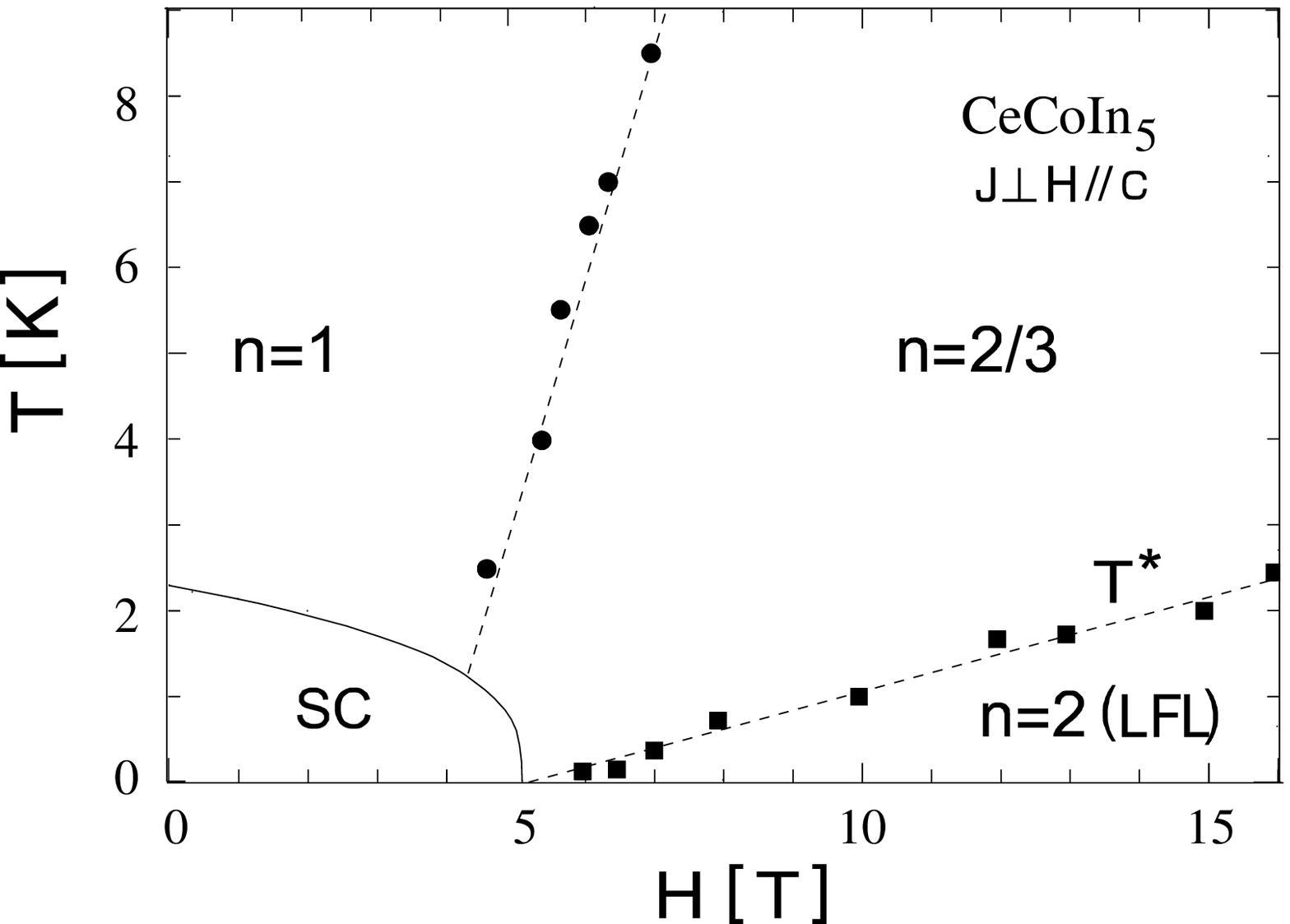}
\caption{Left panel: Field dependence of $\gamma$ = C/T for
\bH~$\parallel$ [001]. Solid lines are fits by Moriya's SCR spin
fluctuation theory \protect\cite{Bianchi03}. Right panel: B-T phase
diagram of \CCI~ showing the evolution of two distinct NFL regimes
(characterised by different exponents n of $\Delta\rho$(T) $\sim$
T$^n$) and the high field LFL phase \protect\cite{Paglione04}.} 
\label{fig:CeCoIn5NFL}
\end{figure}

Pronounced NFL anomalies have been identified in the ambient pressure SC
\CCI~and \CII, pointing to the closeness of the QCP \cite{Kim01}. This
is well illustrated by $\rho$(T) data for \CCI~at ambient pressure and
zero field. They exhibit
linear temperature behaviour above T$_c$ and the logarithmic
$\gamma$(T) variation between 0.4 K and 8 K for fields larger than
H = H$^{[001]}_{c2}$ = 5 T but
smaller than H = 8T (fig.~\ref{fig:CeCoIn5NFL}) \cite{Bianchi03}. From systematic
$\rho(T,H)$ investigations of \CCI~ a B-T phase diagram may be mapped
out showing the development of a field-tuned QCP and associated
with it two different NFL regimes. They are characterised by
$\rho(T)$ = $\rho_0$+BT$^n$
where n=1 or n=$\frac{2}{3}$ and a low temperature crossover at T$^*$
to the LFL regime \cite{Paglione04} is observed. The n=1 regime is
similar to the underdoped regime in high T$_c$ superonductors and
speculations about the existence of a spin pseudo-gap in \CCI~and
\CRI~have been made \cite{Sidorov02}. This is supported by normal
state Hall angle and magnetoresistance (MR) measurements \cite{Nakajima04}
which show that in the n=1 NFL regime the Hall angle
$\Theta_H\sim$ T$^2$ whereas Kohlers rule $\Delta_{xx}(H)\sim H^2$
which holds for LFL systems is strongly violated and a different scaling
law is obeyed. The similarties of transport coefficients in \CCI~to high T$_c$
materials suggests a common origin of NFL behaviour in the critical 2D
AF spin fluctuations in both type of compounds. This idea is also
supported by an observed giant Nernst effect \cite{Bel04} as in the
cuprates. Clear signature of NFL behaviour has
also been seen in thermal expansion measurements \cite{Oeschler03} and
the associated Gr\"uneisen ratio. A compilation of various scaling
laws in the NFL regime of the Ce115 and Ce218 compounds may be found
in \cite{Kim01,Thompson02}. A kind of inverted behaviour compared to
\CCI~ shown in fig.~\ref{fig:CeCoIn5NFL} is seen in \CIIn, because in this
nonmagnetic HF compound LFL behaviour at zero field changes to NFL
behaviour at a field of 13 T, which is interpreted as the appearance
of a field induced magnetic QCP \cite{Kim04}.
%
%
\subsubsection{The superconducting gap function in \CCI}

Heavy Fermion systems are unconventional superconductors 
regarding their pairing mechanism which is of magnetic origin in most
cases and also
with respect to non-s-wave symmetry of the order parameter, generally
belonging  to a non-trivial representation of the
total symmetry group. Often, but not necessarily, this is connected with the
appearance of nodes or zeroes in the gap function $\Delta(\bk)$. \CCI~
is the prime and best studied example of unconventional SC in the new
family. It is a strong coupling SC with very large BCS ratio
2$\Delta$/kT$_c$ = 8.86 at ambient pressure \cite{Yashima04} and the
gap function is believed to exhibit d-wave symmetry. This has
already been suspected from the T$^3$-power law behaviour of NMR and NQR
relaxation rate 1/T$_1$ and the absence of a Hebel-Slichter peak
\cite{Kohori01}. The same result was found for \CII. In addition \CCI~
shows a $^{115}$In- Knight shift reduction K $\sim$ T for field along a and c,
indicating an isotropic reduction of spin susceptibility in the SC
state. The T- dependence of 1/T$_1$ and K consistently suggest a
singlet pair state with line nodes, e.g. a d-wave state for \CCI. 
%
\begin{SCfigure}
\raisebox{-1.1cm}
{\includegraphics[clip,angle=0,width=7cm]{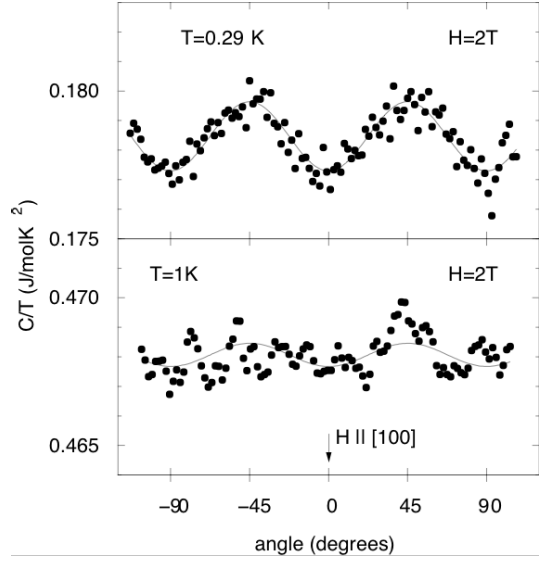}}
\caption{Specific heat $\gamma$-coefficient for \CCI~ as function of field
angle $\phi$
with respect to [100] (\bH~$\perp$ c-axis) in the ab plane. Maxima at
[110] correspond to antinodal direction for \De, suggesting a
d$_{xy}$-gap function. The oscillation amplitude decreases with increasing
temperature \protect\cite{Aoki04}.}  
\label{fig:CeCoIn5specosc}
\end{SCfigure}
%
In subsequent field-angle resolved magnetothermal conductivity
\cite{Izawa01} and   specific heat experiments \cite{Aoki04} the
position of line nodes, in \bk-space
have been investigated to identify which irreducible D$_{4h}$
representation among the five d-wave states is realised.
 In this method a peculiar feature of nodal superconductors in the
vortex phase is used as has been first discussed by Volovik
\cite{Volovik93}. Because of $\Delta(\bk)$ = 0  along certain directions in
\bk-space, quasiparticles can tunnel into the inter-vortex region
where they acquire a Doppler shift energy \bv$_s\cdot$\bk~ due to the
superfluid velocity field  \bv$_s$(\br). This 
leads to a finite, field induced residual quasiparticle DOS. For
a gap with line nodes one has (E $\ll\Delta$):
\begin{equation}
\frac{N_s(E,\bH)}{N_n}\simeq\frac{1}{\Delta}
\la\la|E-\bv_s(\br)\cdot\bk|\ra\ra
\label{RESDOS}
\end{equation}
where a double average over the Fermi surface and the vortex
coordinate has to be taken. This contributes to the specific heat and
most importantly supports a
heat current $\perp$ to the vortices which is not possible in s-wave
superconductors for T $\ll$ T$_c$. For H $\ll$ H$_{c2}$ the residual DOS
and hence $\gamma$(H) has a $\sqrt{H}$ dependence characteristic of
nodal superconductors \cite{Nagai04}. In addition N$_s$(E,\bH) depends
on the field {\em direction} with respect to crystal axes and the
position of node
lines. As a function of polar field angles $\theta$ and $\phi$ the
residual DOS and hence the specific heat and thermal conductivity will
exhibit angular oscillations whose
type and phase allows one to draw direct
conclusions on the \bk-space positions of nodes in
$\Delta$(\bk). The Doppler shift phenomenon has now been exploited
succesfully to clarify the nodal gap structure in many unconventional
superconductors \cite{Izawa01,Izawa01a,Izawa02,Izawa02a,Izawa03}
including \CCI; see also \cite{Thalmeier03} . There both
$\kappa_{xx}$($\theta$,$\phi$) \cite{Izawa01} and
C($\theta$,$\phi$) \cite{Aoki04} exhibit fourfold
oscillations in the azimuthal angle $\phi$ which means that the gap
function $\Delta$(\bk) has four line nodes parallel to the c-axis on the main
cylindrical FS of \CCI. The maximum in the oscillation will be
achieved when each of them gives a Doppler shift contribution in
eq.~(\ref{RESDOS}), i.e. when \bH~ points along the anti-nodal
direction, the minimum
occurs in the nodal direction. Surprisingly the two measurements presently
disagree on the position of the line nodes: Thermal conductivity
results suggest line nodes along c at [110] and
equivalent directions. This means a d$_{x^2-y^2}$ (B$_{1g}$) symmetry of
the SC gap in \CCI~as for high T$_c$ compounds. On the other hand  in
the specific heat results of fig.~\ref{fig:CeCoIn5specosc} the maxima
in the $\phi$- oscillations are shifted by 45$^\circ$, therefore node
lines along c should be situated at the tetragonal [100]
and equivalent positions meaning a d$_{xy}$ (B$_{2g}$)-type symmetry
of the SC gap. This discrepancy is not fully resolved so far.   
 
\subsubsection {Vortex state and FFLO phase in \CCI}

\begin{figure}[tbh]
\raisebox{0.0cm}
{\includegraphics[clip,width=7.5cm]{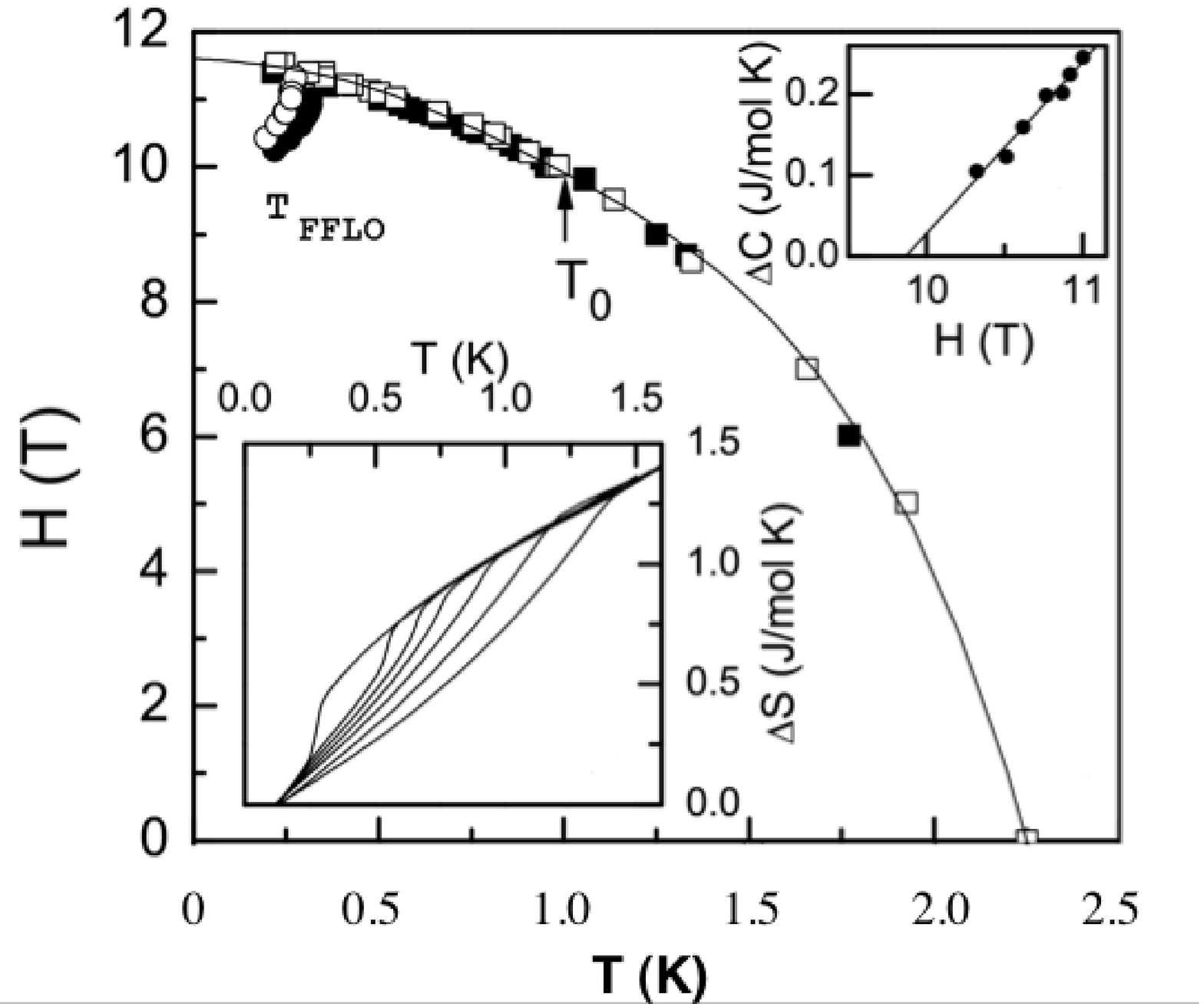}}\hfill
\includegraphics[clip,width=7.5cm,height=6.1cm]{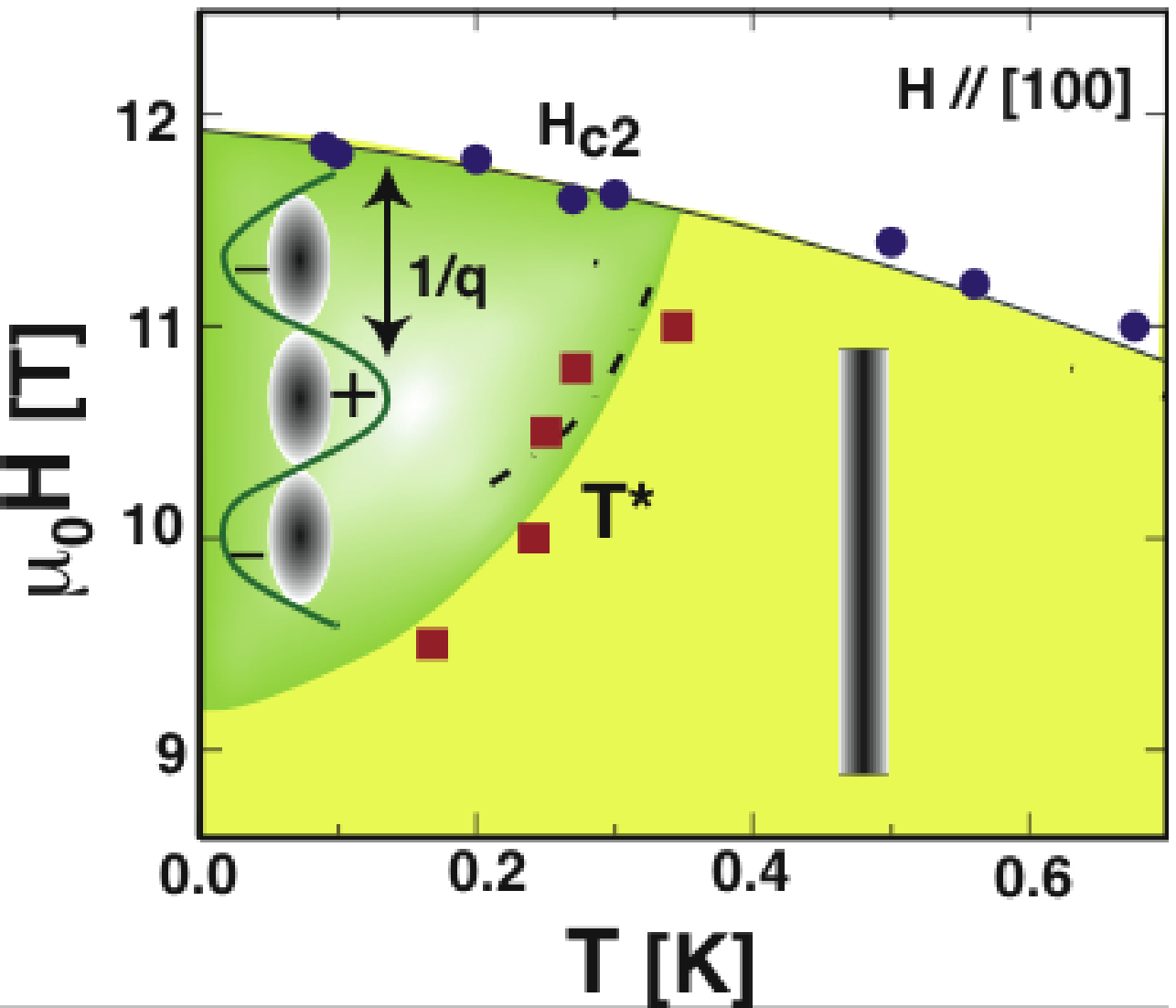}
\caption{Left panel: H$_{c2}$ curve from specific heat results. At
T$_0\simeq$ 1.1 K the SC transition changes from 1$^{st}$ to 2$^{nd}$
order. Lower inset shows entropy gain. Upper inset shows specific heat
jump at T$_{\mbox{FFLO}}$ line \protect\cite{Bianchi03}. Right panel:
enlarged FFLO corner determined from sound velocity
measurements. Segmentation of vortices in FFLO phase with wave length
1/q is indicated \protect\cite{Watanabe03}.}
\label{fig:CeCoIn5Wata}
\end{figure}

Besides providing important clues on gap function symmetry the vortex
state of \CCI~ has proved highly interesting, even unique, in its own
right. It was known already for a long time that for low temperatures
the SC transition at H$_{c2}$ should change from second to first
order. This is due to a competition of orbital- and Pauli-pair breaking
effects characterised by the Maki parameter 
$\alpha$ = $\sqrt{2}$H$_{c2}$/H$_P$ where
H$_P$ = $\Delta_0$/$\sqrt{2}\mu_B$ (for g = 2) is the Pauli limiting field
. Conditions are favorable if the former is weak
(large H$_{c2}$ or $\alpha>$1) compared to the latter and if
spin-orbit scattering by impurities is negligible. Neglect of the
orbital effect ($\alpha\rightarrow\infty$) leads to T$_0$ = 0.55T$_c$
for the appearance of the first order transition in an s-wave SC.
 Apparently these conditions have never been met in
superconductors investigated so far. In \CCI, due to the large effective mass
m$^*\simeq$ 100 m and resulting extremely high
H$^{[110]}_{c2}$(0) = 11.9 T, and due to excellent sample quality they 
are much better fulfilled. Indeed in this compound a change from
second to first order superconducting transition for field \bH~
$\parallel$ [001] has been found at T$_0\simeq$ 0.3T$_c$
(corresponding to $\alpha$ = 3.5) by specific
heat \cite{Bianchi02,Radovan03} and thermal expansion
\cite{Oeschler03b} measurements. The first
order transition was also found for H $\parallel$
[110] from magnetisation \cite{Tayama02} and specific
heat \cite{Bianchi03} measurements below T$_0$ =
0.5T$_c$. (fig~\ref{fig:CeCoIn5Wata}).
%
%
The same favorable circumstances may explain an even more exotic 
observation in \CCI~for \bH~$\parallel$[110]. At very small temperatures
an additional phase appears in the vortex state
(fig.~\ref{fig:CeCoIn5Wata}). It is now considered as the first
realisation of the Fulde-Ferrell-Larkin-Ovchinnikov (FFLO) phase where
the electrons form Cooper pairs with a finite momentum q. It appears when the
Pauli effect largely dominates the orbital effect
($\alpha>$ 1.8 \cite{Gruenberg66}) and spin orbit scattering is very
small. Apparently \CCI~ is the first SC compound where these conditions are
fulfilled leading to the spatially modulated FFLO SC pair state with a
wave number q $\simeq$ 2$\mu_B$H/$\hbar$v$_F$. In this state the order
parameter should have planar nodes perpendicular to the field leading
to a segmentation of vortex lines along the vortex direction. A
definite proof for this conjecture by small angle neutron scattering
is still lacking,
but indirect evidence for the segmentation has been found by sound
velocity measurements \cite{Watanabe03}. The segmentation of vortices
on crossing the T$_{\mbox{FFLO}}$ line changes their pinning properties which
in turn leads to a Lorentz force anomaly for sound modes with
displacement vector \bu~ $\perp$ \bH. The estimated segmentation length is
remarkably close to the calculated modulation length
1/q = 60\AA. Also this method allows an independent
determination of specific heat results and assures that the new phase
is not due to field induced magnetic order since the latter should not
have any effect on the vortex pinning. The FFLO corner is clearly
visible for \bH~ in the ab- plane but hard to identify for \bH~ along
[001] with the smaller $H_{c2}$ in agreement with theoretical
predictions. Evidence for the FFLO state for \bH~$\parallel$[110] has
recently also been found from $^{115}$In-NMR measurements
\cite{Kakuyanagi04} which give
a T$_{\mbox{FFLO}}$(H)-line similar to that from specific heat and
sound velocity results and provide direct evidence for the spatial
texture of the order parameter in the FFLO phase.

\subsection{HF superconductivity without inversion symmetry in \CPSi} 
\label{ssect:CePt3Si}

All previously discussed Ce- intermetallic compounds and in fact all
HF superconductors known possess centrosymmetric space groups
which contain an inversion center. This leads one to a natural classification
of SC pairs into even and odd parity or singlet and triplet
states. For the singlet channnel only time reversal invariance is
required in order to assure the necessary degeneracy of paired electron
states with opposite momenta and spins. For the formation of triplet
pairs however, electron states with opposite momenta but equal spins must also
be degenerate which requires inversion symmetry as an additional
necessary condition \cite{Anderson84}. The lack of an inversion center
may also strongly affect the magnetic properties of a superconductor as
shown in \cite{Bulaevskii76}.

\begin{figure}[htb]
\includegraphics[width=7.5cm]{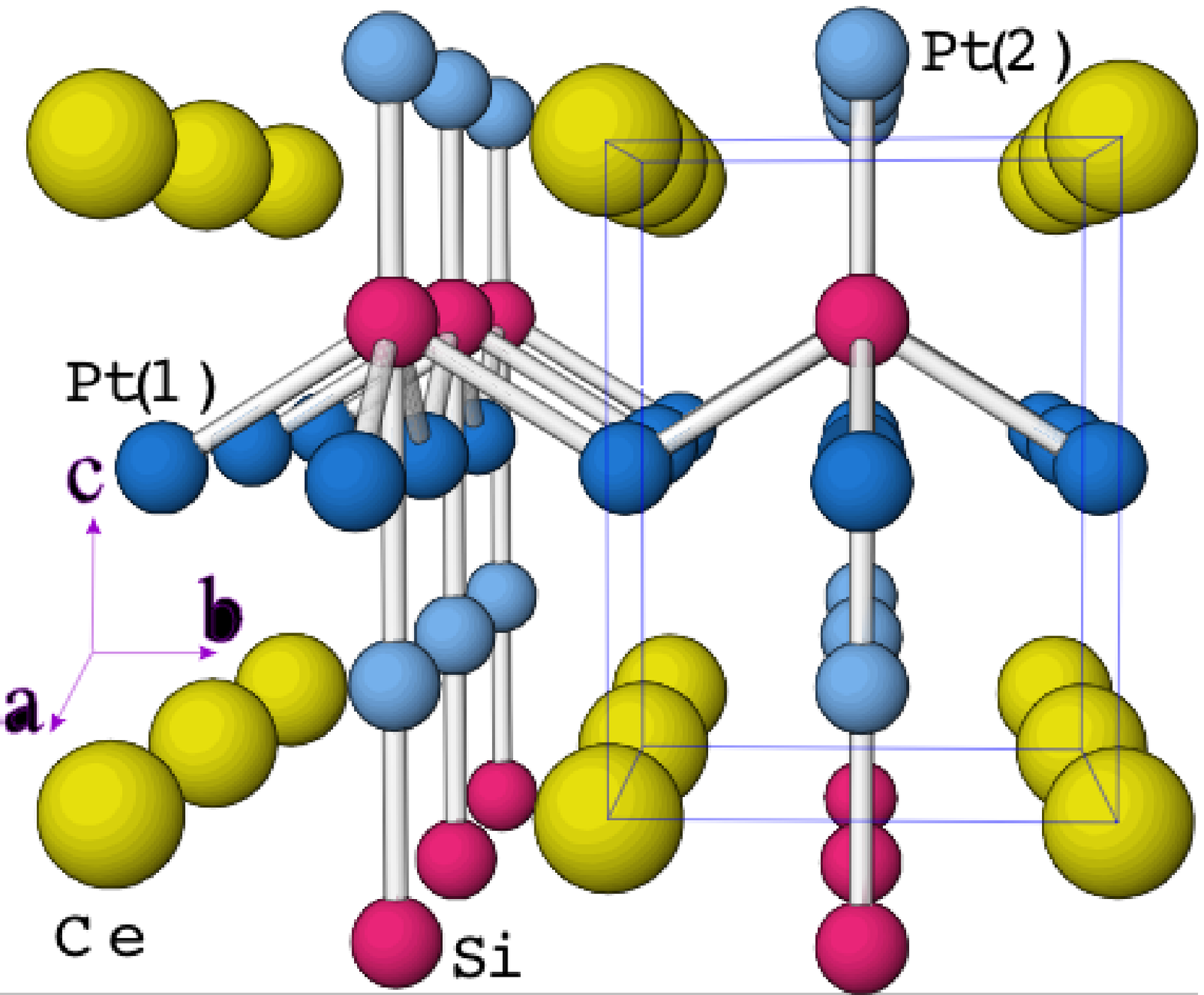}\hfill
\raisebox{6.0cm}
{\includegraphics[angle=-90,width=7.5cm]{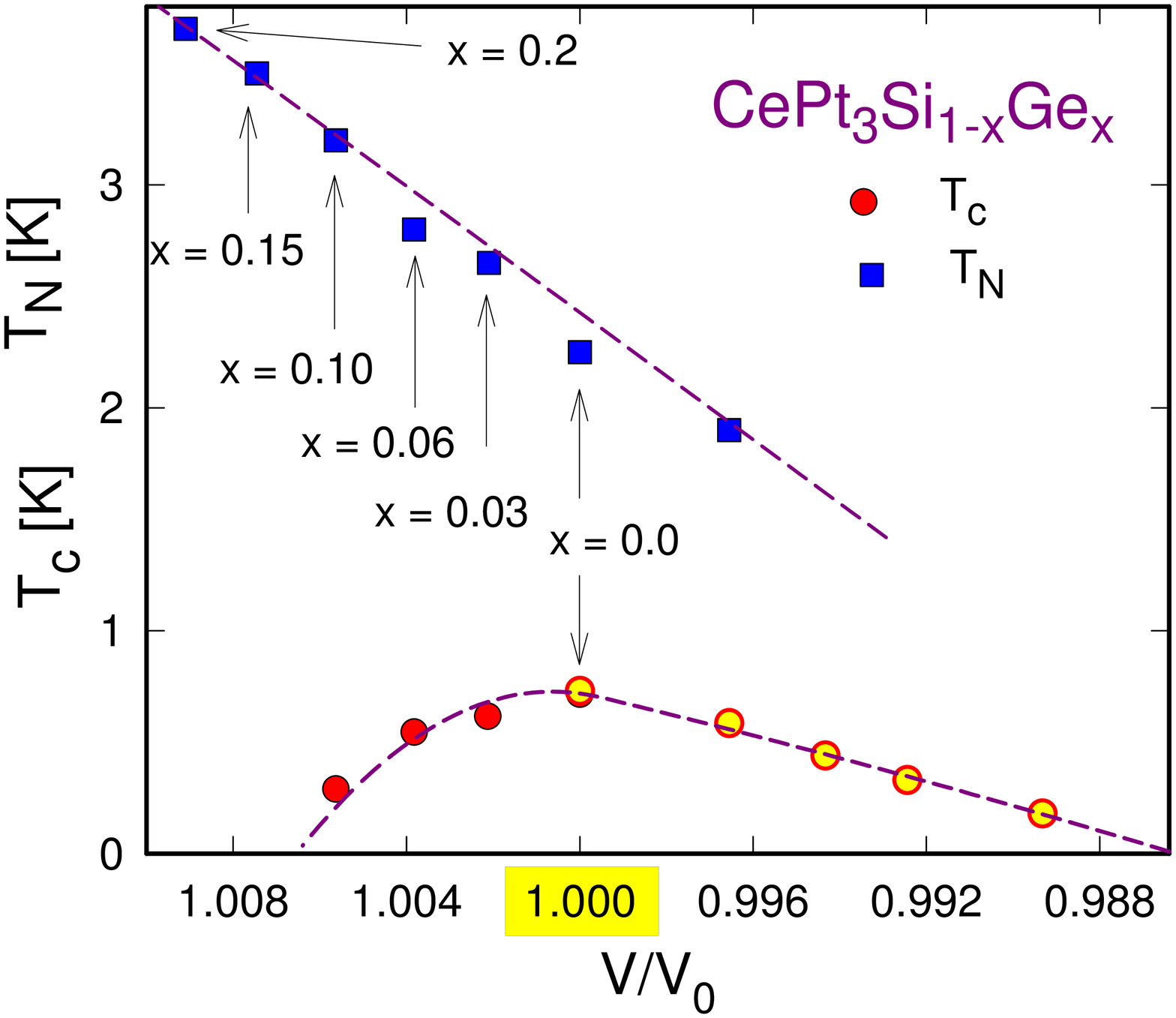}}
\caption{Left panel: Crystal structure of \CPSi~belongs to the tetragonal space
group P4mm, the conventional unit cell (a = 4.072 \AA, c = 5.442 \AA)
is indicated \protect\cite{Bauer04}. It lacks
the reflection symmetry z $\rightarrow$ -z. Right panel: Phase diagram
of CePt$_3$Si$_{1-x}$Ge$_x$ from ambient pressure (mainly T$_N$) and
hydrostatic pressure results (mainly T$_c$). Here 1\% volume reduction
corresponds to 1.5 GPa \protect\cite{Bauer04a}.} 
\label{fig:CePt3Sistruc}
\end{figure}

\begin{figure}[htb]
\raisebox{0.5cm}
{\includegraphics[width=7.5cm,height=8.4cm]{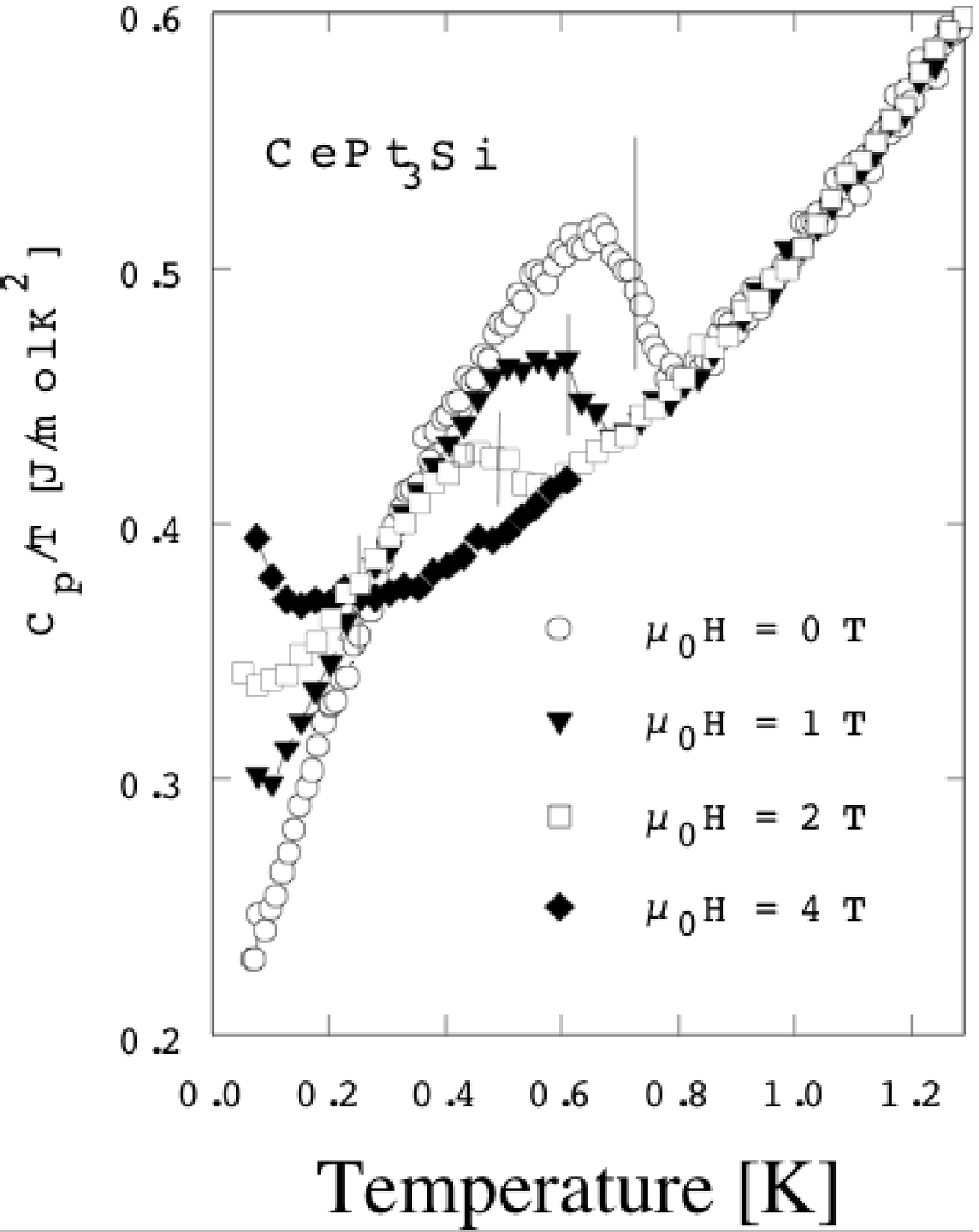}}\hfill
\includegraphics[width=7.0cm,height=9cm]{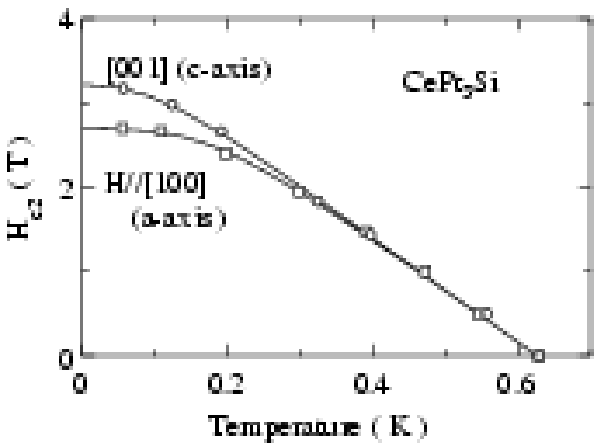}
\caption{Left panel: Field dependence of C/T in polycrystalline \CPSi~with SC
transition indicated by vertical lines \protect\cite{Bauer04}. Right
panel: Upper critical field in single crystals from $\rho$ 
\protect\cite{Yasuda04}.}
\label{fig:CePt3Sispec}
\end{figure}

The discovery of the first HF superconductor \CPSi~ \cite{Bauer04}
without inversion symmetry is therefore of considerable fundamental
interest. \CPSi~ belongs to the tetragonal space group P4/mm
(fig.~\ref{fig:CePt3Sistruc}) where the removal of the inversion
center leads to the
Ce-point group C$_{4v}$ meaning the loss of the basal plane as mirror
plane (z$\rightarrow$-z). The structure can be derived from the
hypothetical cubic CePt$_3$ structure (the same as for \CIN) by tetragonal
distortion with c/a=1.336 and by filling the voids with Si. 
%
%
The Ce$^{3+}$4f states of \CPSi~ are well localised as evident from
sharp CEF excitations found in INS experiments \cite{Metoki04} where
the C$_{4v}$ level scheme was determined as $\Gamma^{(1)}_7$(0),
$\Gamma_6$(1 meV) and $\Gamma^{(2)}_7$(24 meV). At T$_N$ = 2.2 K AF magnetic
order appears where FM planes are staggered along c with
\bQ~= (0,0,$\frac{1}{2}$). Below T$_c$ = 0.75 K superconductivity sets in
(fig.~\ref{fig:CePt3Sispec})
and coexists microscopically with AF order. Only the second time after
A-type \CCS~(fig.~\ref{fig:CeCu2Si2BTPhaseDiag}) this has now been
observed for a stoichiometric Ce-compound at ambient
pressure. The ordered moments $\mu$ = 0.16$\mu_B$ however are much
less than the value of the localised  $\Gamma^{(1)}_7$ ground state doublet
which suggests a large Kondo screening, and indeed there is an
associated mass enhancement of itinerant quasiparticles as is
evident from $\gamma_n$(\CPSi) = 390 mJ/mol K$^2$ as compared to
$\gamma_n$ = 9 mJ/mol K$^2$ for LaPt$_3$Si. A
considerable  residual $\gamma_s$ = 180 mJ/mol K$^2$ and a 
$\Delta$C/$\gamma_n$T$_c$ = 0.25 much smaller than the BCS value 1.43
has been observed. This raises questions on the quality of the
polycrystalline samples although the $\xi$ estimate from H$_{c2}$
indicates that one is not in the dirty limit. Later resistivity
measurements on single crystals \cite{Yasuda04} in a field indeed have
identified a coherence length $\xi\simeq$ 100 \AA~$\ll$ l (mean free
path). There the almost isotropic upper critical fields has been determined as 
H$^a_{c2}$(0) = 3.2 T and H$^c_{c2}$(0) = 2.9 T
(fig.~\ref{fig:CePt3Sispec}) which is considerably larger
than the Pauli-Clogston limiting field estimated by
H$_P\sim$$\Delta/\sqrt{2}\mu_B\sim$ 1.2 T for g = 2. The pressure
depencence of T$_c$ which vanishes around p$_c$ = 1.5 GPa has still
some uncertainty due to the broad appearance of the resistive transition
\cite{Yasuda04}. Nevertheless a combined p(V)-T phase diagram from
Ge-doping and hydrostatic pressure results is shown in
fig.~\ref{fig:CePt3Sistruc}. Whether the AF T$_N$ and the ordered moment
vanish continuously and the SC dome extends on both sides of the magnetic
phase is presently not known.  

The observation of H$_{c2}$(0) $\gg$ H$_P$ would rather advocate for
triplet pairing, this confronts
one with two major problems: i) the lack of inversion symmetry should
disfavor triplet pairing according to the above arguments. ii)
coexistence with fully developed AF order should be destructive for
triplet pairs. Proposals to avoid this contradiction have been made in
theoretical investigations \cite{Frigeri03,Frigeri04} which include
an explicit inversion symmetry breaking but time reversal conserving
term in the single particle part of the BCS Hamiltionian. It has the
form of a Rashba-type antisymmetric spin-orbit
coupling for conduction electrons c$_{\bk s}$ which is of the form
($\boldsigma$ = Pauli matrices)
\begin{equation}
H_P=\alpha\sum_{\bk ,ss'}\bg _{\bk}\cdot\boldsigma_{ss'}
c^\dagger_{\bk s}c_{\bk s'}
\label{IBREAK}
\end{equation}
where \bg$_{\bk}$ = -\bg$_{-\bk}$ is an antisymmetric function which may
be obtained from the periodic potential of Bloch states and is
conveniently normalised to one on the Fermi surface. For finite
$\alpha$ this term lifts the twofold spin degeneracy of conduction
bands (but preserves the \bk, s $\rightarrow$ -\bk, -s symmetry). It leads to
i) modifications for both singlet and triplet gap equations,
ii) a mixing term for singlet and triplet gaps which, however, is
of the order $\alpha/\epsilon_F\ll$ 1 and therefore can be
neglected. 
%
%
Likewise T$_c$ reduction in the singlet case is of order
$(\alpha/\epsilon_F)^2\ll$ 1. The triplet case is more involved and
provides important clues on the effect of H$_P$. For $\alpha$ = 0 all
triplet states of a a given odd $l$ and described by a 
\bd(\bk) vector gap function have the same T$_c$. Turning on the I-
symmetry breaking ($\alpha >$0) lifts this degeneracy, the expected
T$_c$- reduction however does not affect all triplet states
indiscriminantly, those triplet states with \bd(\bk)$\parallel$
\bg$_{\bk}$ are well protected and have no T$_c$ reduction as in the
singlet case. From the C$_{4v}$ point group one may expect
\bg$_{\bk}\cdot\boldsigma$ = k$_x\sigma_y$-k$_y\sigma_x$ and then the
protected triplet
state with unaffected T$_c$ corresponds to the A$_{2u}$ gap function
\bd(\bk) = $\hat{\bx}k_y$-$\hat{\by}k_x$ which has point nodes . Other
triplet states are strongly suppressed for $\alpha\gg$ T$_c$. Thus in
principle favorable triplet states are not destroyed by inversion
symmetry breaking.

On the other hand the loss of inversion symmetry described by
eq.~(\ref{IBREAK}) weakens the Pauli-Clogston limiting argument against
singlet pairing considerably. It is known that spin-orbit impurity
scattering reduces the effect of paramagnetic limiting, the same is
true for the antisymmetric spin-orbit coupling of paired conduction electrons
in eq.(~\ref{IBREAK}). With increasing $\alpha$ the renormalised H$_p$
increases above its starting value $\Delta/\sqrt{2}\mu_B$ for
$\alpha$ = 0. For $\alpha$/T$_c\gg$ 1 it becomes arbitrarily large. As a result
the observation of a large H$_{c2}$(0) $\gg$ $\Delta/\sqrt{2}\mu_B$
does not contradict singlet pairing, which in addition would be more
compatible with the coexisting AF order.

As a preliminary conclusion one may say that there is no compelling
theoretical reason why triplet pairing for SC without inversion
symmetry should
be prohibited. Likewise the upper critical field observations do not
contradict singlet pairing so that the question of gap symmetry in
\CPSi~ remains open. Recent NMR relaxation experiments \cite{Yogi04} on
Pt-sites also did not clarify this question but rather added new puzzles: 
The low temperature 1/T$_1$ behaviour is neither described by the
usual T$^3$ power law generally observed in HF SC, nor by an
exponential decay indicative of an isotropic gap. However the latter
is more compatible with the observation of a small coherence peak
below T$_c$ and a Balian-Werthamer type relaxation rate immediately below it.

\section{\bf Microscopic SC pairing mechanism in HF compounds\rm}
\label{sect:SCmech}

In this section we illustrate to some detail microscopic concepts and 
theories that were proposed to describe the superconducting pairing
in HF systems or, more generally, in strongly correlated electron
compounds including the cuprates and ruthenates. The long history of
the subject of non-phononic superconductivity will not be recounted
and we refer to the literature for this purpose
\cite{Thalmeier03b,Moriya00,Vollhardtbook}.

 Far below the Kondo temperature T$^*$ the ideal HF
compound is in a LFL state with quasiparticles that have an
effective mass m$^*$ much enhanced compared to the band mass m$_b$. However
due to the presence of virtual high energy excitations a residual
on-site quasiparticle repulsion appears. The low energy excitations may then
be described by a spectrum of spin fluctuations characterised by a
dynamic susceptibility matrix $\boldchi(\bq,\omega)$. In the vicinity
of a magnetic instability, for example AF order close to a QCP, overdamped
critical spin fluctuations called antiferromagnetic paramagnons appear whose
spectral density is concentrated around the AF \bQ-vector and which
has a typical energy T$_{sf}<$ T$^*$ at the maximum of
$\boldchi(\bQ,\omega)$. The virtual exchange of these excitations
mediates an effective interaction
between quasiparticles which may lead to a superconducting
instability. In contrast to conventional phonon-mediated
superconductors the pair states have to be of unconventional (non
s-wave) type whose order parameter exhibits a sign change in \bk-space
to avoid the strong on-site repulsion of
quasiparticles. A brief symmetry classification of unconventional SC
order parameters was given in sect.~\ref{ssect:CeCu2Si2}.

Originally the spin fluctuation mechanism in the case of nearly FM systems
with paramagnon exchange was proposed to explain superfluidity in
$^3$He \cite{Nakajima73,Anderson73}. It was later extended to HF
\cite{Miyake86} and cuprate superconductors \cite{Scalapino86}. In the
latter one has simple tight binding (TB) hole bands which allows one to 
carry out fully microscopic calculations based on the fluctuation
exchange (FLEX) approximation \cite{Pao94,Dahm95} for the Hubbard
model. However the theory
in its original single band form is not able to describe the
complications of SC pairing in real HF materials, for the
following reasons: i) Orbital (spin-orbit) and crystal field effects lead to
complicated renormalised band structures with multisheeted FS, therefore the
dynamic susceptibility $\boldchi(\bq,\omega)$ cannot be reliably
calculated, instead empirical forms have to be assumed which also
incorporate the vicinity to a possible QCP frequently present in 
Ce-HF compounds. ii) In the
U-compounds 5f- electron have a dual nature, partly localised and
partly itinerant. The internal CEF excitations (magnetic excitons) of
the former may dominate the effective interactions between itinerant
quasiparticles, which is an alternative to the spin fluctuation
mechanism that involves only the itinerant 5f electrons.

We shall discuss both electronic pairing mechanisms in the
following. While the spin fluctuation mechanism may be
appropriate for the Kondo-like Ce HF superconductors, the magnetic
exciton mechanism is a serious candidate for a number of U-compounds.

\subsection{Spin fluctuation mechanism for C\lowercase{e}-based HF compounds}

If one is not too close to the QCP of magnetic order one may assume that
the frequency dependence of  spin fluctuations is negligible on the
scale of the cutoff energy for SC pairing. Then in
the effective interaction V(\bq,$\omega$) between quasiparticles
$\boldchi(\bq,\omega)$ may be replaced by the nonretarded RPA
expression $\boldchi(\bq)$. For the spin dependent part one obtains
the spin rotationally invariant interaction    
\begin{eqnarray}
\label{EFFINT}
V_{\alpha\beta;\gamma\rho}(\bq)
=V(\bq)\boldsigma_{\alpha\gamma}\cdot\boldsigma_{\beta\rho}
\qquad \mbox{with}\nonumber 
\end{eqnarray}
\begin{eqnarray}
V(\bq)=-\frac{1}{2}\frac{I}{1-I\chi_0(\bq)}
\simeq -\frac{1}{2}I^2\chi(\bq) 
\qquad \mbox{and} \qquad 
\chi(\bq)=\frac{\chi_0(\bq)}{1-I\chi_0(\bq)}
\end{eqnarray} 
Here I is the residual on-site quasiparticle interaction and
$\chi_0({\bf q})\leq$ I$^{-1}$ their static noninteracting susceptibility. The
interaction may be split into singlet (S=0) and triplet (S=1) channels
according to V$_0$ = -3V and V$_1$ = V, the factor 3 is due to the fact
that only longitudinal spin fluctuations contribute in the triplet
channel whereas both longitudinal and transverse parts contribute in
the singlet channel. In this approximation the SC pairing is
completely determined by the static susceptibility $\boldchi(\bq)$
which may in principle be obtained from diffuse neutron
scattering. Its \bq- dependence together with the characterisics of the
Fermi surface determines the dominant angular momentum component
V$^l_S$ of V$_S$(\bq) that leads to the stable SC pair state (\it l\rm~= even for
S=0 and odd for S=1). Obviously the s-wave state is not stable
because V$^{l=0}_0$ is always positive (repulsive), pair states with
\it l\rm~$>$ 0, however, may have attractive V$_l<$ 0 and become stable depending
on the characteristics of $\chi(\bq)$. The model has been
analysed in some detail in \cite{Miyake86} by assuming a spherical FS
and a plausible behaviour for V(\bq) in a cubic lattice with a peak near
the AF wave vector \bQ~ to simulate the AF spin fluctuations
observed in HF compounds. This analysis predicts that the unconventional
even singlet SC state (\it l\rm~$>$ 0), e.g. \it l\rm~= 2 d-wave state, should be
stable because V$_l<$ 0 is attractive in this orbital channel. 

This result led to the general expectation that AF spin
fluctuations favor singlet pairing. While it is true in the simple
case of cuprates, and it may be true for some Ce-based HF compounds
like \CCI~ (sect.~\ref{ssect:CeCoIn5}), it turned out to be misleading
for a number
of U-compounds like \UPT~ or \UND~ and possibly \UBE~ which exhibit
triplet pairing as concluded e.g. from Knight shift results \cite{Tou03}.
 Therefore the above prediction of singlet pairing may be partly an
artefact of the model simplifications and point to the inadequacy
of the spin fluctuation model for U-HF superconductors. Indeed the
model has obvious deficiencies: i) It completely ignores the orbital
structure strongly influenced by spin orbit coupling and CEF effects.
ii) It does not take into account the uniaxial crystal anisotropy
present in most HF superconductors which possess either tetragonal or
hexagonal space groups. iii) It ignores the spin space anisotropy of
the pair potential which is assumed as rotationally invariant in
eq.(\ref{EFFINT}). In case this symmetry is reduced the triplet state
will be less disfavored.

From the low temperature 'power laws' in various thermodynamic and
spectroscopic quantities it is also concluded that most HF SC have
gap functions with nodes \De~= 0 along lines on the FS. However if one
assumes the strong spin orbit coupling case for quasiparticles,
i.e. the crystal double group as symmetry group, 'Blount's theorem'
states that triplet pair states can have only point nodes in the SC gap
\cite{Volovik85}. Therefore the observed power laws which support line
nodes have
also been used as an argument for singlet pairing. This argument is
misleading however, since the spin orbit coupling is so large that it
has to be included already in the orbital basis for band structure
calculations within the jj- coupling scheme. The scale of SC pairing
energies is much smaller than band energies and therefore it is the
'pseudo-spin' degrees of freedom connected with Kramers degeneracy of
quasiparticle band states which are involved in the singlet and triplet
pair wave functions. Therefore the pseudo-spins may be assumed to have an
effectively weak (pseudo-)spin orbit coupling, and then no 
contradiction between the existence of line nodes and SC pairing in
triplet states for U-compounds appears. We note however that ironically, in
\UPD, which is the one U compound where the spin fluctuation model is
known to be inappropriate probably a singlet SC pair state is realised.

In a series of papers the spin fluctuation mechanism has also been
studied within the non-retarded strong-coupling Eliashberg
approach \cite{Monthoux99,Monthoux01,Monthoux02,McHale03} including the
frequency dependence of the pairing interaction and assuming a simple
tight binding band with n.n. and n.n.n hopping t and t' respectively
for conduction electrons. The frequency dependence is especially
important close to a magnetic QCP when the typical energy of spin
fluctuations becomes comparable to 2$\pi$T$_c$. Instead of
starting from a microscopic model Hamiltionian, a phenomenological
effective interaction kernel V(\bq,$\omega$) was used as input for the
Eliashberg equation which determines the quasiparticle self energy
$\Sigma(\bk,i\omega_n)$ and anomalous self energy ('gap function')
$\Delta(\bk,i\omega_n)$. For singlet (S=0) and triplet (S=1) pairing
it is given in the AF case by 
\begin{eqnarray}
\label{SFLkernel}
V_S(\bq,\omega)&=&a_SI^2\chi(\bq,\omega)\qquad \mbox{with}~
a_0=-1~ and~ a_1=\frac{1}{3}\nonumber\\
\chi({\bq},\omega) &=& \frac{\chi_0\kappa_0^2}
{\kappa^2+{\hat q}^2 - i\frac{\omega}{\eta({\hat q})}}\qquad
\mbox{where}~\eta(\hat q)=T_{sf}{\hat q}_-,\;\;{\hat q}= {\hat q}_+ \\
\hat q^2_{\pm} &=&(4+2\alpha_m)\pm 2[\cos q_x +\cos q_y + \alpha_m\cos q_z]
\nonumber
\end{eqnarray}
Here $\kappa$ and $\kappa_0$ denote the nearly critical and background
inverse magnetic correlation lengths  respectively (in units of the
inverse lattice spacing a$^{-1}$). In Stoner-RPA theory
$\kappa_0^2/\kappa^2$ = (1-I$\chi_0)^{-1}$ and $\kappa_0\sim$ 2k$_F$a. 
This interaction is mediated by \emph{overdamped} spin fluctuations
since the pole of $\chi$(\bq,$\omega$) is at a purely imaginary
frequency, i.e. the spin excitation does not have a real group
velocity and therefore is not propagating. This is a natural
assumption close to
the QCP where the restoring force for spin excitations due to the
molecular field breaks down. It is completely different from the
phonon case and also the magnetic exciton mechanism discussed later
which are both characterised by the exchange of propagating bosons
which have a real frequency bigger than the line width. In
eq.(~\ref{SFLkernel}) $\alpha_m$ determines the crystal anisotropy of
the spectrum, it is of 2D character when $\alpha_m$ = 0 and 3D character
for $\alpha_m$ = 1. It is still fully isotropic in (pseudo-) spin space
as evident from the singlet and triplet classification. According to
eq.(~\ref{SFLkernel}) the interaction in the AF case peaks at the zone
boundaray vector \bQ~ with a spread given by $\kappa$
($\kappa$ = inverse magnetic correlation length $\xi_m^{-1}$) and a frequency 
maximum at $\omega_{max}$ = $\eta(\hat q)\kappa^2$. 
The same type of expression holds for FM spin fluctuations where
one has to replace  ${\hat q}= {\hat q}_-$. Using the effective
pairing in eq.~(\ref{SFLkernel}) the Eliashberg equation
for the gap function $\Delta(\bk,i\omega_n)$
($\omega_n$ = Matsubara frequencies) is 
\begin{eqnarray}
\Lambda(T)\Delta(\bk,i\omega_n)=a_S\frac{T}{N}\sum_{\omega_m}\sum_{\bk}
\chi(\bk-\bq,i\omega_n-i\omega_m)|G(\bq,i\omega_m)|^2\Delta(\bq,i\omega_m)
\end{eqnarray}
where G(\bq,i$\omega_m$) is the quasiparticle Green's function
renormalised by the normal self energy  $\Sigma(\bq,i\omega_m)$ that
satisfies a similar equation. The solution of this set of self
consistent equations determines T$_c$ via the condition that the
eigenvalue $\Lambda$(T$_c$) = 1. The fixed parameters of the model are
t,t' (hopping elements), $\kappa_0$ (inverse background correlation length),
and T$_{sf}$ (spin fluctuation energy scale). The critical temperature
T$_c$($\lambda$,$\kappa$) is studied as function of the dimensionless
coupling constant $\lambda$ = I$^2\chi_0$/t and the inverse magnetic
correlation lenght $\kappa$ which determines the width of the AF spin
fluctuation peak in
\bk-space. In reality this parameter may be varied experimentally by
applying pressure, i.e. tuning the distance to the magnetic QCP where
$\kappa$ = 0. In the spirit of Ginzburg-Landau theory,
at T$_c$ the gap function should belong to a single representation
$\Gamma$ of the spatial symmetry group whose \bk-dependence
is characterised by the basis function $\Phi_\Gamma(\bk)$. The stable SC
state is the one which has the highest T$_c$ among the representations
$\Gamma$ and pair spins S = 0,1. The T$_c$($\lambda$,$\kappa$) variation of
the stable state may then be calculated. This has been done e.g. for
the d$_{x^2-y^2}$- state in the 2D case appropriate for cuprates, or,
with suitable redefinition of bandwidth  and T$_{sf}$ also for
\CCI. Without going to great detail the essential results with the
above procedure can be summarised as follows
\cite{Monthoux99,Monthoux01,Monthoux02}:
\begin{itemize}
\item
When \bQ~= 0 (FM) the p-wave state (S=1) is favored (has the largest
T$_c$) and for the AF \bQ~ = ($\frac{1}{2}$,$\frac{1}{2}$) (2D) or
\bQ~ = ($\frac{1}{2}$,$\frac{1}{2}$,$\frac{1}{2}$) (3D) the d-wave
state (S=0) is prefered.
\item
In both cases T$_c$ increases monotonously with coupling
constant $\lambda$ for
any $\kappa$. For constant $\lambda$ it first increases with
$\kappa$ and reaches a maximum at $\kappa^2\simeq$ 0.5 and then shows
slow monotonic decrease
with $\kappa$. When $\lambda$ decreases the optimum T$_c$ position moves
towards $\kappa$ = 0, i.e. to the AF QCP. 
\item
Generally the achievable T$_c$'s under equal conditions are larger in
2D than in 3D, and they are much larger for AF spin fluctuations
leading to even pair states as compared to FM spin fluctuations
associated with the odd SC states. The latter also depend more
strongly on parameters, e.g. on $\kappa$. 
\end{itemize}
This result is somewhat
counter-intuitive: In the FM case the effective interaction in real space
is purely attractive while it is oscillating in the AF case. However
as mentioned above, for triplet pairing only longitudinal spin
fluctuations contribute, leading to the reduction factor $a_1$=1/3 in
eq.~(\ref{SFLkernel}) whereas in the AF singlet pairing case the
additional transverse contributions lead to $|a_0|$=1. This advantage
of singlet pairing depends on the isotropy of the effective interaction
in spin space. In the opposite limit with maximal Ising type
anisotropy in spin space, the disadvantage of triplet pairing is
removed as will be discussed in sect.~\ref{ssect:UPd2Al3}.
%
\begin{SCfigure}
\raisebox{-1cm}
{\includegraphics[clip,width=7.5cm]{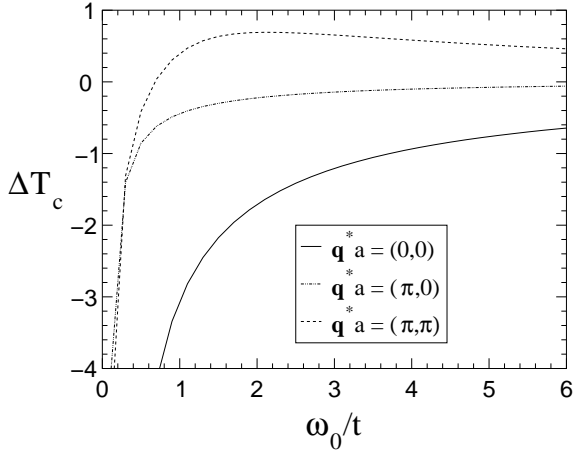}}
\caption{Dependence of $\Delta$T$_c$ on spin fluctuation frequency
$\omega_0$ (t = hopping energy) for various momenta \bq$^*$ in the 2D AF
case with \bQ~ = (0.5,0.5) (in r.l.u.). Coupling constant
I$^2\chi_0$/t = 10; $\kappa^2$ = $\xi_m^{-2}$ = 0.5, $\kappa_0^2$ = 12 and
T$_{sf}$ = 0.67t. Only spin fluctuations at \bq$^*$
= ($\pi$/a,$\pi$/a) $\equiv$ \bQ~ or close to \bQ~contribute positively
to T$_c$ above the crossover frequency $\omega_{cr}$(\bq$^*$) $\simeq$
0.5t. Below $\omega_{cr}$(\bq$^*$) they are pair breaking.
\protect\cite{McHale03}.} 
\label{fig:TCvar}
\end{SCfigure}
%
\subsection{T$_{\lowercase{\mbox{c}}}$-dependence on spectral properties
of spin fluctuations}

In the above considerations the change of T$_c$ under variation of
global interaction parameters and characteristics is considered. This
does not answer the question how the variation of the spectrum of spin
fluctuations for a specific \bq$^*$ and $\omega_0$ influences T$_c$,
specifically
one would like to know which parts of the spectrum are most favorable
for achieving a large T$_c$ for a given pair state. For the
electron-phonon superconductor this has been first investigated in
\cite{Bergmann73}. It was found that every part of the e-ph
interaction function
$\alpha^2$F($\omega$) contributes positively with 
2$\pi$T$_c$ being the optimum phonon frequency while phonons with
frequencies much larger or much smaller frequencies contribute little
to T$_c$.

This analysis has been extended later \cite{Millis88} to the
spin fluctuation model for HF systems, albeit using an unrealistic
effective interaction that factorises with respect to frequency and
momentum. As a main result it was found that, for any momentum \bq, adding
spectral weight below a crossover frequency $\omega_{cr}$ reduces
T$_c$, i.e. low energy spin fluctuations act pair breaking. However a
subsequent analysis by McHale and Monthoux \cite{McHale03} using the
more realistic effective interaction in eq.~(\ref{SFLkernel}) has
modified this picture considerably. By adding an infinitesimal amount
of spectral weight to $\chi(\bq,\omega)$ at fixed wave vector \bq$^*$
and frequency $\omega_0$ and -$\omega_0$, i.e. adding two
delta-functions, the change in the critical
temperature $\Delta T_c(\bq^*,\omega_0)$ was computed numerically.
A representative result in 2D for three different \bq$^*$ is shown in
fig.~\ref{fig:TCvar} for the nearly AF spin fluctuation spectrum
with 2D incipient
ordering vector \bQ~= ($\frac{1}{2}$,$\frac{1}{2}$). It clearly shows
that the effect of
the added intensity to the spectrum also depends strongly on
\bq$^*$, i.e. where it is added in the BZ, in addition to the
$\omega_0$ dependence. For \bq$^*$ = (0,0) or ($\frac{1}{2}$,0) which are far
away from the AF \bQ~= ($\frac{1}{2}$,$\frac{1}{2}$) the added intensity
reduces T$_c$
for all frequencies $\omega_0$. Only when it is added at the incipient
ordering wave vector \bQ~or close to it one has $\Delta T_c>$ 0 for
frequencies $\omega >$ $\omega_{cr}(\bq^*)$. The crossover frequency
$\omega_{cr}$ from pair breaking to pair formation therefore strongly
depends on the wave vector \bq* of the added intensity relative to
\bQ. Sufficiently
far away from their spectral maximum at \bQ~spin fluctuations
with \emph{all} frequencies $\omega_0$ are pair breaking ($\Delta
T_c(\bq^*,\omega_0)<$ 0). The pair forming fluctuations are constrained
to the region around \bQ~ which means that the pairing occurs mostly
through the short range magnetic fluctuations, except very close to
the QCP ($\kappa\rightarrow$ 0). The analysis again shows that $\Delta
T_c(\bq^*,\omega_0)$ is much more robust in the AF case as compared to
FM case (\bQ~ = 0). From the $\Delta T_c(\bq^*,\omega_0)$ curves in the AF
case it is also found that the optimum $\omega_0$ where $\Delta T_c$
is largest lies above the spectral maximum of $\chi(\bq,\omega)$ for
$\kappa^2\ll$ 1 (nearly critical case) and vice versa for
$\kappa^2\gg$ 1. Despite the complications the optimum $\omega_0$ is of
the same order of magnitude given by 2$\pi$T$_c$ as for the
electron-phonon mechanism.

These results show that the formation of the SC state is not
determined by the very low energy critical spin fluctuations which are
responsible for the NFL anomalies in compounds like \CPS. It explains
qualitatively why compounds like \CCG~ or \CRS~ which exhibit LFL
behaviour around the critical pressure for destruction of AF order may
nevertheless be spin fluctuation mediated superconductors (see also
sect.~\ref{ssect:CePd2Si2}).

The previous analysis also suggests another reason why triplet pairing should
be relatively scarce. In strongly correlated elecron systems not too
close to a QCP dominant spin fluctuations involve mostly nearest
neighbors, i.e. they have wave vectors \bQ~ at the zone boundary, these
are however triplet-pair breaking for all frequencies in the spin
fluctuation mechanism. Considering this fact and the observation that
most  U-based HF superconductors seem to have triplet pairs it seems
doubtful whether the spin fluctuation model is appropriate for them
and one has to think about other possibilities. 

\subsection{Magnetic exciton mechanism for superconductivity in U-
based HF compounds}

The origin of mass enhancement in Ce- and U-based compounds is not
the same as explained in sect.~\ref{sect:QPmech}. This is due to the
multiple 5f-shell occupation and strongly orbital dependent
hybridisation in U compounds which, in some cases like \UPD, \UPT~and
possibly \UBE~leads to a partly itinerant and partly localised nature
of 5f electrons. In a strong-coupling theory for the quasiparticles,
mass enhancement and effective pair interactions are two sides of the
same coin. Therefore we discuss now the origin of SC pairing within the
dual model of 5f electrons. This idea has been proposed and explored
to some extent in \cite{Sato01,Thalmeier02,McHale04} for \UPD~ as
discussed further in sect.~\ref{ssect:UPd2Al3}.

In the dual model the localised 5f$^2$ electrons occupy specific total
angular momentum orbitals which are split by the CEF. For simplicity
we consider only the two lowest CEF states which are assumed to be
nonmagnetic singlets. They are then described by a pseudo-spin
(S=$\frac{1}{2}$)
variable where S$_z$ = $\pm\frac{1}{2}$ correspond to singlet CEF states
with energies $\pm\frac{\delta}{2}$. They have an effective inter-
site coupling (superexchange) as well as a coupling to the itinerant
5f electrons of the conduction band $\epsilon_{\bk}$. The latter
couples only with the $\sigma_z$- conduction electron spin component due to
the (Ising-type) anisotropy introduced by the singlet
CEF states which have only transition matrix elements for the local
S$_x$ pseudo-spin operator. This dual 5f-model is described
by \cite{McHale04} 
\begin{equation}
H = \sum_{\bk\sigma} \epsilon_{\bk} c_{\bk\sigma}^\dagger c_{\bk\sigma} 
+ \delta \sum_i S_{iz} - J_{ff} \sum_{\langle ij\rangle} S^i_x S^j_x
- I \sum_i \sigma_{iz} S_{ix} .
\label{HfirstVers}
\end{equation}
The localised part may be diagonalised separately. Due to the effective
inter-site interaction J(\bq) which contains the superexchange
J$_{ff}$ and a RKKY contribution from the last term the local CEF
excitations at an energy $\delta$ evolve into propagating magnetic
exciton modes with a dispersion given by 
\begin{eqnarray}
\label{DI1}
\omega_E(\bq)&=&\delta[1-\frac{J(\bq)}{2\delta}
\tanh\frac{\beta}{2}\delta]
\end{eqnarray}
If J(\bq) has its maximum J$_e$ = J(\bQ) at an AF  zone boundary
vector \bQ~the mode
frequency becomes soft at \bQ~at the N\'eel temperature where induced AF
appears; 
\begin{eqnarray}
\label{CRINEEL}
T_N&=&\frac{\Delta}{2\tanh^{-1}(\frac{1}{\xi})}\qquad
\xi=\frac{J_e}{2\Delta}
\end{eqnarray}
provided one has $\xi>$1 for the control parameter. In reality the
softening will be
arrested at a finite magnetic exciton gap due to the effect of higher
lying CEF states which may shift T$_N$ slightly to higher
temperatures. The magnetic excitons associated with the localised 5f
system may be seen in INS as discussed for \UPD~ in sect.~\ref{ssect:UPd2Al3}.
For T $\ll\delta$ they are bosonic modes
like phonons and then a canonical transformation on
eq.~(\ref{HfirstVers}) leads to a more convenient dual model Hamiltonian
\cite{McHale04}
\begin{equation}
H = \sum_{\bk\sigma} \epsilon_{\bk} c_{\bk\sigma}^\dagger c_{\bk\sigma}
+ \sum_{\bq} \omega_{E}(\bq)(\alpha_{\bq}^\dagger \alpha_{\bq} + \frac{1}{2}) 
- I \int d\br \psi_\alpha^\dagger(\br) \sigma^z_{\alpha\beta}
\psi_\beta (\br) \phi(\br)
\label{eqHam}
\end{equation}
where  c$_{\bk\sigma}^\dagger$ and $\alpha_{\bq}^\dagger$ create
conduction electrons and magnetic excitons respectively and 
$\psi_\alpha^\dagger(\br)$, $\phi(\br)$ are the associated field
operators.
It is indeed very similar to the electron-phonon Hamiltonian, except
that magnetic excitons do not couple to the electronic density but (in
this simplified model)
only to the z-component of the itinerant spin density. The resulting
normal self energy of conduction electrons due to dressing by virtual magnetic
excitons has been discussed before. Complementary virtual exchange
processes involving magnetic excitons lead to an effective
quasiparticle interaction given by ($\nu_n$ = 2$\pi$nT) \cite{McHale04}
\begin{eqnarray}
\label{EFFEX}
\hat{V}(\bq,i\nu_n) =  
-I^2 D^0 (\bq,i\nu_n )\hat{\sigma^z}\hat{\sigma^z}
=\left(\frac{I^2\Delta}{2}\right)\frac{\hat{\sigma^z}\hat{\sigma^z}}
{\nu_n^2 +\omega_E(\bq)^2}
\end{eqnarray}
This effective pairing breaks spin rotational symmetry in a maximal
(Ising type) manner. This is enforced by the anistropy of the CEF
states and has important consequences for the classification of SC
pair states, one now has to use the equal spin pairing (ESP) states
$ |\chi\rangle = |\!\!\uparrow\uparrow\rangle\;\mbox{and}\;
|\!\!\downarrow\downarrow\rangle$,
and opposite spin pairing (OSP) states, given by
\begin{equation}
|\chi\rangle = 
\left\{ 
\begin{array}{l}
\frac{1}{\sqrt{2}} \left(
|\!\uparrow\downarrow\rangle - |\!\downarrow\uparrow\rangle \right)\\
\frac{1}{\sqrt{2}} \left( 
|\!\uparrow\downarrow\rangle + |\!\downarrow\uparrow\rangle \right)
\end{array} 
\right.
\end{equation}
This aspect is completely different from the previous 
spin fluctuation theory which leads to a spin
rotationally invariant  pairing potential and the pair states are
classified as (pseudo-) spin singlet (S = 0) and triplet (S = 1) states. As a
consequence, because p =$ \langle\chi|  \hat{\sigma}^z\hat{\sigma}^z
|\chi\rangle$ is +1 for ESP and -1 for OSP respectively, the coupling
strength of the effective potential in eq.~(\ref{EFFEX}) is equal for OSP
and ESP states, in contrast to the enhancement factor 3 of singlet vs.
triplet pair interaction  in the spin fluctuation model. Thus the present
mechanism with its Ising type spin anisotropy in eq.~(\ref{EFFEX}) does not
disfavor the odd parity states over even parity states, in fact they
may be degenerate as will be discussed for a concrete case in \UPD.

The magnetic exciton mediated pairing mechanism is due to the
interaction between itinerant 5f electrons caused by the exchange of excitations
within the localised 5f-CEF states. On one hand it is similar to the
electron-phonon mechanism because propagating modes are involved and
not overdamped spin fluctuations, on the other hand it is repulsive in the
s-wave channel and contains the conduction electron spin variables
leading to the possibility of unconventional or nodal pair states.

\section{U- based HF superconductors}
\label{sect:ursc}

Most ambient pressure HF supercoductors are intermetallic U
compounds like \UBE~\cite{Ott83}, \UPT~\cite{Stewart84b},
\URU~\cite{Schlabitz84}, \UPD~\cite{Geibel91a} and
\UND~\cite{Geibel91b}. This may partly be
caused by the enhanced delocalisation of
5f as compared to 4f electrons. Due to the multiply  occupied 5f
shell in U which hosts between two and three electrons, the origin of
heavy electron mass enhancement in U compounds is in fact quite
different as compared to Ce compounds with nearly singly occupied 4f
orbitals. In
the latter, heavy quasiparticles originate from coherent resonant
scattering in the Kondo lattice \cite{Hewsonbook}. For some U-HF systems
however the dual, i.e. localised and itinerant nature of 5f electrons
in different orbitals generates the mass enhancement
\cite{Zwicknagl02,Zwicknagl03} (sect.~\ref{sect:QPmech}) and may also dominate
the SC pairing mechanism. 
The AF QCP scenario so prevalent in Ce-compounds is not of general importance
in U-based HF superconductors with the possible exception of \UBE. One
rather observes that SC phases in U-compounds are
embedded in a region of reduced moment AF order. In the more recently
discovered FM U-based superconductors like UGe$_2$ or
URhGe however QCP of FM or even hidden order may  indeed be associated
with the observed unconventional SC 
phases. On the other hand these superconductors are not really HF
materials as witnessed by their low $\gamma$-values and they will not
be discussed here.

In this review we shall not give a complete overview on U-based HF
superconductors but shall focus on two examples, \UBT~and \UPD~where
new results have been obtained recently. Some of the other classical
U-compounds, notably the unique multicomponent superconductor \UPT~
are reviewed in \cite{Sauls94,Joynt02,Thalmeier03b} while \URU~ is reviewed in
\cite{Thalmeier03b}.

\subsection{Multiphase superconductivity in the non-Fermi liquid
compound \UBT}  
\label{ssect:UBe13}

The cubic Be-cage compound \UBE~ was the second HF superconductor
discovered \cite{Ott83,Ott84} but remains only partly understood. The
SC transition takes place in a normal state with
pronounced NFL behaviour, as can be seen from the C/T behaviour for
fields above H$_{c2}$. The thorated \UBE~has a complicated x-T
phase diagram with two distinct SC phases and possibly a coexisting
magnetic phase. In fact it is surprising that SC survives up to about
x = 0.06 since the potential scattering of normal Th impurities should
lead to a strong pair breaking for an unconventional SC state and an
ensuing a rapid reduction of T$_c$. Instead a pronounced nonmonotonic
behaviour of T$_c$(x) is found. The x-T
phase boundaries are still partly a conjecture and may have to be
revised according to recent flux creep experiments \cite{Mota03}. The
most interesting issue is doubtlessly the origin of small magnetic
moments in the low temperature SC phase (C) in the regime
0.02 $<$ x $<$ 0.045. They may either be caused by a non-unitary SC
C-phase which breaks time reversal symmetry as suggested by $\mu$SR
experiments \cite{Heffner90} or they may be associated with the
formation of a long range SDW order suggested by thermal expansion
experiments \cite{Kromer98,Kromer00b,Kromer02}. Theoretically the
former scenario has been
explored within the context of a Ginzburg-Landau approach assuming that the
T$_{c}$(x) of two distinct SC phases cross \cite{Sigrist91}. The
SC/SDW scenario on the
other hand has been theoretically proposed in \cite{Kato87}. It
should be noted however that assumptions on the gap function symmetry
remain purely speculative. Even in the pure \UBE~ it is not known
although the presence of nodes in the gap function and hence its
unconventional nature is assured, e.g. from the existence of a zero-bias
conductance peak in tunneling experiments \cite{Waelti00} which provides
evidence for a sign change of the gap function \De.

\subsubsection{Electronic structure and NFL-like normal state}

The 5f states in \UBE~ show both signatures of localisation and itineracy. On
one hand the 5f- orbital energy is close to the Fermi level according
to photoemission results. This disfavors the Kondo 
picture with a localised 5f$^3$ configuration as origin of the
extremely large mass enhancement characterised by $\gamma$ = 1100 mJ/mol
K$^2$ corresponding to a quasiparticle band width of T$^*\simeq$ 8
K. LDA calculations \cite{Maehira02} with fully itinerant 5f
electrons however lead to only $\gamma_b$ = 13 mJ/mol K$^2$ implying
a large many body enhancement factor of m$^*$/m$_b$ = 85. The Fermi
surface consists of two closed hole sheets with nesting properties and
one closed electron sheet.

On the other hand Schottky-type specific heat anomalies \cite{Felten86}
suggest the presence of CEF excitations $\simeq$ 180 K out of a
$\Gamma_6$(5f$^3$) Kramers doublet ground state. An alternative model
\cite{Cox87} assuming a nonmagnetic $\Gamma_3$(5f$^2$) ground state
can be ruled out because it predicts the wrong sign of the nonlinear
suscetibility $\chi^{(3)}$(T) at low temperatures \cite{Ramirez94}. A mixed
valent model involving both nonmagnetic $\Gamma_3$ and
magnetic $\Gamma_6$ doublets has also lead to the conclusion that the
ground state must be the magnetic $\Gamma_6$ to obtain qualitative
agreement with the observed  $\chi^{(3)}$(T) behaviour \cite{Schiller98}.
In the pure $\Gamma_6$
picture the Kondo effect would lead to a LFL state with enhanced
m$^*$. However a further temperature scale below T$^*$
given by T$_m\simeq$ 2 K has been found from a clear maximum structure
in resistivity, specific heat and thermal expansion
\cite{Kromer00b,Kromer02}. This indicates that \UBE~ has not yet reached a true
LFL state when SC sets in at T$_c\simeq$ 0.9 K.
Indeed for fields above H$_{c2}$ a pronounced
NFL behaviour with logarithmic increase in $\gamma(T)$ and no
saturation to the lowest temperatures has been observed
\cite{Helfrich98,Gegenwart04} in \UBE. 
In addition a universal field-independent $\rho(T)/\rho_{1K}$
behaviour is observed for $T\geq 1.2K$ and $4T\leq H\leq
10T$ in the normal state. One finds $\rho(T)\sim T$ at elevated and
$(\rho(T)-\rho_0)\sim T^{3/2}$ at lower T \cite{Gegenwart97,Gegenwart04}.
The origin of NFL behaviour may reside in the vicinity of
a QCP for SDW or possibly hidden order, this would be in accordance with the
nesting properties of electronic bands in \UBE~ and with the evolution
of T$_m$(x) which hits the maximum of T$_c$(x) in \UBT~ as concluded from
resistivity measurements \cite{Kromer00b,Kromer02}. Another
possibility is provided by the T$_m$(B$^*$) dependence, implying
T$_m\rightarrow$ 0 for B$^*\rightarrow 5T$
(fig.~\ref{fig:UBe13BTphase}), i.e. a field-induced QCP of short-range
AF correlations \cite{Gegenwart04}.

\subsubsection{The superconducting phase diagram of \UBT}

\begin{figure}[tbh]
\raisebox{-0.5cm}
{\includegraphics[clip,width=7.5cm,height=8.6cm]{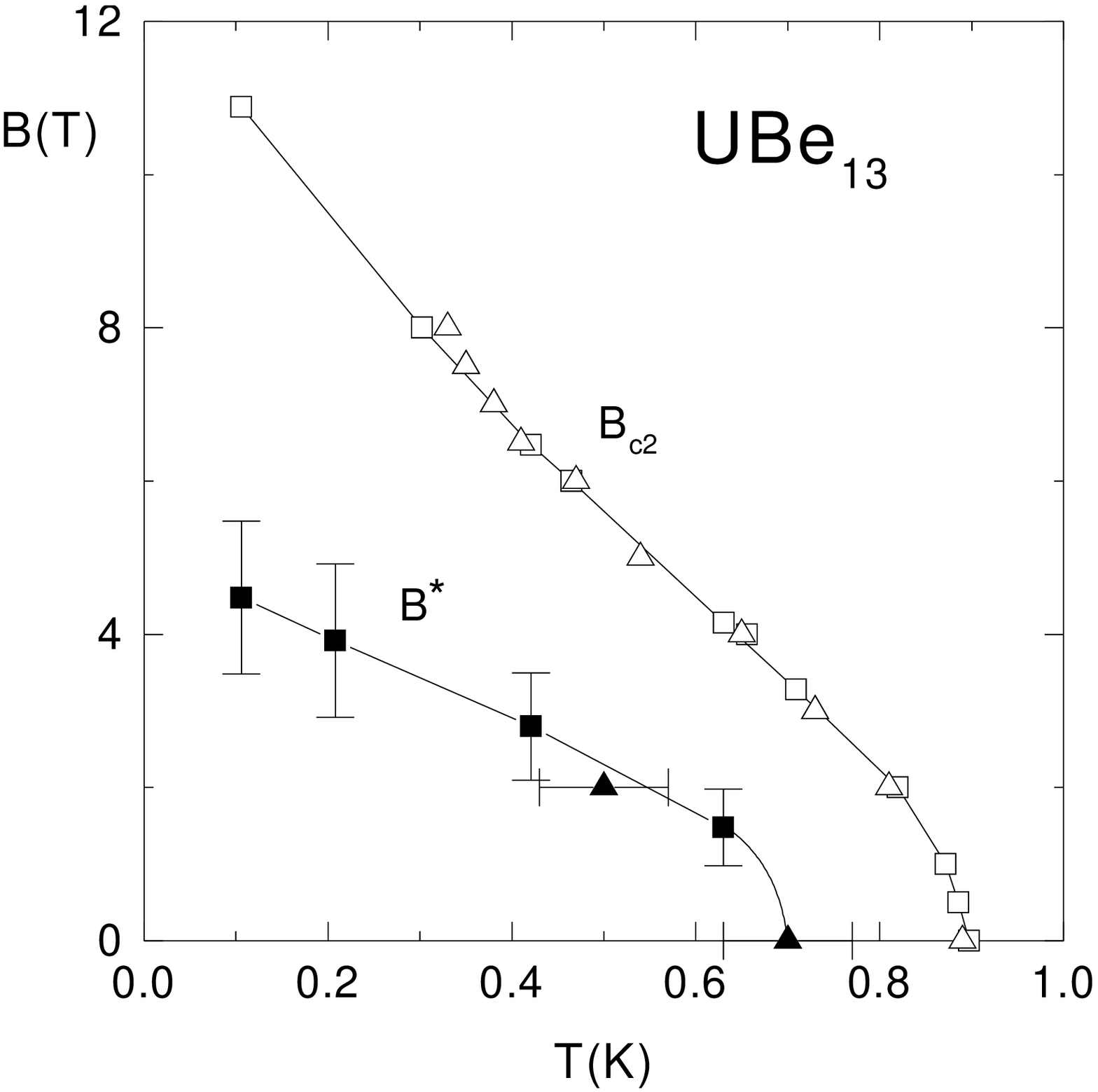}}\hfill
\includegraphics[clip,width=8cm,height=8cm]{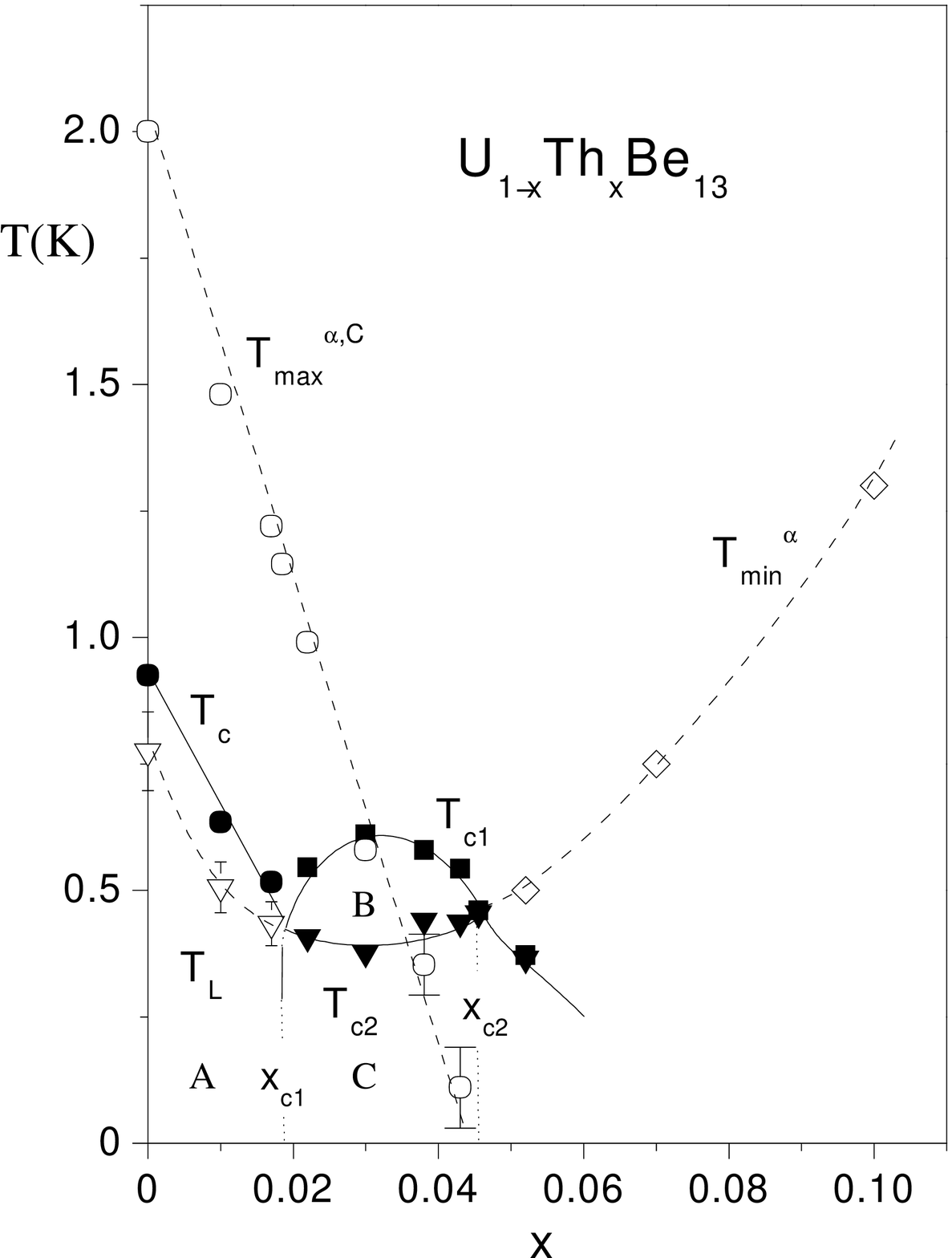}
\caption{Left panel: B-T phase diagram of \UBE~with additional anomaly
line B$^*$(T) obtained from specific heat (squares) and thermal
expansion (triangles) experiments. Right panel: SC phase diagram of
\UBT~. Full lines and symbols denote
boundaries of different thermodynamic phases, broken lines and open
symbols denote lines of anomalies. T$_L$ and T$^\alpha_{min}$ are
anomaly lines obtained from the minimum in thermal expansion
$\alpha$(T). T$_{max}^{\alpha,C}\equiv$ T$_m$ is the line of anomalies
from the maximum in C(T) and $\alpha$(T). A,B and C are the distinct
SC phases in the T$_c$-crossing
model. \protect\cite{Kromer00a,Kromer00b,Kromer02}} 
\label{fig:UBe13BTphase}
\end{figure}

The symmetry of the SC state in pure \UBE~is not known with any confidence,
but its anisotropic nature is suggested by power law behaviour of
specific heat, penetration depth and NMR relaxation and by 
tunneling results mentioned before. As usual from the
power law exponents no consenus has emerged whether the gap function
is characterised by point or line nodes and therefore we do not
discuss these results. A strong indication of
nonconventional SC is obtained from a giant ultrasonic absorption
anomaly immediately below T$_c$ \cite{Golding85,Mueller86} which was
attributed to damping by domain walls formed by a multicomponent SC
order parameter \cite{Sigrist91}. Also the B-T phase diagram
(fig.~\ref{fig:UBe13BTphase}) is quite anomalous, it exhibits a line of anomaly
B$^*$(T) deep in the SC regime which starts at T$_L$(x=0) = 0.7 K
signifying the onset of short range AF correlations. As
function of x they eventually develop into a long range SDW order
discussed below.
%
%
The SC phase diagram of thoriated \UBT~ (x $\leq$ 0.1) has required a
great effort to unravel and the task is still not finished. It
displays a bewildering richnes of different phases and lines of
anomalies which have been obtained mostly from specific heat,
resistivity and thermal expansion (magnetostriction) measurements
\cite{Kromer00b,Kromer02}.
The salient feature of the phase diagram is a nonmonotonic variation
of the T$_c$(x) curve with a cusp at x$_{c1}\approx$ 0.02 and
associated with it a splitting in two SC transitions T$_{c1}$(x) and
T$_{c2}$(x) in the
range x$_{c1}<$ x $<$ x$_{c2}$ with x$_{c2}\approx$ 0.045
\cite{Ott85}. This leads to three
distinct SC regions A,B and C in the x-T plane. When crossing from B to C,
H$_{c1}$ exhibits a sudden change in slope \cite{Rauchschwalbe87} and
small magnetic moments $\mu\simeq$ 10$^{-3}\mu_B$ seen in $\mu SR$
experiments \cite{Heffner90} appear which are absent below x$_{c1}$ (A)
and above x$_{c2}$. There are presently two distinct scenarios,
denoted by I and II, to explain these observations.\\

\noindent
(I) \emph{The T$_c$-crossing model}\\ 
The most widely accepted interpretation has been
given quite early in \cite{Sigrist89,Sigrist91} in terms of a
Ginzburg-Landau model involving only SC order parameters. The
fundamental assumption is that two SC phases with gap functions
belonging to different 
representations, e.g. $\Gamma_1$ and $\Gamma_5$ of the cubic symmetry
group O$_h$, have close transition temperatures T$_c$($\Gamma_1$,x)
and  T$_c$($\Gamma_5$,x) which cross at a critical concentration
x$_{c1}$. Below x$_{c1}$  T$_c$($\Gamma_5$,x) $>$ T$_c$($\Gamma_1$,x)
and the unconventional $\Gamma_5$ (A) phase which has gap nodes is
stable. Above x$_{c1}$ the inequality is reversed and the conventional
(fully symmetric) $\Gamma_1$ (B) phase is stable immediately below
T$_{c1}$ = T$_c$($\Gamma_1$,x). For lower temperatures (T $<$ T$_{c2}$(x))
higher order
mixing terms in the GL free energy functional stabilise a mixed
$\Gamma_1\bigotimes\Gamma_5$ (C) SC phase with a different H$_{c1}$,
hence the observed kink. The C- phase gap function may be non-unitary and
time reversal symmetry breaking, thus leading to condensate magnetic
moments induced around impurities and defects which allegedly have been observed
in the $\mu$SR experiments. For even larger concentrations enhanced
paibreaking or the reduction of the spin excitation energy scale T$_m$(x)
involved in pair formation leads to a decrease in
T$_c$($\Gamma_1$,x) = T$_{c1}$ and a second crossing point
appears. However a GL expansion around x$_{c1}$ cannot naturally
explain the second crossing.\\

\noindent
(II) \emph{The SC/SDW coexistence model}\\
An alternative scenario to explain the phase diagram in
fig.~\ref{fig:UBe13BTphase} (right panel) has
appeared as a result of investigations of the differential thermal expansion
$\alpha=l^{-1}\partial l/\partial T$
\cite{Kromer00b,Kromer02}. In this explanation the SC phases A and B are
assumed as identical unconventional SC phases as infered from the
almost equal pressure dependence away from the critical x$_{c1}$. In
phase C the SC order parameter is unchanged by a SDW modulation which appears
within the unconventional SC phase. This was previously already proposed
theoretically in
\cite{Kato87} within the context of a model calculation for
coexistence of a necessarily unconventional SC and a SDW phase. The
moments seen in
the C phase are then associated with the SDW. The main reason for
this alternative interpretation of the x-T phase diagram is the
discovery of a new line of anomaly T$_L$(x) in $\alpha$(T) for subcritical
x $<$ x$_{c1}$ which starts at a T$_L$(0) identical to the temperature
for which B$^*$(T) = 0 as seen in the B-T phase diagram. With growing x the
$\alpha$(T) anomaly becomes increasingly sharp and T$_L$(x) exactly
joins the T$_{c2}$ line at x$_{c2}$. This suggests to interpret
T$_L$(x) as the onset temperature of short range magnetic correlations
which turn into the long range SDW state at x$_{c1}$. The $\alpha$(T)
anomaly line is seen to continue even far above x$_{c2}$. The shape of the
$\alpha$(T) anomaly above  x$_{c1}$ is indeed familiar from other
magnetic transitions. It is also in accord with the
interpretation of a huge (Cr-type) anomaly in the ultrasonic
attenuation \cite{Batlogg85}. The SC/SDW coexistence scenario lacks however a natural
explanation for the observed kink in H$_{c1}$ at T$_{c2}$.

Recent flux-creep experiments \cite{Dumont02,Mota03} have 
further complicated the picture. A steep drop in the flux creep rate was
observed below T$_{c2}$ which may naturally be explained within the
T$_c$- crossing model because the nonunitary C-phase may have domain
walls which pin vortices very effectively
\cite{Sigrist99}. Using this effect for mapping out the C/A-phase
boundary it was found that contrary to the original view shown in
fig.~\ref{fig:UBe13BTphase} the C-phase boundary is not perpendicular
at x$_{c1}$ but
rather has a low temperature tail extending down to x = 0, i.e. the
drop in flux creep indicating a transition into the C-phase is even
seen in pure \UBE.

\subsection{Nodal superconductivity in \UPD~ mediated by magnetic excitons}
\label{ssect:UPd2Al3}

Since its discovery in 1991 by Geibel et al. \cite{Geibel91a} the extensive
experimental studies on this antiferromagnetically ordered  moderate HF
superconductor (see table~\ref{tab:UPd2Al3}) have
lead to two important conclusions: Firstly the 5f electrons have a dual
nature, partly localised and partly itinerant where the former carry
the magnetic moments and the latter become superconducting
\cite{Caspary93,Feyerherm94}. Secondly
there is now compelling evidence that the internal low energy
excitations of the localised 5f subsystem mediate an effective pairing
potential between the itinerant electrons. This leads to a
superconducting state which has an anisotropic gap
function with line nodes \cite{Sato01}. Moreover the virtual low energy
excitations of localised 5f electrons lead to a quasiparticle mass
enhancement
factor m$^*$/m$_b$ which is thought to be the origin of the normal HF
state of \UPD~\cite{Zwicknagl03}. Indeed this new mechanism gives
consistent T$_c$ and mass enhancement
factor within a strong coupling approach based on a dual 5f electron model 
Hamiltonian \cite{McHale04}. As discussed in sect.~\ref{sect:SCmech}, HF
superconductivity is assumed to exhibit a non-phononic SC
pairing mechanism leading to nodal gap functions  because the strong
on-site repulsion of quasi-particles prevents the (nearly) isotropic
SC state most favored by the electron-phonon mechanism. However, \UPD~
is the only case known so far where this theoretical conjecture has
actually been proven by experiment. This was achieved by complementary
INS \cite{Sato97,Sato97a,Bernhoeft98,Bernhoeft00} and quasiparticle
tunneling experiments \cite{Jourdan99} which, respectively, probe the
two aspects of the
dual nature of 5f electrons: (1) The internal singlet-singlet CEF
excitations ($\delta\simeq$ 6 meV) of localised 5f electrons 
which form
propagating magnetic exciton bands in the range of 1-8 meV due to
inter-site interactions. (2) The tunneling current which probes the
superconducting gap of itinerant 5f electrons in epitaxially grown
UPd$_2$Al$_3$-AlO$_x$-Pb tunneling devices. There is a considerable
interaction between localised and itinerant 5f electrons signified on
one hand  by the appearance of a resonance in the INS scattering
function associated with the SC gap. On the other hand, even more
compelling is the presence of typical 'strong coupling' signatures in the
tunneling DOS at about the magnetic exciton energy in the center
of the AF BZ as determined in
INS. This is a direct proof that the exchange of these magnetic bosons
mediates the SC pairing in \UPD. This important new mechanism is
distinctly different from both the electron-phonon and the spin fluctuation
mechanism (sect.\ref{sect:SCmech}).  

\subsubsection{Physical properties}

Further experimental evidence for the dual nature of 5f electrons
comes from susecptibility measurements \cite{Grauel92} above T$_c$ where
$\chi$(T) for field perpendicular to c shows a pronounced maximum
below T = 50 K, typical for the effect of CEF states which originate
from the 5f$^2$ configuration of the U$^{4+}$ ions. The presence of
localised states is also seen in Knight shift measurements
\cite{Feyerherm94} and optical experiments \cite{Dressel02}. There in
addition the formation of heavy itinerant quasiparticles below T$^*$
was concluded from an analysis of the Drude peak in the optical
conductivity. The low lying CEF states of localised 5f electrons are
of singlet-singlet (doublet) type with a splitting $\delta\simeq$ 6
meV as obtained from a fit to the overall magnetic
exciton dispersion \cite{Thalmeier02}. This means that the AF ordering
(table~\ref{tab:UPd2Al3}) found below T$_N$ is of the induced moment
type. The FM ordered ab planes are stacked along c with an AF wave
vector \bQ~ = (0,0,0.5) (r.l.u.) \cite{Krimmel92,Kita94} and an easy
axis [100]. Moment reorientation can only be observed for fields in
the ab plane and the phase diagram was determined in \cite{Kita94}.

The superconducting state below T$_c$ was investigated in numerous
experiments, but a definite conclusion on the symmetry of the pair
state has not been achieved yet. Upper critical field measurements
show a flattening of H$_{c2}$(T) for low temperatures \cite{Hessert97}
which is interpreted as the effect of Pauli limiting in a spin singlet pair
state. This hypothesis was further investigated by $^{27}$Al Knight shift
experiments \cite{Tou95} which shows a considerable reduction below
T$_c$ again in favor of singlet pairing, however, the interpretation is
not unambiguous due to a large local moment contributions to
K$_s$. The $^{27}$Al NMR relaxation rate T$_1^{-1}$ was found to
exhibit clear T$^3$ power law \cite{Tou95}, naively interpreted as
evidence for line nodes in the gap function $\Delta(\bk)$ leading to
a quasiparticle DOS $\sim |E|$. This was also suggested by $^{105}$Pd
NMR/NQR \cite{Matsuda97}. Conclusions from the low temperature
specific heat C(T) which should be $\sim$ T$^2$ for line nodes are
hampered by the difficult subtraction of the nuclear
contribution. A C(T) = $\gamma_0$T + aT$^3$ behaviour was found where
the residual value $\gamma_0$ scales with the width of the SC
transition \cite{Steglich96} that characterises sample quality. 
\begin{table}
\begin{center}
\begin{tabular}{c|c|c|c|c|c|c}
\hline
~ &$\gamma \bigl[\frac{mJ}{molK^2}\bigr]$ & 
T$_N$ [K] & $\mu$ [$\mu_B$] & T$_c$ [K] &
$\frac{\Delta C(T_c)}{\gamma T_c}$ [K] & $\frac{\Delta S(T_N)}{Rln2}$\\
\hline
\UPD     & 140 & 14.3  & 0.85  & 1.8  & 1.2     & 0.67\\
\UND     & 120 & 4.6   & 0.20  & 1.2  & 0.2-0.4 & 0.12 \\
\hline
\end{tabular}
\end{center}
\caption{Material parameters of \UPD~ and its isostructural sister
compound \UND.}
\label{tab:UPd2Al3}
\end{table}
The isostructural sister compound \UND~is also a superconductor with
coexisting incommensurate magnetic order
(table~\ref{tab:UPd2Al3}). However, contrary to \UPD~its 5f states have all
a delocalised character as is obvious from the much smaller moment,
the incommmensurate ordering wave vector and the small entropy release
$\Delta$S at T$_N$ which is only one sixth of the value in \UPD. Therefore,
superconductivity in this compound is thought to be mediated by spin
fluctuations of conduction electrons rather than localised 5f
excitations as in \UPD. However, as in UPt$_3$, a spin triplet gap
function has been
proposed for \UND~\cite{Ishida02} which according to NMR results
should have a line node. Note that this is in conflict with
the simple theory of isotropic AF spin fluctuations proposed for U-HF
compounds since they would rather prefer spin singlet pairing
(sect.~\ref{sect:SCmech}) \cite{Miyake86}. Also it suggests that  Blount's
theorem which predicts the absence of line nodes for strong spin-orbit
coupling cannot naively be applied to real U-HF
superconductors (sect.~\ref{sect:SCmech}).
Finally we note that from recent
H$_{c2}$-measurements on epitaxially grown thin films \cite{Jourdan04}
a considerable paramagnetic Pauli-limiting effect was deduced for both
a and c directions which led the authors to the opposite conclusion
that \UND~should also have a SC singlet pair state.

\subsubsection{Electronic structure, the dual model}

Within conventional LDA type band structure calculations for \UPD~
\cite{Sandratskii94,Knoepfle96,Inada99} there is no way to
treat the partly localised and partly itinerant character of 5f
electrons properly, all 5f orbitals are incorporated in the basis set for the
band states. While the FS topology corresponds reasonably to
dHvA results \cite{Inada95} the effective masses of the various sheets
are far too small. Even within a modified LSDA (local spin density
approximation) treatment including self interaction corrections
\cite{Petit03} the effective masses are still too small  by an overall
factor of ten. The dual model opens an attractive way to remove this
discrepancy \cite{Zwicknagl03}. The coupling of itinerant 5f- electrons
to the low lying discrete CEF excitations of the localised subsystem
offers an effective mechanism of mass enhancement
(sect.~\ref{sect:QPmech}). In this
approach the calculaction of heavy bands proceeds in three steps: 1)
5f orbitals in the jj-coupling limit (j = $\frac{5}{2}$) are used and
those with j$_z$ = $\pm\frac{5}{2}$,$\pm\frac{1}{2}$ are excluded from
the basis set of LDA bands which comprises only j$_z$ = $\pm\frac{3}{2}$
states. The band center is fixed to obain the correct 5f count. 2) The
multiplet structure of localised states 5f is calculated using the proper
intra-shell Coulomb interactions. The ground state is a doublet
$|J=4,J_z=\pm 3\ra$ which is further split by the CEF potential into
two singlets
\begin{eqnarray}
|\Gamma _{3,4}\rangle  = \frac{1}{\sqrt{2}}(|J=4;J_{z}=3\rangle 
\pm|J=4;J_{z}=-3\rangle )
\end{eqnarray}
The splitting energy $\delta\simeq$ 6 meV is taken as an empirical
parameter obtained from INS. 3) The scattering of localised band
states from CEF excitations leads to the mass renormalisation of the
former. In the simplest case, without including the dispersion
which turns the localised CEF transitions with energy $\delta$ into a
band of magnetic excitons and neglecting strong coupling effects the
mass enhancement would be given by
\begin{eqnarray}
\frac{m^*}{m_b}=1+2\frac{I^2N(\epsilon_F)}{\delta}
\end{eqnarray}
%
\begin{figure}
\raisebox{0.5cm}
{\includegraphics[width=85mm]{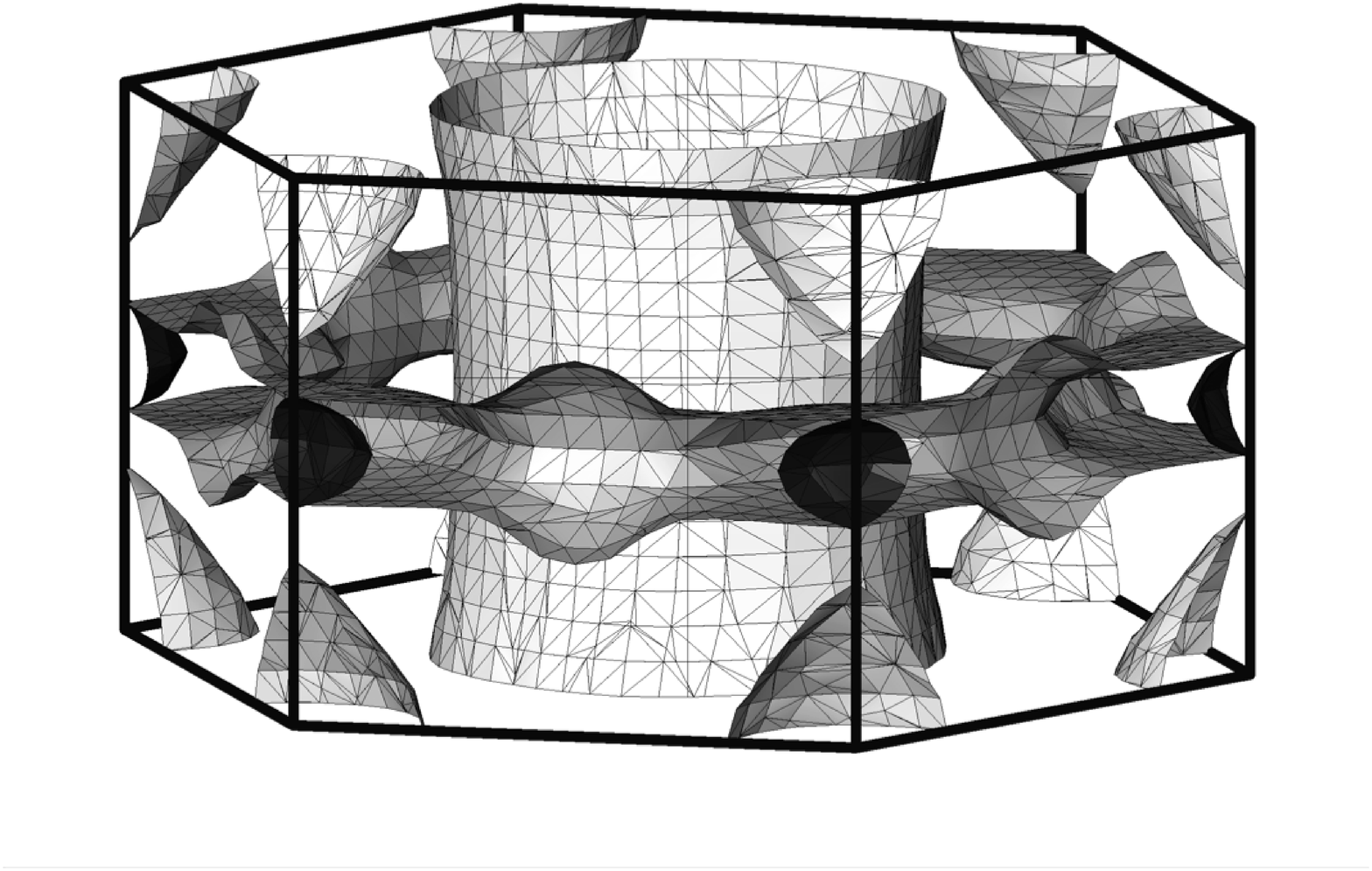}}\hfill
\includegraphics[width=75mm]{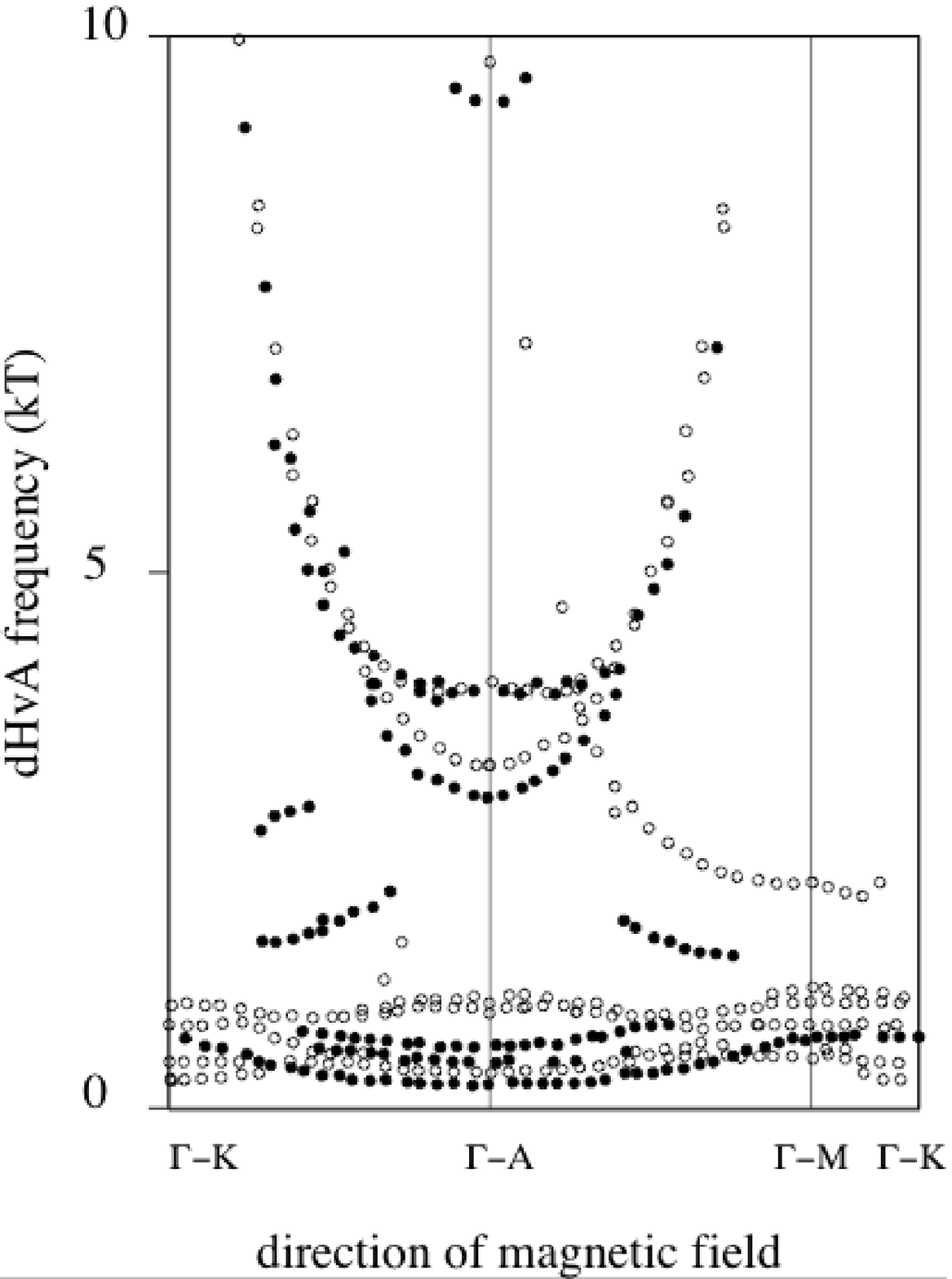}
\caption{Left panel: Fermi surface of \UPD~calculated within the dual
model \protect\cite{Zwicknagl03}. The main cylinder part has
a heavy mass with m$^*$ = 19 - 33 m and the torus (`crown') has m$^*$
= 65 m. Right panel: Comparison of experimental dHvA frequencies
(black symbols)
from \protect\cite{Inada99} and calculated frequencies (open symbols)
from the dual model \protect\cite{Zwicknagl03}. Large parabolas
correspond to the corrugated main FS cylinder. Small arc on top of
$\Gamma$-A  corresponds to the crown sheet.}
\label{FIGfermi}
\end{figure}
%
with I defined in eqs.~(\ref{HfirstVers}) and (\ref{eqHam}). A more
refined treatment for $\frac{m^*}{m_b}$ appopriate for \UPD~is described below. 
In this approach the different treatment of 5f orbitals with
j=$\frac{5}{2}$ and different j$_z$ cannot be explained on the single
particle level
because their hybridisation matrix elements are rather similar. The
amplification of the orbital dependence of hybridisation due to
many body effects has been proposed in \cite{Efremov04} and is described in
detail in sect.~\ref{sect:DUALtheory}.

Although the starting point is indeed very different from standard
LDA, with only a single delocalised 5f electron, the main FS sheets
are reproduced quite well as is obvious from fig.~\ref{FIGfermi}.
The most prominent FS sheet
has the form of a slightly corrugated cylinder oriented along c. It
has also a large mass enhancement of m$^*$/m $\simeq$ 33 (m = free
electron mass). In the following model discussion one may therefore
restrict to this main FS sheet.

\subsubsection{Mass enhancement and SC pairing due to magnetic excitons}

\begin{figure}
\raisebox{-0.1cm}
{\includegraphics[width=7cm,height=6.7cm]{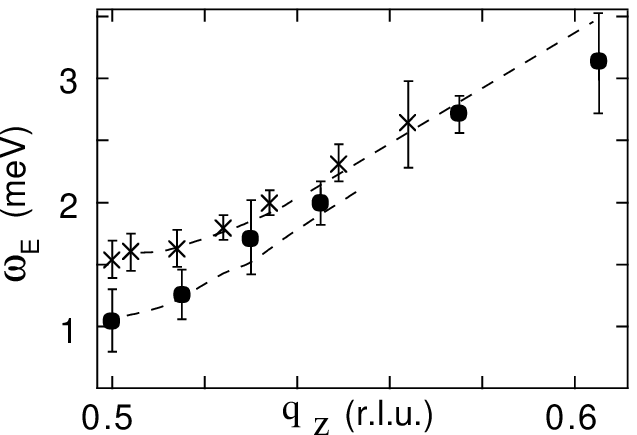}}\hfill
\includegraphics[width=7cm]{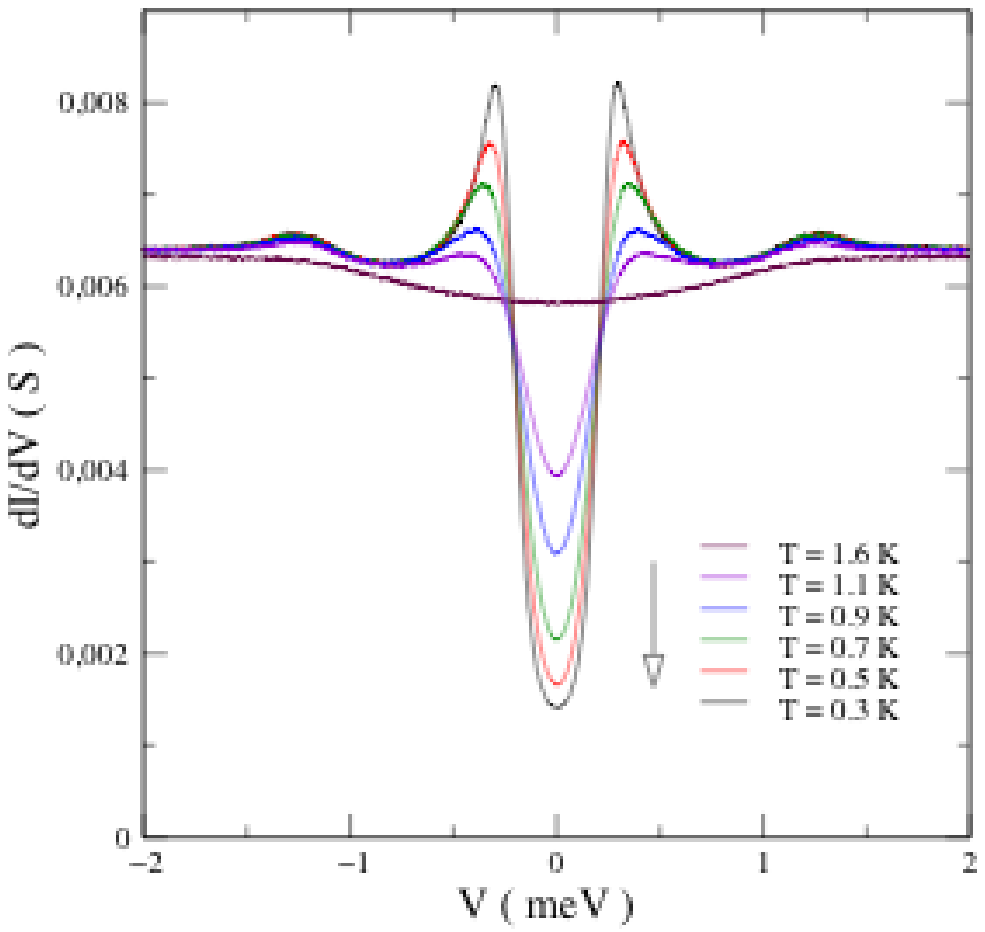}
\caption{Left panel: Magnetic exciton dispersion $\omega_E$(\bq) (\bq~=(0,0,q$_z$))
derived from low energy INS \protect\cite{Sato01} for temperatures
above (crosses) and below (circles) T$_c$. Right panel: Differential
conductivity from tunneling for various temperatures (the arrow
indicates the sequence of curves). The hump at 1
meV is a strong coupling signature of the magnetic exciton at \bQ~ =
(0,0,0.5) (cf. left panel) which mediates Cooper pairing
\protect\cite{Jourdan99}.}
\label{FIGdual}
\end{figure}

The great advantage of the dual model as compared to the pure LDA
approach is its natural explanation of 
the mass enhancement factor of m$^*$/m$_b$ $\simeq$ 10 (m$_b$ = LDA
band mass) which is due to the
contribution of virtual singlet-singlet CEF excitations to the
conduction electron self energy. Assuming the value of $\delta$ = 6 meV from
INS the proper mass enhancement is predicted without further
adjustable parameters. In addition to the mass enhancement the
interaction between the subsystems leads to induced AF magnetic order
\cite{Sato01,Thalmeier02} of localised 5f moments and superconductivity in
the itinerant part \cite{Sato01,Thalmeier02,McHale04}. In its most
rudimentary form a model Hamiltonian is given by eq.~(\ref{eqHam}). 
In \UPD~ the conduction electrons corresponding to the main FS
cylinder in fig.~\ref{FIGfermi} have only dispersion $\perp$ c with
$\epsilon_{\perp\bk}$ = $\epsilon_\perp$(\bk$_\perp$/k$_0$)$^2$.
The magnetic exciton dispersion $\omega_E$(\bq) which originates from the
$\Gamma_3\leftrightarrow\Gamma_4$ CEF transitions given by
eq.~(\ref{DI1})
is of central importance. For large T eq.~(\ref{DI1}) describes isolated
CEF transitions at an energy $\delta$ which develop a dispersion
at lower temperature due to an effective exchange J(\bq). In this RPA
expression a complete softening at the N\`eel temperature T$_N$ in
eq.~(\ref{CRINEEL}) for induced AF order is expected at the AF
ordering wave vector \bQ~= (0,0,$\frac{1}{2}$). Note that, unlike spin
waves the magnetic excitons which originate in singlet-singlet CEF
transitions may exist already above T$_N$ and their softening signifies
the approaching induced AF order.
The experimental investigation of magnetic excitons was undertaken in
many INS studies \cite{Mason97,Bernhoeft98,Sato01,Hiess04}. In the former the
overall dispersion up to the maximum $\simeq$ 8 meV was determined
and theoretically analysed in \cite{Thalmeier02}. The latter focused
on high resolution analysis of the low
energy excitations around the AF wave vector \bQ. The result is shown
in the left panel of fig.~\ref{FIGdual}. Indeed, contrary to the above simple
RPA singlet-singlet result a magnetic exciton gap of about 1 meV
appears. This may be
caused by the effect of higher magnetic CEF states whose contribution
to the staggered susceptibility leads to AF order slightly before
complete softening is achieved and also by self energy effects beyond
RPA.
%
%
The dispersion along c is much larger than in the hexagonal ab-
plane which may lead to a considerable DOS for the low
energy magnetic excitons around 1 meV. The spectral shape of magnetic
excitons is Lorentzian above T$_c$ and evolves into a double peak
structure where the lower has the appeareance of a sharp resonance
within the SC gap region \cite{Sato01,Hiess04}. This is a signature of the
strong residual interaction between itinerant (SC) electrons and the
excitations within the localised 5f subsystem. A model analysis of the
INS spectra at \bQ~leads to conclude that
$\Delta$/$\omega_E(\bQ)\simeq$ 1, i.e. SC gap amplitude and magnetic
exciton gap are nearly degenerate.

Complementary evidence for this interaction was found in the
quasiparticle tunneling spectra of c-axis oriented epitaxially grown
\UPD~ films
\cite{Jourdan99} which probe the SC electrons. The differential
conductivity dI/dV of the \UPD-AlO$_x$-Pb tunneling barrier is proportional
to the quasiparticle DOS N$_S(\omega)$. In the weak coupling BCS limit
it shows the monotonic square root singular behaviour above
$\Delta$(\bk$\parallel$ c). However,
retardation effects in the strong coupling limit for the effective
pairing interaction lead to characteristic signatures in dI/dV $\sim$
N$_S$($\omega$) at an energy which corresponds to typical energies of
the bosons which mediate the interaction. This is well know in the
case of strong-coupling electron-phonon superconductors. 
In their breakthrough tunneling experiment in \UPD~ Jourdan et
al. \cite{Jourdan99} have for the first time seen evidence of the boson
that mediates
superconductivity in a HF metal. Its signature can clearly be seen in
the tunneling spectra in the right panel of fig.~\ref{FIGdual} at an
energy slightly above 1 meV. This is far too small for characteristic
phonon frequencies, since the Debye energy is of the order 13
meV. Instead it corresponds directly to the magnetic exciton energy at
the AF vector \bQ~for T $<$ T$_c$ in the left panel. This strongly suggests
that magnetic excitons mediate SC pair formation in \UPD. 
The coupling between the itinerant heavy quasiparticles and the
magnetic exciton in the center of the AF BZ (q$_z$ = 0.5) shows substantial
retardation because the group velocity ($\partial\omega_E/\partial
q_z$) of the latter is much smaller than the Fermi velocity v$_F^*$ of
heavy quasiparticles due to the flatness of $\omega_E(q_z)$ over a
considerable part of the BZ along c$^*$ \cite{Hiess04}.
Within Eliashberg theory
the quasiparticle DOS can be explained by using a phenomenological
retarded potential centered at $\omega_E(\bQ)$, the analysis
\cite{Sato01} leads to the conclusion 2$\Delta$/T$_c$ = 5.6 in
agreement with the gap estimate above and also with NMR results \cite{Kyogaku93}.
Together the two complementary results shown in fig.~\ref{FIGdual}
present the first direct evidence for a non-phononic SC pairing
mechanism in a HF superconductor. The magnetic exciton-mediated
pairing identified here is distinctly different from both the
conventional electron-phonon mechanism and the common spin-fluctuation
mechanism (sect.~\ref{sect:SCmech}) which does not involve any
localised electron component.
\subsubsection{Symmetry of the SC gap function and Eliashberg theory}
\begin{figure}
\raisebox{-0.6cm}
{\includegraphics[clip,width=7.5cm,height=7.7cm]{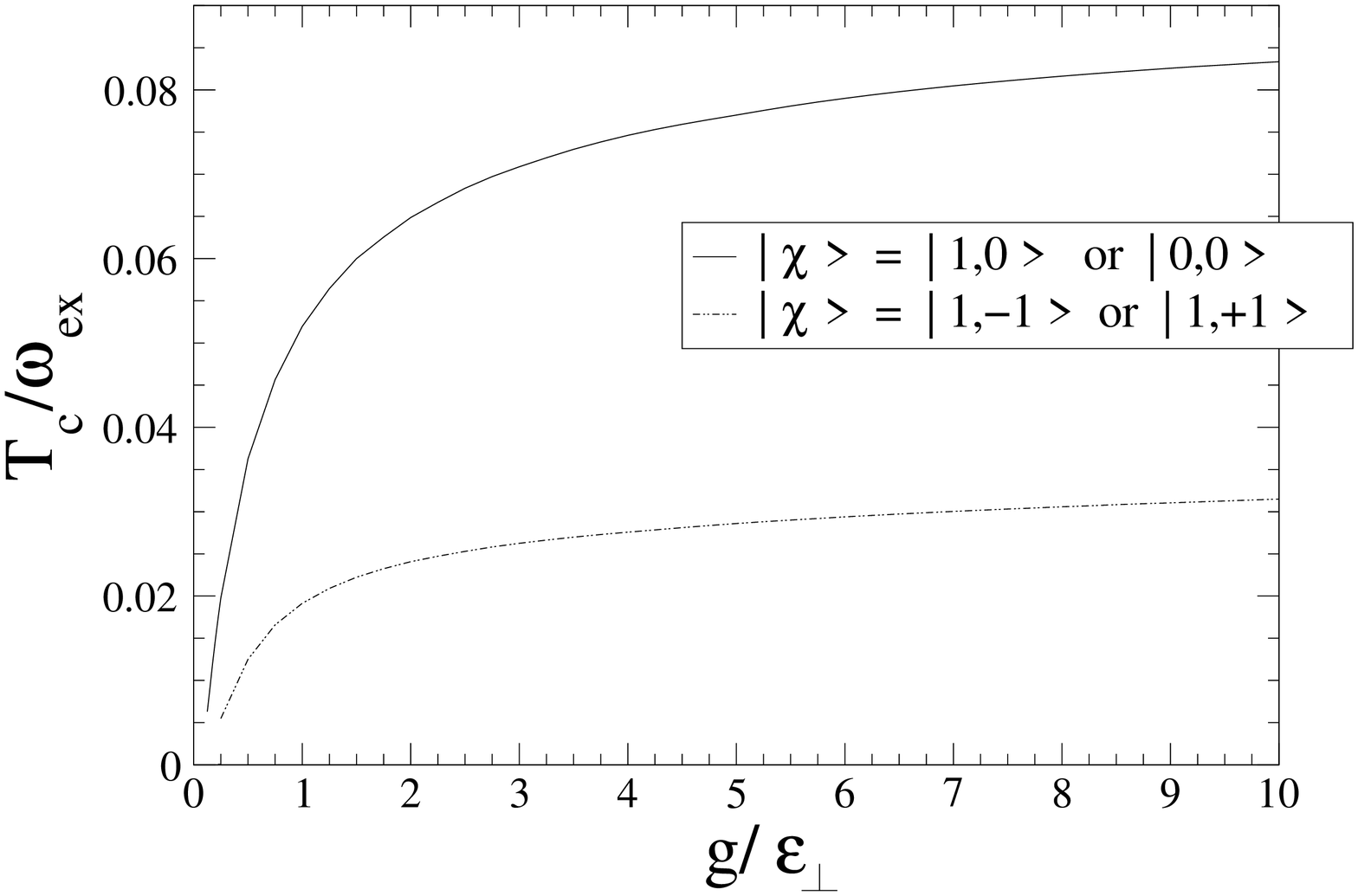}}\hfill
\includegraphics[clip,width=7.5cm]{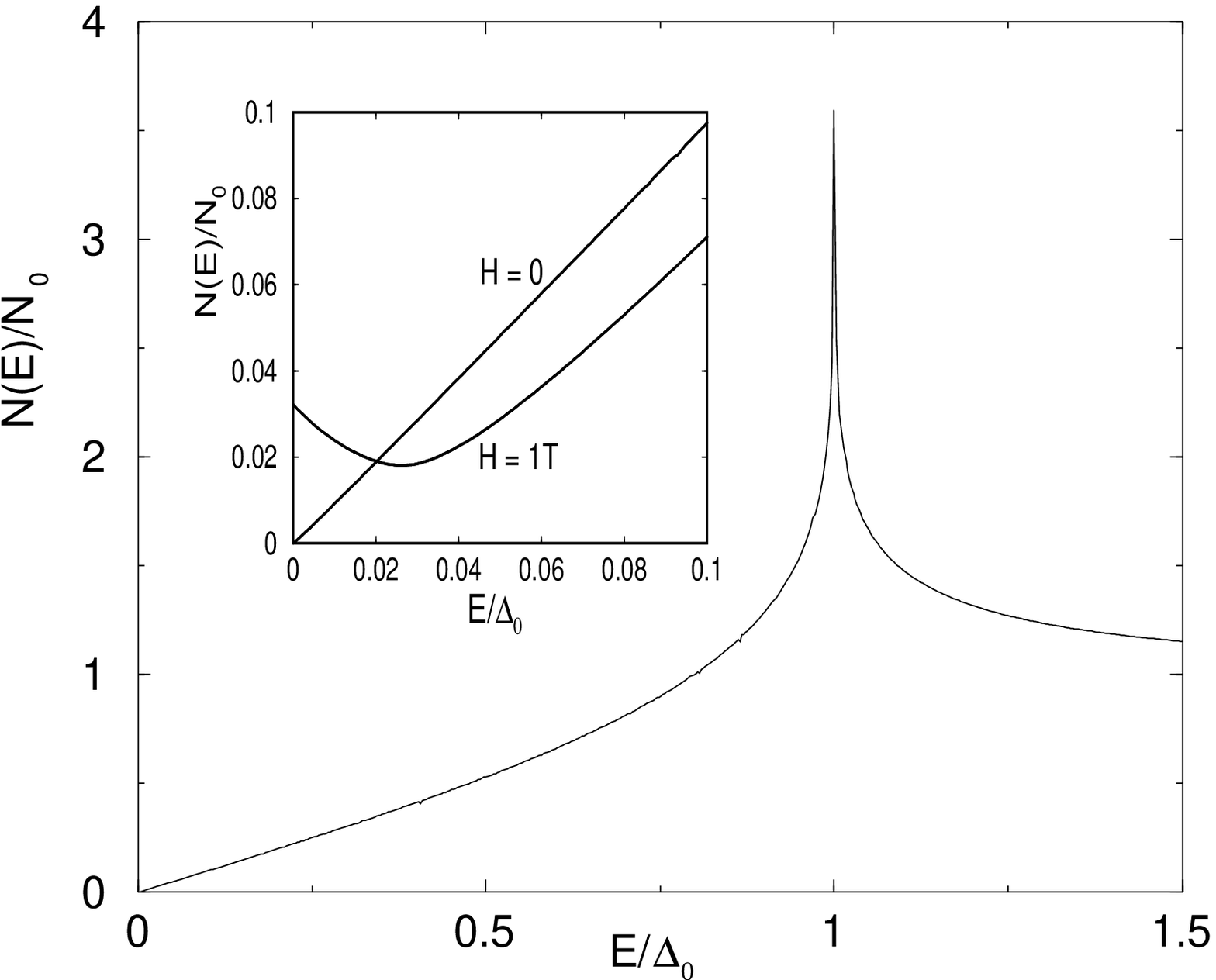}
\caption{Left panel: Critical temperature as function of the dimensionless
coupling constant $\hat{g}$ = g/$\epsilon_\perp$ = $\frac{I^2\Delta}{2}
\frac{N(\epsilon_F)}{\omega_{E}^{02}}$. Model parameters for
$\omega_E$(\bq) are
$\omega^0_{E}=0.01\epsilon_\perp$ ($\epsilon_\perp$= conduction band
width) and $\beta$ = 0.8. Full (dashed) curve: OSP (ESP)
SC states (see table~\ref{tab:gap}) \protect\cite{McHale04}. Right
panel: DOS in the A$_{1g}$
state of table~\ref{tab:gap} with N(E) $\sim$ E for E
$\ll\Delta_0$. Inset shows the DOS for field of 1T and
orientation $\parallel$ c ($\theta$ = 0). Finite N(0) is induced by
the Doppler shift of quasiparticles and contributes to specific heat
and transport \protect\cite{Thalmeier04}.}
\label{FIGTc}
\end{figure}

The question of the symmetry of the gap function $\Delta(\bk)$ is most
difficult to resolve
in unconventional superconductors, and perhaps with the exception of
UPt$_3$ (see \cite{Joynt02,Thalmeier03b} for a review)  has not been
unambiguously achieved for any of the HF superconductors. And yet this
is very important because the nodal structure of the gap function and
the associated low temperature thermodynamic and transport
properties are directly determined by the symmetry class of the gap
function. It was suggested \cite{Bernhoeft00,Bernhoeft04} that the INS
double peak structure for T $\ll$ T$_c$ mentioned previously requires the
translational symmetry property 
$\Delta(\bk+\bQ)$ = -$\Delta(\bk)$. Furthermore Knight shift and
H$_{c2}$ results mentioned above have lead to the proposal of a
spin singlet gap with $\Delta(-\bk)$ = $\Delta(\bk)$. Together this
suggest a gap function with node lines $\Delta(\bk)$ = 0 perpendicular
to c at the AF Bragg planes $\bk =\pm\frac{1}{2}\bQ$ (A$_{1g}$ in
table~\ref{tab:gap}).
To investigate the consistency of the magnetic exciton model strong
coupling calculations based on the Hamiltonian in eq.~(\ref{eqHam}) have been
performed \cite{McHale04}. Such model calculations cannot predict reliably the
symmetry of the gap function, but they may decide which one is
the most favorable within a restricted class of possible gap functions
. In addition one may at the same time obtain the mass enhancement
(m$^*$/m$_b$) of
normal quasiparticles from the self energy and T$_c$ from the
Eliashberg equations. Both quantities are determined by the effective retarded
potential due to magnetic exchange given in eq.~(\ref{EFFEX}). For
simplicity we use an empirical form for $\omega_E$(\bq) with
a finite gap and no dispersion $\perp$ c, given by 
\begin{eqnarray}
\omega_E(q_z) &=&\omega_E^0[1 + \beta \cos ( q_z )]
\end{eqnarray}
The parameters $\omega_E^0$, $\alpha$ and $\beta$ (fig.~\ref{FIGTc}) are
chosen to describe the experimental dispersion along c obtained from
INS \cite{Mason97}.
%
%
The effective interaction of eq.~(\ref{EFFEX}) breaks spin rotational
symmetry in a maximal (Ising type) manner. As explained in
sect.~\ref{sect:SCmech} this leads to a classification into ESP and OSP pair
states with spin projection factors p = $\pm$1. The Eliashberg
equation for the gap function is then given by
\begin{equation}
\Lambda(T)\Delta(\bk,i\omega_n)  =  p \, \frac{T}{N} \sum_{\bk'\omega_n'}
V(\bk-\bk',i\omega_n-i\omega_n')
|G(\bk',i\omega_n')|^2\, \Delta(\bk',i\omega_n').
\label{gapEq}
\end{equation}
where G(\bk,i$\omega_{n}$) is the renormalised conduction electron
Green's function and T$_c$ is obtained from $\Lambda$(T$_c$) = 1.
The result for
the T$_c$ of various gap function candidates as function of the
dimensionless coupling parameter $\hat{g}$ are shown in
fig.~\ref{FIGTc}. Likewise the dependence of m$^*$/m$_b$ on $\hat{g}$ which is
roughly linear may be calculated. From dHvA results the enhancement factor for
the main FS cylinder in fig.~\ref{FIGfermi} is m$^*$/m$_b\simeq$ 10. This
would imply $\hat{g}$ = 2. The corresponding theoretical value T$_c$ =
2.9 K for the degenerate A$_{1g}$ and A$_{1u}$ states from
fig.~\ref{FIGTc} is a factor 1.6 larger than the
experimental value of 1.8 K. Given that the latter is reduced due to the
action of static AF order, this result shows that the magnetic exciton
mechanism gives both mass enhancement and a T$_c$ which are
consistent. This supports the previous empirical conjecture for this
new mechanism of HF superconductivity.

\begin{figure}
\raisebox{-0.5cm}
{\includegraphics[clip,width=7.5cm,height=7.0cm]{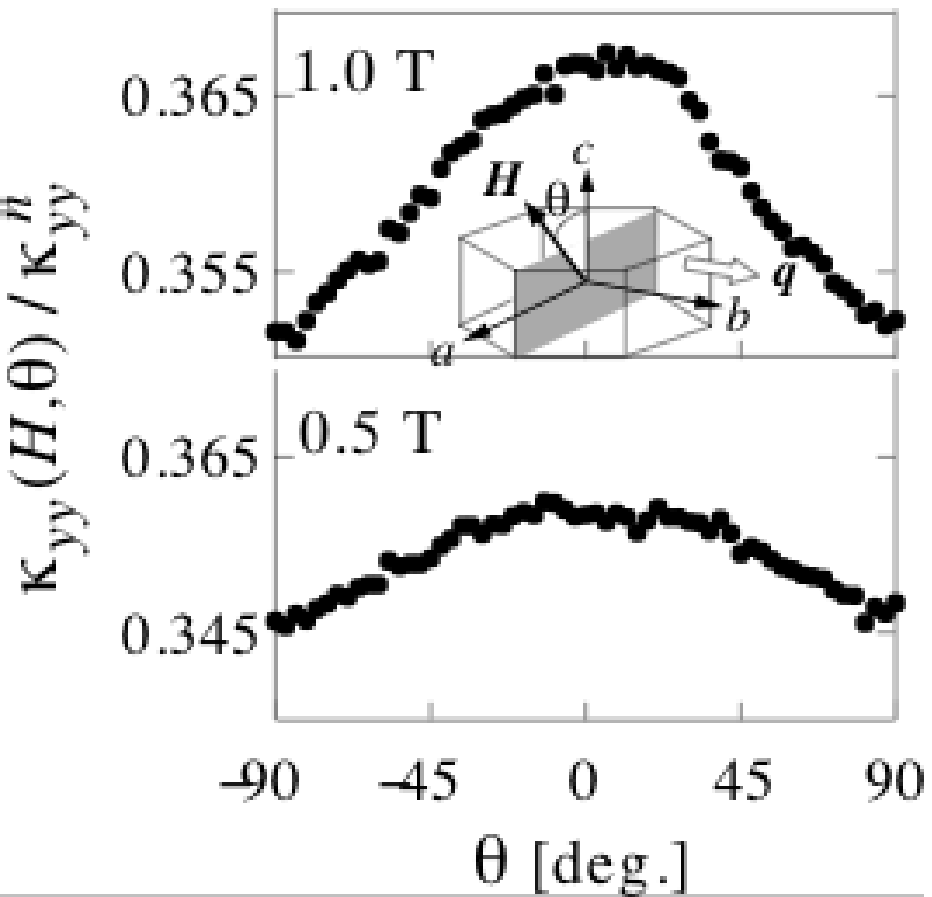}}
\hfill
\includegraphics[clip,width=7.5cm]{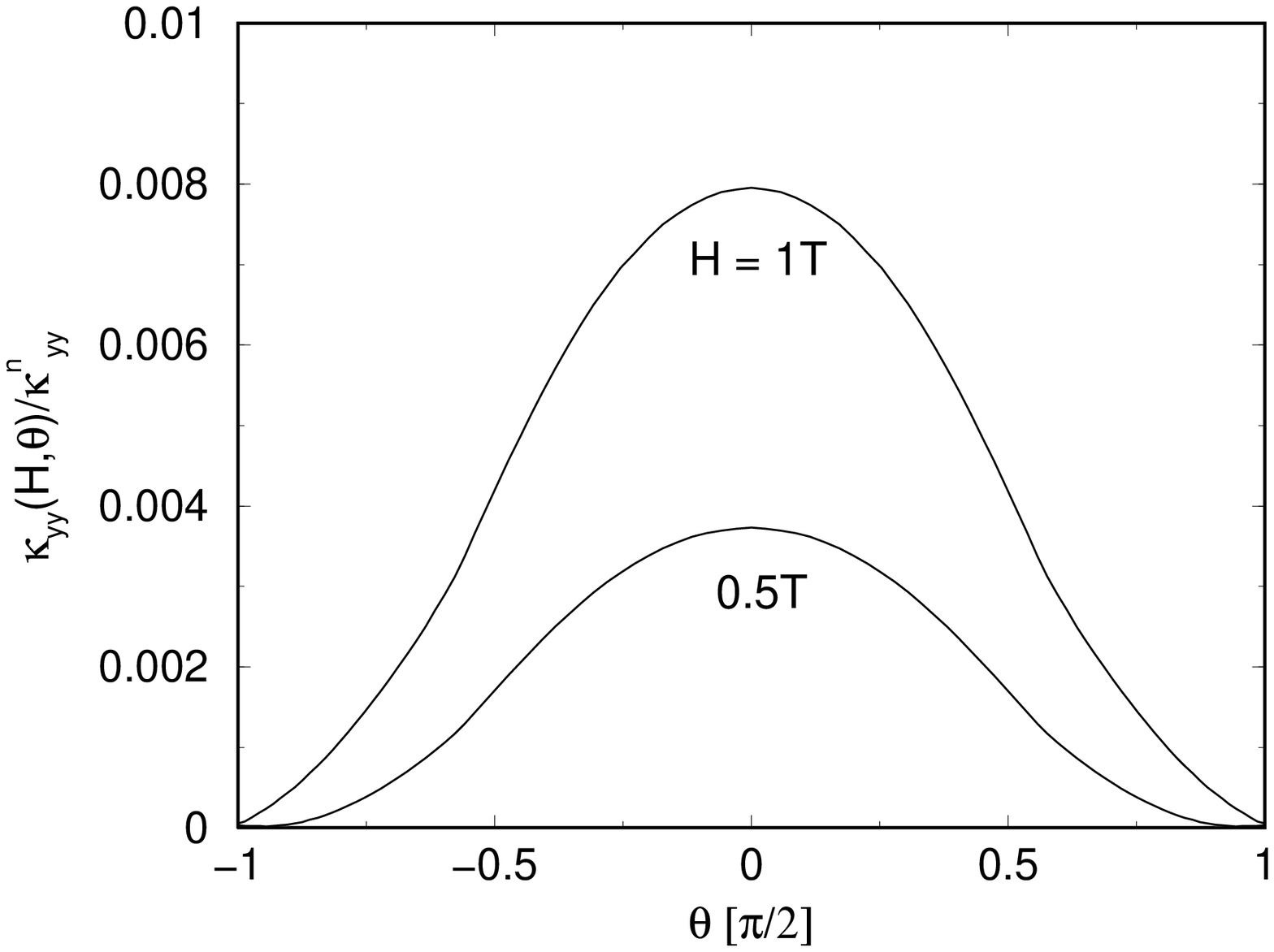}
\caption{Left panel: Angular variation of the normalised b-axis thermal
conductivity $\kappa_{yy}(\bH,\theta)$. Field is rotated in the
ac-plane $\perp$ to the heat current
\bq~\protect\cite{Watanabe04}. Right panel: Theoretical
calculation of $\kappa_{yy}(\bH,\theta)$ according to
eq.~(\ref{KAPPATH}) for the corrugated FS cylinder of \UPD~and the
A$_{1g}$ gap function. Twofold oscillations in $\theta$ proves the
existence of a node line in the ab plane \protect\cite{Thalmeier04}.}
\label{fig:KAPPOSC}
\end{figure}

The dispersion of $\omega_E$(\bk) and hence the \bk-dependence of the
interaction is strongest along c. Therefore one may restrict to SC
candidate states of the type $\Delta(\bk)$ = $\Delta$ $\Phi$(k$_z$) with a
few possible form factors  $\Phi$(k$_z$) given in table~\ref{tab:gap}.
It is seen in fig.~\ref{FIGTc} that the A$_{1g}$ and A$_{1u}$ OSP spin
pairing states have the highest and equal T$_c$'s. This degeneracy
between even and odd parity states is due to the Ising type anisotropy
of the effective interaction in spin space signified by their equal
spin projection factors p = -1 (table~\ref{tab:gap}).
It will be lifted by the
interaction of the SC order parameters with the background AF
order. The respective node lines for $\Phi$(k$_z$) would be at k$_z = \pm\pi/c$
(zone boundary) for  A$_{1g}$ and k$_z$ = 0 for  A$_{1u}$ (zone center).
Note that both states have S$_z$ = 0 and therefore both should
show Pauli limiting of H$_{c2}$ and a Knight shift reduction below
T$_c$. For the odd parity state this is due to the fact that the other
'triplet' components with S$_z=\pm 1$ are not stable, i.e. the \bd~-
vector is pinned along c. Note that although A$_{1g}$ with $\Delta(\bk)$ =
$\Delta_0\cos k_z$ has a node line, it is not an unconventional order
parameter in the strict sense since it has the full D$_{6h}$
symmetry. The quasiparticle DOS of the A$_{1g}$ state is shown in
fig.~\ref{FIGTc} is linear for E $\ll\Delta_0$ due to the node line at
k$_z=\pm\frac{\pi}{2}$. The inset shows that an external field
induces a finite residual DOS at E = 0 according to the Doppler shift
expression of eq.~(\ref{RESDOS}). The residual DOS depends on the field
direction relative to the nodal position. This leads to a polar field-angle
dependent residual $\gamma(\theta)$-coefficient and thermal
conductivity $\kappa_{ii}(\theta)$. There is no azimuthal
$\phi$-dependence due to cylindrical FS symmetry. The residual
$\kappa_{ii}$ normalised to the normal state $\kappa_n$ is given by
\begin{table}
\caption{Spin and orbital structure of the possible gap functions
which are solutions of the Eliashberg equations for the dual model of
UPd$_2$Al$_3$. The A$_{1g}$ OSP state is fully symmetric and its node line
($k_z=\pm\pi/c$) is not enforced by symmetry. \label{tab:gap}}
\vspace{0.5cm}
\begin{tabular}{llccl}
\hline
$p$ & $|\chi\rangle = |S,\,S_z\rangle$ & D$_{6h}$ repres. & spin
pairing & $\Phi(k_z)$ \\
\hline
-1 & $|0,0\rangle$ = $\frac{1}{\sqrt{2}}\left(
|\!\uparrow\downarrow\rangle -|\!\downarrow\uparrow\rangle\right)$ 
&$\Gamma_1^+$ (A$_{1g}$) & OSP & $\cos(ck_z)$ \\
-1 & $|1,0\rangle$ = $\frac{1}{\sqrt{2}} \left(|\!\uparrow\downarrow\rangle +
|\!\downarrow\uparrow\rangle \right)$
&$\Gamma_1^-$ (A$_{1u}$)  & OSP & $\sin(ck_z)$ \\
+1 & $|1,\pm1\rangle$ = $|\!\uparrow\uparrow\rangle,
|\!\downarrow\downarrow\rangle$ 
&$\Gamma_1^-$ (A$_{1u}$) & ESP & $\sin(2ck_z)$\\
\hline
\end{tabular}
\end{table}
\begin{eqnarray}
\kappa_{ii}(T,\theta)/\kappa_n &=& \frac{3}{4\pi^2}\frac{1}{T^3}
\int_0^\infty \frac{d\omega}{\cosh^2(\omega/2T)}
\hat{\tau}(\omega)\langle\hat{v}_{i\bk}^2\rangle_{FS}^{-1}
\langle\langle\hat{v}_{i\bk}^2
K(\omega,\hat{\bk}\hat{\br})\rangle\rangle_{FS,V}
\nonumber\\
K(\omega,\hat{\bk}\hat{\br}) &=& \frac{2}{\hat{\omega}}
[\hat{\omega}^2-\Delta(\hat{\bk})^2]^\frac{1}{2}
\Theta_H(\hat{\omega}^2-\Delta(\hat{\bk}^2))
\label{KAPPATH}
\end{eqnarray}
Here $\hat{\omega}=\omega-\bv_s(\br,\hat{\theta})\cdot\bk$ is the Doppler shifted
energy of the quasiparticle, \bk~and \br~are its momentum and its
position with respect to the vortex core oriented along
$\hat{\theta}$. The double average is performed over the FS and the
inter-vortex region and $\Theta_H$ is the Heaviside function. Furthermore
$\hat{\tau}(\omega)$ is the effective quasiparticle lifetime. 
As already mentioned in sect.~\ref{ssect:CeCoIn5} for \CCI~ angle-resolved
measurements of thermal conductivity give important information the
position of the node lines of \De~with respect to the crystal axes. This
is of great help in the determination of gap symmetry. One
should have H $\ll$ H$_{c2}$ to minimise the background effect of the
H$_{c2}$ anisotropy. 
These experiments, which were proposed in \cite{Thalmeier02}, have
recently been performed on \UPD~single crystals \cite{Watanabe04} in a
rotating field geometry with heat current perpendicular to the field (vortex)
direction to probe the itinerant quasiparticle contribution
(fig.~\ref{fig:KAPPOSC}). For 
fields rotated in the hexagonal ab-plane no oscillations have been
found excluding the presence of node lines perpendicular to
the ab plane as has been proposed in the model of \cite{Nishikawa02}. On the
other hand field rotation in the ac-plane clearly leads to
twofold oscillations for H $\ll$ H$_{c2}$ proving the existence of a
node line which is parallel to the ab-plane as is predicted by the
magnetic exciton model of SC in \UPD~\cite{McHale04}
(fig.~\ref{fig:KAPPOSC}). Because of the
cylindrical FS geometry the oscillations are qualitatively similar for
the three gap functions of table~\ref{tab:gap}. Thus thermal transport
measurements alone are in this case not able to fix the position of
the horizontal node line. However if one accepts that the Knight
shift results point to singlet (OSP) pairing, then the A$_{1g}$-gap
function in table~\ref{tab:gap} is the proper one for \UPD.
%
%

\section{Summary and Outlook}
\label{sect:summary}

The understanding of the heavy fermion state in Ce- and U-
intermetallics and its SC phases  has made great progress in recent
years. In the normal state it has
become clear that the origin of quasiparticle mass enhancement may be
profoundly different. It is described by the Kondo lattice mechanism in
4f-Ce compounds and by a dual (multiorbital) model of localised and
itinerant 5f electrons in U compounds. For Ce sytems the heavy electron
bands may be described by the renormalised band theory where the
single 4f electron is treated as itinerant with a resonant phase
shift. Within the dual model for U-compounds, two electrons are
treated as being localised in CEF states. The remaining 5f electron forms
itinerant conduction bands and its coupling to magnetic excitons,
i.e. propagating CEF excitations,
leads to the mass enhancement. These different approaches lead to
realistic Fermi surfaces in cases like \CCS, \UPT, \UPD~and the
mass enhancement may be incorporated naturally.   

The Kondo lattice mechanism in Ce-compounds implies the competition
between HF state and AF ordering as illustrated in the Doniach phase
diagram. When AF order is destroyed by variaton of a control
parameter, pressure or chemical substitution, the normal state
behaviour around the QCP shows distinct NFL behaviour. It is
characterised  by anomalous scaling exponents in the temperature and
field dependences of specific heat, resistivity, thermal expansion and
other properties. In Ce-compounds these NFL exponents may largely be
understood by
considering the dressing of quasiparticle with a spectrum of soft
magnetic spin fluctuations, this picture seems especially appropriate
for \CNG~but also for \CCS~and \CPS. Simultaneously effective
quasiparticle interactions mediated by soft spin
fluctuations favor SC pair formation as predicted by Eliashberg
theory. In fact in many Ce-compounds SC domes exist around the QCP and
frequently but not necessarily is associated with NFL behaviour.   

Mass enhancement and SC pairing are closely related. The
importance of virtual magnetic exciton exchange for SC pair formation
in HF-U compounds with dual 5f electron behaviour may therefore be
supposed. Indeed this novel mechanism has been identified in the model
compound \UPD~ by complementary tunneling and INS measurements. The
former exhibit a characteristic strong coupling feature at the energy
of the zone center magnetic exciton mode which identifies the latter
as the exchanged boson that mediates pairing. This observation is
quite unique for HF-compounds, it has not been achieved for the spin
fluctutation Ce-compounds which is explicable since the latter are
overdamped modes whereas magnetic excitons are propagating modes with
a real group velocity. Whether SC in other U-HF compounds may be
explained with a similar mechanism is a matter of debate. \UBE~is the
only SC U compound which possibly exhibits a magnetic QCP and may
therefore be more
similar to Ce-based SC, while \UPT~and \URU~where SC is embedded in a
magnetic (or hidden order) phase are more plausible candidates for the
magnetic exciton mechanism. Also the AF spin fluctuation model in its
simple versions strongly favors singlet pairing whereas a number of U-HF
compounds, notably \UPT, exhibit triplet pairing. The latter
is not disfavored in the magnetic exciton model due to an inherent
spin space anisotropy of the effective interaction.

The most important property of unconventional HF SC is the symmetry
class of the gap function. Progress in understanding these symmetries
for specific compounds has been
slow in the past since the traditional method, i.e. interpreting
'power law' behaviour of thermodynamic and transport quantities
etc. is ambiguous and often contradictory. Recently the situation has
improved considerably with the advent of angle-resolved specific heat
and magnetotransport measurements that can identify or restrict nodal
positions of the gap functions in \bk-space. Together with Knight
shift measurements they are the most powerful tool to investigate the
gap symmetry in ultra-pure stoichiometric samples. Up to now the SC
gap is known with some certainty only for three cases: f-wave triplet
(E$_{2u}$) in \UPT, nodal singlet (A$_{1g}$) in \UPD~and d-wave
singlet (B$_{1g}$ or B$_{2g}$ ) in \CCI.

An issue of permanent interest in HF SC is the coexistence/competition
with long range AF or SDW order away from the QCP. This is now fairly
well understood for the 'A-phase' SDW state in \CCS~and \CSG, both
experimentally and within theoretical models involving a
realistic FS geometry. A similar approach may be possible for Ce115
materials like \CRCI.

The vortex phase can give further insight into the SC state. For
\UPT~the appearance of three distinct SC regions in the B-T plane was
a sure sign of a multicomponent SC order parameter. More recently the
observation of a distinct SC vortex phase in the low temperature and
high field corner close to H$_{c2}$ for \CCI~has been interpreted as
the long-sought FFLO phase where SC pairs have finite momentum wich 
leads to a segmentation of vortices.

Many fundamental issues in HF materials, like the importance of
unconventional density wave states for NFL behaviour, QCP scenarios
with a local breakup of quasiparticles, the effect of
time reversal and inversion symmetry breaking in unconventional SC
states, the vortex and FFLO phase for nodal SC gap functions and many
others, are still incompletely understood. This guarantees that heavy fermion
physics will be a thriving field in the future.\\[1cm] 

\noindent
\emph{Acknowledgements}\\
The authors are grateful for collaboration and discussion with
numerous colleagues, especially T. Dahm, E. Faulhaber, P. Fulde,
P. Gegenwart, C. Geibel, K. Izawa, K. Maki,
Y. Matsuda, P. McHale, M. Neef, F. Pollmann N. Sato, B. Schmidt,
G. Varelogiannis, T. Watanabe, A. Yaresko, H.Q. Yuan and Q. Yuan. 
\nopagebreak
\begin{table}[hbtp]
{\em List of acronyms}\\[0.5cm]
\begin{tabular}{ll}
AF     &   antiferromagnet, antiferromagnetism, antiferromagnetic\\
AFBZ   &   antiferromagnetic Brillouin zone\\
ARPES  &   angle resolved photoemission spectroscopy\\
BIS    &   bremsstrahlung isochromat spectroscopy\\
BZ     &   Brillouin zone\\
CDW    &   charge density wave\\
CEF    &   crystalline electric field\\
CPT    &   cluster perturbation theory\\
dHvA   &   de Haas-van Alphen\\
DLRO   &   diagonal long range order\\
DOS    &   density of states\\
ESP    &   equal spin pairing\\
FFLO   &   Fulde-Ferrell-Larkin-Ovchinnikov\\
GL     &   Ginzburg-Landau\\
LFL    &   Landau-Fermi liquid\\
FLEX   &   fluctuation exchange\\
FM     &   ferromagnet, ferromagnetism, ferromagnetic\\
FS     &   Fermi surface\\
HF     &   heavy fermion\\
IC     &   incommensurate\\
INS    &   inelastic neutron scattering\\
LDA    &   local density approximation\\
LSDA   &   local spin density approximation\\
mf     &   mean field\\
NFL    &   non-Fermi liquid\\
n.n.   &   next neighbor\\ 
n.n.n. &   next nearest neighbor\\
NMR    &   nuclear magnetic resonance\\
NQR    &   nuclear quadrupole resonance\\
ODLRO  &   off-diagonal long range order\\
OSP    &   opposite spin pairing\\
PES    &   photoemission spectroscopy\\
QCP    &   quantum critical point\\
RPA    &   random phase approximation\\
RKKY   &   Ruderman-Kittel-Kasuya-Yoshida\\
SC     &   superconductor, superconductivity, superconducting\\
SIC-LDA&   self interaction corrected local density approximation\\
SDW    &   spin density wave\\
s.o.   &   spin orbit\\
TB     &   tight binding
\end{tabular}
\end{table}
\newpage

\bibliographystyle{camera-ready}
\bibliography{TReferencesSTB,ZReferencesSTB2,ZReferencesSTB3}

\end{document}